\documentclass[oneside]{article}
\usepackage[a4paper]{geometry}
\usepackage[latin1]{inputenc} 
\usepackage[T1]{fontenc} 
\usepackage{RR}
\usepackage{hyperref}
\usepackage{float}
\usepackage{algorithm}
\usepackage[noend]{algorithmic}
\usepackage[geometry]{ifsym}
\usepackage {ifsym}
\usepackage{amssymb}
\usepackage{yhmath}
\usepackage{amscd}
\usepackage{dsfont}
\usepackage{amsmath}
\usepackage{epsfig}
\usepackage{graphics}
\usepackage{psfrag}
\usepackage{rotating}
\usepackage{amsmath}
\usepackage{amsfonts}
\usepackage{url}
\usepackage{color}
\usepackage{float}
\usepackage{balance} 
\usepackage{IEEEtrantools}
\usepackage[toc,page]{appendix}

\newtheorem{theorem}{Theorem}
\newtheorem{definition}{Definition}

\newcommand{\bs}{\boldsymbol}

\newcommand{\ds}{\displaystyle}

\newcommand{\pr}[1]{\mathrm{Pr} \left(#1\right)}

\newcommand{\sfT}{\textsf{T}}

\newcommand{\Cldicnfb}{\mathcal{C} }

\RRNo{456}
\RRdate{January 2015}
\RRauthor{
Victor Quintero
 \and Samir M. Perlaza
\and Jean-Marie Gorce 
}

\authorhead{Quintero \& Perlaza \& Gorce}
\RRtitle{Capacit\'e du Canal Lin\'eaire D\'{e}terministe \`{a} Interf\'{e}rences avec R\'{e}troalimentation Degrad\'{e}e par Bruit Additif.}
\RRetitle{Noisy Channel-Output Feedback Capacity of the Linear Deterministic Interference Channel}
\titlehead{Noisy Channel-Output Feedback Capacity of the Linear Deterministic Interference Channel}
\RRnote{Victor Quintero, Samir M. Perlaza  and Jean-Marie Gorce are with the CITI Laboratory of the Institut National de Recherche en Informatique et en Automatique (INRIA), Universit\'{e} de Lyon, and Institut National de Sciences Apliqu\'ees (INSA) de Lyon. 6 Av. des Arts 69621 Villeurbanne, France. ($\lbrace$victor.quintero-florez, samir.perlaza, jean-marie.gorce$\rbrace$@inria.fr).

Victor Quintero is also with Universidad del Cauca, Popay\'{a}n, Colombia.

Samir M. Perlaza is also with the Department of Electrical Engineering at Princeton University, Princeton, NJ.

This research was supported in part by the European Commission under Marie Sklodowska-Curie Individual Fellowship No. 659316 (CYBERNETS); and the Administrative Department of Science, Technology, and Innovation of Colombia (Colciencias), fellowship No. 617-2013.

Parts of this work were presented at the IEEE International Workshop on Information Theory (ITW), Jeju Island, South Korea, October, 2015. This work was also submitted to the IEEE Transactions on Information Theory in November 10 2016.}

\RRresume{
Dans ce rapport, la r\'{e}gion de capacit\'{e} du canal lin\'{e}aire d\'{e}terministe \`{a} interf\'{e}rences avec r\'{e}troalimentation degrad\'{e}e entre les r\'{e}cepteurs et leurs \'{e}metteurs correspondants est caract\'{e}ris\'{e}e. 
Ce r\'{e}sultat permet l'identification de plusieurs scenarios asym\'{e}triques dans lesquels la r\'{e}troalimentation  dans un seul couple  r\'{e}cepteur-\'{e}metteur montre autant de b\'{e}n\'{e}fices que des r\'{e}troalimentations dans les deux couples r\'{e}cepteurs-\'{e}metteurs. 
Ces b\'{e}n\'{e}fices sont mis en \'{e}vidence par l'am\'{e}lioration des taux de transmission individuels et de leur somme par rapport aux cas o\`{u} il n'y a aucune r\'{e}troalimentation. 
D'autres scenarios montrent qu'une r\'{e}troalimentation dans un des couple \'{e}metteur-r\'{e}cepteur am\'eliore le taux individuel d'un des deux couples \'{e}metteurs-r\'{e}cepteurs. 
D'ailleurs, il existe d'autres scenarios o\`{u} l'utilisation d'un ou plusieurs liens de  r\'{e}troalimentation ne montre aucun b\'{e}n\'{e}fice ni pour les taux individuels ni pour leur somme.  Dans ces scenarios, cela montre que les r\'{e}gions de capacit\'{e} avec et sans  r\'{e}troalimentation sont identiques.  
}

\RRabstract{
In this technical report, the capacity region of the two-user linear deterministic (LD) interference channel with noisy output feedback (IC-NOF) is fully characterized.
This result allows the identification of several asymmetric scenarios in which implementing channel-output feedback in only one of the transmitter-receiver pairs is as beneficial as implementing it in both links, in terms of achievable individual rate and sum-rate improvements w.r.t. the case without feedback. In other scenarios, the use of channel-output feedback in any of the transmitter-receiver pairs benefits only one of the two pairs in terms of achievable individual rate improvements or simply, it turns out to be useless, i.e., the capacity regions with and without feedback turn out to be identical even in the full absence of noise in the feedback links.
 }
\RRmotcle{R\'{e}gion de Capacit\'{e}, Mod\`{e}le lin\'{e}aire d\'{e}terministe, canal \`{a} interf\'{e}rences, r\'{e}troalimentation degrad\'{e}e.}
\RRkeyword{Capacity, Linear Deterministic Interference Channel, Noisy Channel-Output Feedback}
\RRprojet{Socrate}  
\RCGrenoble 
\begin{document}

\makeRT 

\clearpage
\tableofcontents
\clearpage

\section{Notation}

Throughout this technical report, sets are denoted with uppercase calligraphic letters, e.g. $\mathcal{X}$. Random variables are denoted by uppercase letters, e.g., $X$. The realizations and the set of events from which the random variable $X$ takes values are respectively denoted by  $x$ and $\mathcal{X}$. The probability distribution of $X$ over the set $\mathcal{X}$ is denoted $P_{X}$. Whenever a second random variable $Y$ is involved, $P_{X \, Y}$ and $P_{Y|X}$ denote respectively the joint probability distribution of $(X, Y)$ and the conditional probability distribution of $Y$ given $X$. Let $N$ be a fixed natural number. An $N$-dimensional vector of random variables is denoted by ${\bf X} = (X_{1}, X_{2}, ..., X_{N})^\sfT$ and a corresponding realization is denoted by ${\bf x}= (x_{1}, x_{2}, ..., x_{N})^\sfT \in \mathcal{X}^{N}$.  Given ${\bf X} = (X_{1}, X_{2}, ..., X_{N})^\sfT$ and $(a,b) \in \mathds{N}^2$, with $a < b \leqslant N$,  the $(b-a+1)$-dimensional vector of random variables formed by the components $a$ to $b$ of $\bs{X}$ is denoted by ${{\bf X}_{(a:b)} = (X_a, X_{a+1}, \ldots, X_b)^\sfT}$.  The notation $(\cdot)^+$ denotes the positive part operator, i.e., $(\cdot)^+ = \max(\cdot, 0)$ and $\mathbb{E}_{X}[ \cdot ]$ denotes the expectation with respect to the distribution of the random variable $X$. The logarithm function $\log$ is assumed to be base $2$.
\section{Problem Formulation} \label{SecLDICNOF}

\begin{figure}[t!]
 \centerline{\epsfig{figure=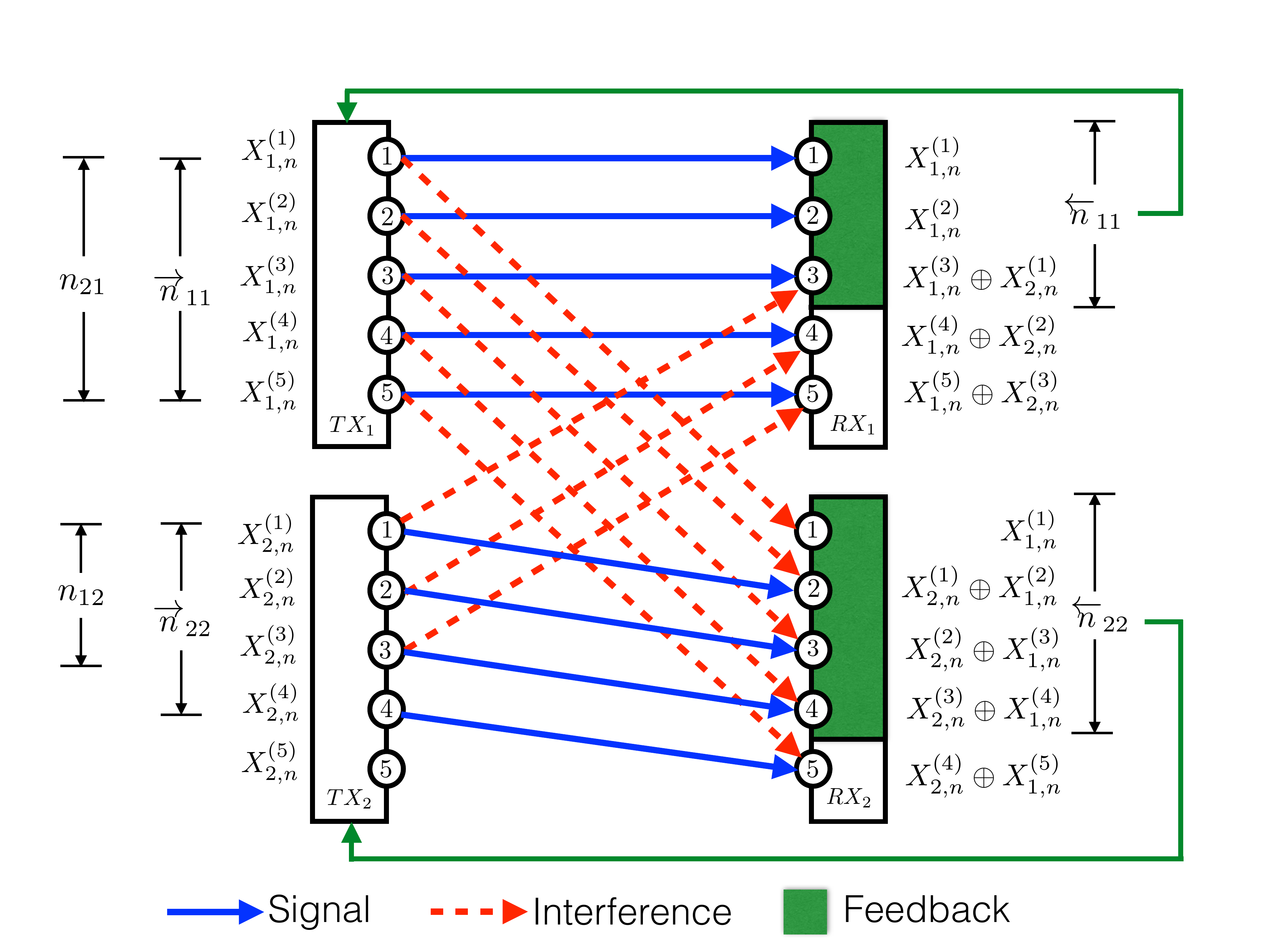,width=0.8\textwidth}}
\caption{Two-user linear deterministic interference channel with noisy channel-output feedback at channel use $n$.} 
  \label{FigLD-IC-NOF}
\end{figure}

Consider the two-user linear deterministic interference channel with noisy channel-output feedback (LD-IC-NOF) described in Figure~\ref{FigLD-IC-NOF}. 
For all $i \in \lbrace 1, 2 \rbrace$, with $j\in \lbrace 1, 2 \rbrace\setminus \lbrace i \rbrace$, the number of bit-pipes between transmitter $i$ and its corresponding intended receiver  is denoted by $\overrightarrow{n}_{ii}$; the number of bit-pipes between transmitter $i$ and its corresponding non-intended receiver is denoted by $n_{ji}$; and the number of bit-pipes between receiver $i$ and its corresponding transmitter is denoted by  $\overleftarrow{n}_{ii}$. These six integer non-negative parameters fully describe the LD-IC-NOF in Figure~\ref{FigLD-IC-NOF}.

\noindent
At transmitter $i$, the channel-input $\bs{X}_{i,n}$ at channel use $n$, with $n \in \lbrace 1, 2,  \ldots, N \rbrace$, is a $q$-dimensional binary vector ${\bs{X}_{i,n} = \left(X_{i,n}^{(1)}, X_{i,n}^{(2)}, \ldots, X_{i,n}^{(q)}\right)^{\sfT}}$, with  
\begin{equation}\label{Eqq}
q=\ds\max \left(\overrightarrow{n}_{11}, \overrightarrow{n}_{22}, n_{12}, n_{21}\right),
\end{equation}
and $N$ the block-length. 
At receiver $i$, the channel-output $\overrightarrow{\bs{Y}}_{i,n}$ at channel use $n$ is also a $q$-dimensional binary vector ${\overrightarrow{\bs{Y}}_{i,n} = \left(\overrightarrow{Y}_{i,n}^{(1)}, \overrightarrow{Y}_{i,n}^{(2)}, \ldots, \overrightarrow{Y}_{i,n}^{(q)}\right)^{\sfT}}$. 
The input-output relation during channel use $n$ is given by 
\begin{IEEEeqnarray}{lcl}
\label{EqLDICsignals}
\overrightarrow{\bs{Y}}_{i,n}&=& \bs{S}^{q - \overrightarrow{n}_{ii}} \bs{X}_{i,n} + \bs{S}^{q - n_{ij}} \bs{X}_{j,n},
\end{IEEEeqnarray}
and the feedback signal $\bs{\overleftarrow{Y}}_{i,n}$ available at transmitter $i$ at the end of channel use $n$ satisfies:
\begin{IEEEeqnarray}{lcl}\label{EqLDICsignalsc}
\left(\left(0, \ldots, 0\right), \bs{\overleftarrow{Y}}_{i,n}^{\sfT} \right)^{\sfT} &=& \bs{S}^{\left(\max(\overrightarrow{n}_{ii},n_{ij})-\overleftarrow{n}_{ii}\right)^+} \, \bs{\overrightarrow{Y}}_{i,n-d},
\end{IEEEeqnarray}
 where $d$ is a finite delay, additions and multiplications are defined over the binary field, and $\bs{S}$ is a $q\times q$ lower shift matrix of the form:
\begin{align}
\bs{S}=
\left[
  \begin{array}{cccccc}
    0 & 0 &0 & \cdots & 0 \\
    1 & 0 & 0 &\cdots & 0\\
    0 & 1 & 0 &\cdots  & \vdots\\
    \vdots & \ddots & \ddots  & \ddots & 0\\
    0 & \cdots & 0 & 1& 0
  \end{array}
\right].
\end{align}

\noindent
The dimension of the vector $\left(0,  \ldots,  0\right)$ in \eqref{EqLDICsignalsc} is $q  -  \min \big(\overleftarrow{n}_{ii}, \max(\overrightarrow{n}_{ii}, n_{ij}) \big)$ and the vector $\bs{\overleftarrow{Y}}_{i,n}$ represents the $\min \big(\overleftarrow{n}_{ii}, \max(\overrightarrow{n}_{ii},n_{ij}) \big)$ least significant bits of $\bs{S}^{\left(\max(\overrightarrow{n}_{ii},n_{ij})-\overleftarrow{n}_{ii}\right)^+} \, \bs{\overrightarrow{Y}}_{i,n-d}$.

\noindent
Without any loss of generality, the feedback delay is assumed to be equal to $1$ channel use, i.e., $d=1$. 
Transmitter $i$ sends the message index $W_i$ by sending the codeword $\bs{X}_{i}=\left(\bs{X}_{i,1}, \bs{X}_{i,2}, \ldots, \bs{X}_{i,N}\right) \in \mathcal{X}_{i}^{q \times N}$. 
The encoder of transmitter $i$ can be modeled as a set of deterministic mappings $f_i^{(1)}$, $f_i^{(2)}, \ldots$, $f_i^{(N)}$, with $f_i^{(1)}: \mathcal{W}_i \rightarrow \lbrace 0, 1 \rbrace^{q}$ and for all $n \in \lbrace 2, 3, \ldots, N\rbrace$, $f_i^{(n)}: \mathcal{W}_i  \times \lbrace 0, 1 \rbrace^{q(n-1)} \rightarrow \lbrace 0, 1 \rbrace^{q}$, such that 
\begin{IEEEeqnarray}{lcl}\label{EqEnconderf}
\bs{X}_{i,1} &=& f_i^{(1)}\big(W_i \big) \mbox{ and }\\
\bs{X}_{i,n} &=& f_i^{(n)}\big(W_i, \bs{\overleftarrow{Y}}_{i,1}, \bs{\overleftarrow{Y}}_{i,2}, \ldots, \bs{\overleftarrow{Y}}_{i,n-1} \big).
\end{IEEEeqnarray}
Let $T\in \mathds{N}$ be fixed. Assume that during a given communication, $T$ blocks are transmitted. Hence, the decoder of receiver $i$ is defined by a deterministic function $\psi_i: \lbrace 0,1 \rbrace^{q \times N \times T} \rightarrow \mathcal{W}_i^{T}$.
At the end of the communication, receiver $i$ uses the sequence $\left(\overrightarrow{\bs{Y}}_{i,1}, \overrightarrow{\bs{Y}}_{i,2}, \ldots, \overrightarrow{\bs{Y}}_{i,N \,T}\right)$ to obtain an estimate of the message indices:

\begin{IEEEeqnarray}{rcl}
\label{Eqdecoder}
\left(\widehat{W}_i^{(1)}, \widehat{W}_i^{(2)}, \ldots, \widehat{W}_i^{(T)}\right) &=& \psi_i \left(\overrightarrow{\bs{Y}}_{i,1}, \overrightarrow{\bs{Y}}_{i,2}, \ldots, \overrightarrow{\bs{Y}}_{i,N \, T}\right), \quad
\end{IEEEeqnarray} 
where $\widehat{W}_i^{(t)}$ is an estimate of the message index sent during block $t \in \lbrace 1, 2, \ldots, T \rbrace$.
The decoding error probability in the two-user G-IC-NOF during block $t$, denoted by $P_{e}^{(t)}(N)$, is given by   
\begin{IEEEeqnarray}{rcl}
\label{EqDecErrorProb}
P_{e}^{(t)} ( N ) &=&   \max   \Bigg(  \pr{ \widehat{W_1}^{(t)}   \neq   W_1^{(t)}  }  ,  \pr{ \widehat{W_2}^{(t)}   \neq   W_2^{(t)} }\Bigg). 
\end{IEEEeqnarray}

\noindent
The definition of an achievable rate pair $(R_1,R_2) \in \mathds{R}_+^{2}$ is given below. 
\begin{definition}[Achievable Rate Pairs]\label{DefAchievableRatePairs}\emph{
A rate pair $(R_1,R_2) \in \mathds{R}_+^{2}$ is achievable if there exists at least one pair of codebooks $\mathcal{X}_1^{N}$ and $\mathcal{X}_2^{N}$ with codewords of length $N$, and the corresponding encoding functions $f_1^{(1)}, f_1^{(2)}, \ldots,f_1^{(N)}$ and $f_2^{(1)}, f_2^{(2)}, \ldots, f_2^{(N)}$ such that the decoding error probability $P_{e}^{(t)}(N)$ can be made arbitrarily small by letting the block-length $N$ grow to infinity, for all blocks $t \in \lbrace 1, 2,  \ldots, T \rbrace$.
 }
\end{definition}

\noindent
The following section determines the set of all the rate pairs $(R_1,R_2)$ that are achievable in the LD-IC-NOF with parameters $\overrightarrow{n}_{11}$, $\overrightarrow{n}_{22}$, $n_{12}$, $n_{21}$, $\overleftarrow{n}_{11}$ and $\overleftarrow{n}_{22}$.

\section{Main Results}

Denote by $\Cldicnfb(\overrightarrow{n}_{11}, \overrightarrow{n}_{22}, n_{12}, n_{21}$, $ \overleftarrow{n}_{11} , \overleftarrow{n}_{22})$ the capacity region of the LD-IC-NOF with parameters $\overrightarrow{n}_{11}$, $\overrightarrow{n}_{22}$, $n_{12}$, $n_{21}$, $\overleftarrow{n}_{11}$, and $\overleftarrow{n}_{22}$. Theorem~\ref{TheoremANFBLDMCap} fully characterizes this capacity region. 

\begin{theorem}\label{TheoremANFBLDMCap} \emph{
The capacity region $\Cldicnfb(\overrightarrow{n}_{11}, \overrightarrow{n}_{22}, n_{12}, n_{21}, \overleftarrow{n}_{11} , \overleftarrow{n}_{22})$ of the two-user LD-IC-NOF is the set of non-negative rate pairs $(R_1,R_2)$ that satisfy for all $i \in \lbrace 1, 2 \rbrace$, with $j\in\lbrace 1, 2 \rbrace\setminus\lbrace i \rbrace$:
\begin{subequations}
\begin{IEEEeqnarray}{lcl}
\label{EqRiV2}
R_{i}& \leqslant & \min\left(\max\left(\overrightarrow{n}_{ii},n_{ji}\right),\max\left(\overrightarrow{n}_{ii},n_{ij}\right)\right), \\
\label{EqRi-2-V2}
R_i& \leqslant &\min\left(\max\left(\overrightarrow{n}_{ii},n_{ji}\right),\max\left(\overrightarrow{n}_{ii},\overleftarrow{n}_{jj}-\left(\overrightarrow{n}_{jj}-n_{ji}\right)^+\right)\right),\\
\label{EqRi+Rj-1-V2}
R_1+R_2  &\leqslant& \min \left(\max\left(\overrightarrow{n}_{22},n_{12}\right)+\left(\overrightarrow{n}_{11}-n_{12}\right)^+, \max\left(\overrightarrow{n}_{11},n_{21}\right)+\left(\overrightarrow{n}_{22}-n_{21}\right)^+\right), \\
\nonumber
R_1+R_2 &\leqslant& \max\Big(\left(\overrightarrow{n}_{11}-{n}_{12} \right)^+, n_{21}, \overrightarrow{n}_{11}-\left(\max\left(\overrightarrow{n}_{11},n_{12}\right)-\overleftarrow{n}_{11}\right)^+\Big)\\
  \label{EqRi+Rj-2-V2}
 & &+\max\Big(\left(\overrightarrow{n}_{22}-{n}_{21} \right)^+, n_{12}, \overrightarrow{n}_{22}-\left(\max\left(\overrightarrow{n}_{22},n_{21}\right)-\overleftarrow{n}_{22}\right)^+\Big),\\
\nonumber
2R_i+R_j  &\leqslant& \max\left(\overrightarrow{n}_{ii},{n}_{ji} \right)+\left(\overrightarrow{n}_{ii}-{n}_{ij} \right)^+\\
\label{Eq2Ri+Rj-V2}
& & +\max\Big(\left(\overrightarrow{n}_{jj}-{n}_{ji} \right)^+, n_{ij}, \overrightarrow{n}_{jj}-\left(\max\left(\overrightarrow{n}_{jj},{n}_{ji} \right)-\overleftarrow{n}_{jj}\right)^+\Big).
\end{IEEEeqnarray}
\end{subequations}
}
\end{theorem} 
The proof of Theorem~\ref{TheoremANFBLDMCap} is divided into two parts. The first part describes the achievable region and is presented in Appendix~\ref{AppAch-IC-NOF}. The second part describes the converse region and is presented in  Appendix~\ref{App-C-LD-IC-NOF}. 

\noindent
Theorem~\ref{TheoremANFBLDMCap} generalizes previous results regarding the capacity region of the LD-IC with channel-output feedback. For instance, when $\overleftarrow{n}_{11} = 0$ and $\overleftarrow{n}_{22} = 0$, Theorem~\ref{TheoremANFBLDMCap} describes the capacity region of the LD-IC without feedback (Lemma $4$ in \cite{Bresler-ETT-2008}); when  $\overleftarrow{n}_{11} \geqslant \max\left( \overrightarrow{n}_{11}, n_{12} \right)$ and  $\overleftarrow{n}_{22} \geqslant \max\left( \overrightarrow{n}_{22}, n_{21} \right)$, Theorem~\ref{TheoremANFBLDMCap} describes the capacity region of the LD-IC with perfect channel output feedback (Corollary $1$ in \cite{Suh-TIT-2011}); when  $\overrightarrow{n}_{11}=\overrightarrow{n}_{22}$, $ n_{12}=n_{21}$ and $\overleftarrow{n}_{11}=\overleftarrow{n}_{22}$,  Theorem~\ref{TheoremANFBLDMCap} describes the capacity region of the symmetric LD-IC with noisy channel output feedback (Theorem $1$ in \cite{SyQuoc-TIT-2015} and Theorem $4.1$, case $1001$ in \cite{Sahai-TIT-2013}); and when $\overrightarrow{n}_{11}=\overrightarrow{n}_{22}$, $ n_{12}=n_{21}$, $\overleftarrow{n}_{ii}\geqslant \max\left( \overrightarrow{n}_{ii}, n_{ij} \right)$ and $\overleftarrow{n}_{jj}=0$, with $i \in \lbrace1, 2 \rbrace$ and $j \in \lbrace 1, 2 \rbrace\setminus\lbrace i \rbrace$, Theorem~\ref{TheoremANFBLDMCap} describes the capacity region of the symmetric LD-IC with only one perfect channel output feedback (Theorem $4.1$, cases $1000$ and $0001$ in \cite{Sahai-TIT-2013}).

\subsection*{Comments on the Achievability Scheme}
The achievable region is obtained using a coding scheme that combines classical tools such as rate splitting, superposition coding, and backward decoding. This coding scheme is described in Appendix~\ref{AppAch-IC-NOF}. In the following, an intuitive description of this coding scheme is presented.
Let the message index sent by transmitter $i$ during the $t$-th block be denoted by $W_i^{(t)} \in \lbrace1, 2,  \ldots, 2^{N R_i}\rbrace$. Following a rate-splitting argument, assume that $W_i^{(t)}$ is represented by three subindices $(W_{i,C1}^{(t)}, W_{i,C2}^{(t)}, W_{i,P}^{(t)}) \in \lbrace 1, 2,  \ldots, 2^{NR_{i,C1}} \rbrace \times \lbrace 1, 2,  \ldots, 2^{NR_{i,C2}} \rbrace \times \lbrace 1, 2, \ldots, 2^{NR_{i,P}} \rbrace$, where $R_{i,C1} + R_{i,C2}+R_{i,P} = R_{i}$.
The codeword generation from $(W_{i,C1}^{(t)}, W_{i,C2}^{(t)}, W_{i,P}^{(t)})$ follows a four-level superposition coding scheme.
The index  $W_{i,C1}^{(t-1)}$ is assumed to be decoded at transmitter $j$ via the feedback link of transmitter-receiver pair $j$ at the end of the transmission of block $t-1$. Therefore, at the beginning of block $t$, each transmitter possesses the knowledge of the indices $W_{1,C1}^{(t-1)}$ and $W_{2,C1}^{(t-1)}$. In the case of the first block $t = 1$, the indices  $W_{1,C1}^{(0)}$ and $W_{2,C1}^{(0)}$ correspond to two indices assumed to be known by all transmitters and receivers.
Using these indices both transmitters are able to identify the same codeword in the first code-layer. This first code-layer is a sub-codebook of $2^{N(R_{1,C1} + R_{2,C1})}$ codewords (see Figure~\ref{FigSuperpos}). Denote by $\bs{u}\left(W_{1,C1}^{(t-1)},W_{2,C1}^{(t-1)}\right)$ the corresponding codeword in the first code-layer.  
The second codeword is chosen by transmitter $i$ using  $W_{i,C1}^{(t)}$ from the second code-layer, which is a sub-codebook of $2^{N\,R_{i,C1}}$ codewords corresponding at $\bs{u}\left(W_{1,C1}^{(t-1)},W_{2,C1}^{(t-1)}\right)$ as shown in Figure~\ref{FigSuperpos}. Denote by $\bs{u}_i\left(W_{1,C1}^{(t-1)},W_{2,C1}^{(t-1)},W_{i,C1}^{(t)}\right)$ the corresponding codeword in the second code-layer.  
The third codeword is chosen by transmitter $i$ using  $W_{i,C2}^{(t)}$ from the third code-layer, which is a sub-codebook of $2^{N\,R_{i,C2}}$ codewords corresponding at $\bs{u}_i\left(W_{1,C1}^{(t-1)},W_{2,C1}^{(t-1)},W_{t,C1}^{(t)}\right)$ as shown in Figure~\ref{FigSuperpos}. Denote by $\bs{v}_i\left(W_{1,C1}^{(t-1)},W_{2,C1}^{(t-1)},W_{i,C1}^{(t)},W_{i,C2}^{(t)}\right)$ the corresponding codeword in the third code-layer.  
The fourth codeword is chosen by transmitter $i$ using  $W_{i,P}^{(t)}$ from the fourth code-layer, which is a sub-codebook of $2^{N\,R_{i,P}}$ codewords corresponding at $\bs{v}_i\left(W_{1,C1}^{(t-1)},W_{2,C1}^{(t-1)},W_{i,C1}^{(t)},W_{i,C2}^{(t)}\right)$ as shown in Figure~\ref{FigSuperpos}. Denote by $\bs{x}_{i,P}\left(W_{1,C1}^{(t-1)},W_{2,C1}^{(t-1)},W_{i,C1}^{(t)},W_{i,C2}^{(t)},W_{i,P}^{(t)}\right)$ the corresponding codeword in the fourth code-layer.   
Finally, the generation of the codeword $\bs{x}_i = \left( \bs{x}_{i,1}, \bs{x}_{i,2}, \ldots, \bs{x}_{i,N}\right) \in \mathcal{X}_{i}^{N}$ during block $t \in \lbrace 1, 2,  \ldots, T \rbrace$ is a simple concatenation of the codewords $\bs{u}_i\Big(W_{1,C1}^{(t-1)}$, $W_{2,C1}^{(t-1)}$, $W_{i,C1}^{(t)}\Big)$, $\bs{v}_i\Big(W_{1,C1}^{(t-1)},W_{2,C1}^{(t-1)},W_{i,C1}^{(t)},W_{i,C2}^{(t)}\Big)$ and $\bs{x}_{i,P}\Big(W_{1,C1}^{(t-1)}$, $W_{2,C1}^{(t-1)}$, $W_{i,C1}^{(t)}$, $W_{i,C2}^{(t)}$, $W_{i,P}^{(t)}\Big)$, i.e., ${\bs{x}_i = \left( \bs{u}_i^\sfT, \bs{v}_i^\sfT, \bs{x}_{i,P}^\sfT\right)^\sfT}$, where the message indices have been dropped for ease of notation. 

\noindent
The intuition to build this code structure follows from the identification of three types of bit-pipes that start at transmitter $i$: 
$(a)$ The set of bit-pipes that are observed by receiver $j$ but not necessarily by receiver $i$ and are above the (feedback) noise level; 
$(b)$ The set of bit-pipes that are observed by receiver $j$ but not necessarily by receiver $i$ and are below the (feedback) noise level; and
$(c)$ The set of bit-pipes that are exclusively observed by receiver $i$.
The first set of bit-pipes can be used to convey message index $W_{i,C1}^{(t)}$ from transmitter $i$ to receiver $j$ and to transmitter $j$ during block $t$. 
The second set of bit-pipes can be used to convey message index $W_{i,C2}^{(t)}$ from transmitter $i$ to receiver $j$ and not to transmitter $j$ during block $t$.  
The third set of bit-pipes can be used to convey message index $W_{i,P}^{(t)}$ from transmitter $i$ to receiver $i$ during block $t$. 

\noindent
These three types of bit-pipes justify the three code-layers super-posed over a common layer, which is justified by the fact that feedback allows both transmitters to decode part of the message sent by each other. 
The decoder follows a classical backward decoding scheme.
This coding/decoding scheme is described in Appendix~\ref{AppAch-IC-NOF}. 

\noindent
Other achievable schemes, as reported in \cite{SyQuoc-TIT-2015}, can also be obtained as special cases of the more general scheme presented in \cite{Tuninetti-ISIT-2007}. However, in this more general case, the resulting code for the IC-NOF counts with a handful of unnecessary superposing code-layers, which complicates the error probability analysis.

\subsection*{Comments on the Converse Region}

The outer bounds \eqref{EqRiV2} and \eqref{EqRi+Rj-1-V2} are  cut-set bounds and were first reported in \cite{Bresler-ETT-2008} for the case without feedback. These outer bounds are still useful in the case of perfect channel-output feedback \cite{Suh-TIT-2011}. 
The outer bounds  \eqref{EqRi-2-V2}, \eqref{EqRi+Rj-2-V2} and \eqref{Eq2Ri+Rj-V2} are new and generalize those presented in \cite{SyQuoc-TIT-2015} for the symmetric case. 
These new outer-bounds were obtained using genie-aided models. 
 A complete proof of \eqref{EqRi-2-V2} is presented in Appendix~\ref{App-C-LD-IC-NOF}.

 \subsection*{Discussion}

This section provides a set of examples in which particular scenarios are highlighted to show that channel-output feedback can be strongly beneficial for enlarging the capacity region of the two-user LD-IC. However, these benefits strongly depend on the noise present in the feedback link. This section also highlights other examples in which channel-output feedback does not bring any benefit in terms of the capacity region. These benefits are given in terms of the following metrics: $(a)$ individual rate improvements $\Delta_1$ and $\Delta_2$; and $(b)$ sum-rate improvement $\Sigma$.  \\

\noindent
In order to formally define $\Delta_1$, $\Delta_2$ and $\Sigma$, consider an LD-IC-NOF with parameters $\overrightarrow{n}_{11}$, $\overrightarrow{n}_{22}$, $n_{12}$, $n_{21}$, $\overleftarrow{n}_{11}$ and $\overleftarrow{n}_{22}$.
The maximum improvement $\Delta_i(\overrightarrow{n}_{11}, \overrightarrow{n}_{22}, n_{12}, n_{21},\overleftarrow{n}_{11} , \overleftarrow{n}_{22})$ of the individual rate $R_i$ due to the effect of channel-output feedback with respect to the case without feedback is
\begin{IEEEeqnarray}{lcl}
\Delta_{i}(\overrightarrow{n}_{11}, \overrightarrow{n}_{22}, n_{12}, n_{21}, \overleftarrow{n}_{11} , \overleftarrow{n}_{22} )  = 
& \max_{R_{j} > 0} & \left \lbrace \sup_{(R_i, R_j) \in \Cldicnfb_{1}} \lbrace R_{i} \rbrace - \sup_{(R_i^{\dagger}, R_j)  \in \Cldicnfb_{2}} \lbrace R_i^{\dagger} \rbrace \right \rbrace,
\end{IEEEeqnarray}
and the maximum sum rate improvement $\Sigma(\overrightarrow{n}_{11}, \overrightarrow{n}_{22}, n_{12}, n_{21},\overleftarrow{n}_{11} , \overleftarrow{n}_{22})$ with respect to the case without feedback is

\begin{IEEEeqnarray}{lcl}
\Sigma(\overrightarrow{n}_{11}, \overrightarrow{n}_{22}, n_{12}, n_{21}, \overleftarrow{n}_{11}, \overleftarrow{n}_{22} )  = 
& & \sup_{ (R_1, R_2) \in \Cldicnfb_{1}} \Bigg \lbrace R_{1}+R_{2} \Bigg \rbrace - \sup_{ (R_1^{\dagger}, R_2^{\dagger})  \in \Cldicnfb_{2}} \Bigg \lbrace R_1^{\dagger}+R_2^{\dagger} \Bigg \rbrace,
\end{IEEEeqnarray}
where $\Cldicnfb_{1} = \Cldicnfb(\overrightarrow{n}_{11}, \overrightarrow{n}_{22}, n_{12}, n_{21},\overleftarrow{n}_{11} , \overleftarrow{n}_{22}) $ and $\Cldicnfb_{2} = \Cldicnfb(\overrightarrow{n}_{11}, \overrightarrow{n}_{22}, n_{12}, n_{21},0,0)$ are the capacity region with noisy channel-output feedback and without feedback, respectively. The following describes particular scenarios that highlight some interesting observations.

\subsubsection*{Example 1: only one channel-output feedback link allows simultaneous maximum improvement of both individual rates}\label{SecExample1}
\begin{figure}[t!]
\centerline{\epsfig{figure=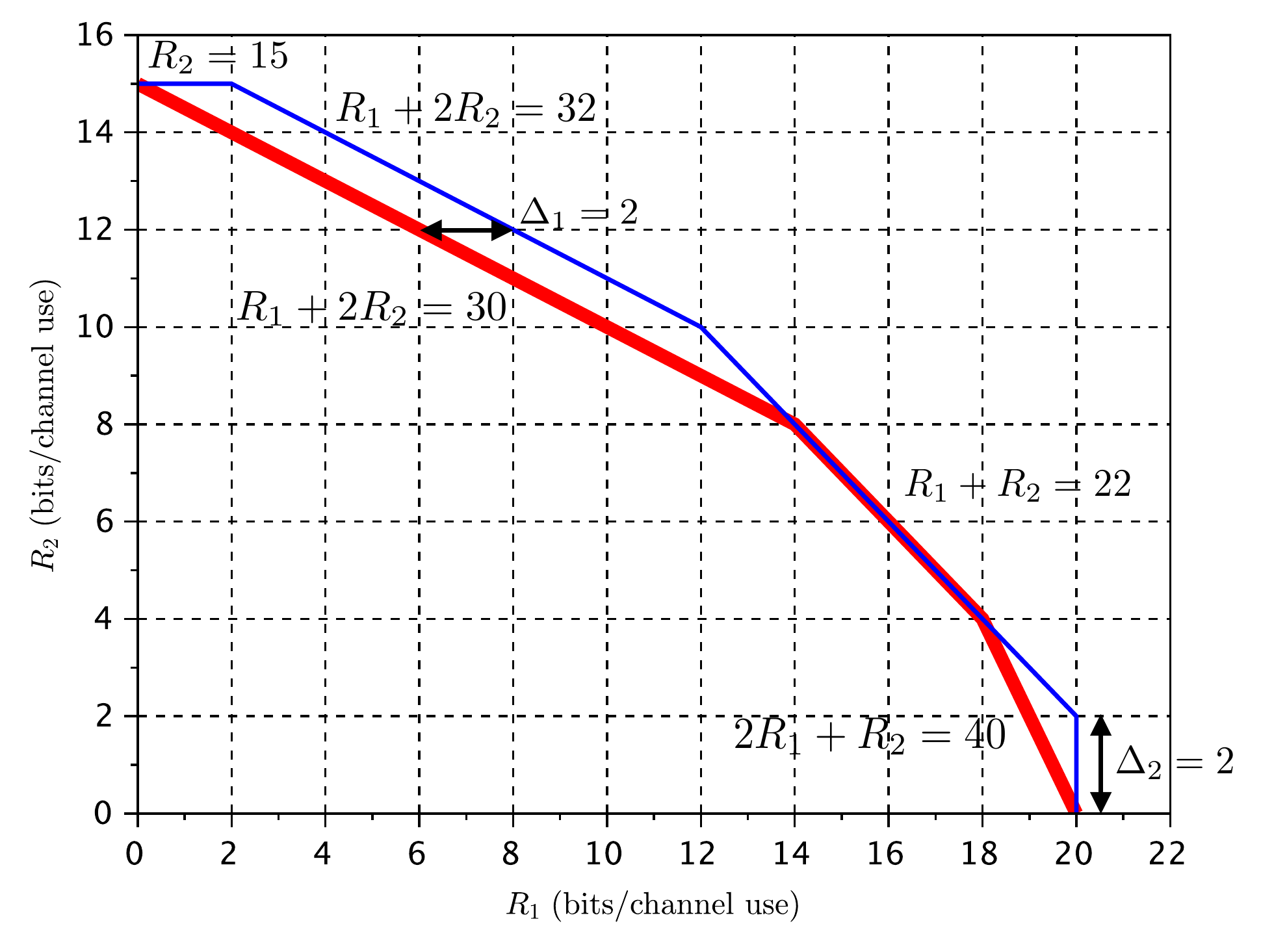,width=0.66\textwidth}}
\caption{Capacity region $\Cldicnfb(20,15,12,13,0,0)$ without feedback  (thick red line) and $\Cldicnfb(20,15,12,13,15,14)$ with noisy channel-output feedback (thin blue line) of the Example $1$. Note that $\Delta_{1}(20,15,12,13,15,14) = 2$ bits/ch.use, $\Delta_{2}(20,15,12,13,15,14) = 2$ bits/ch.use and $\Sigma(20,15,12,13,15,14) = 0$ bits/ch.use.} 
\label{FigExample1capb}
\end{figure}

\begin{figure}[t!]
\centerline{\epsfig{figure=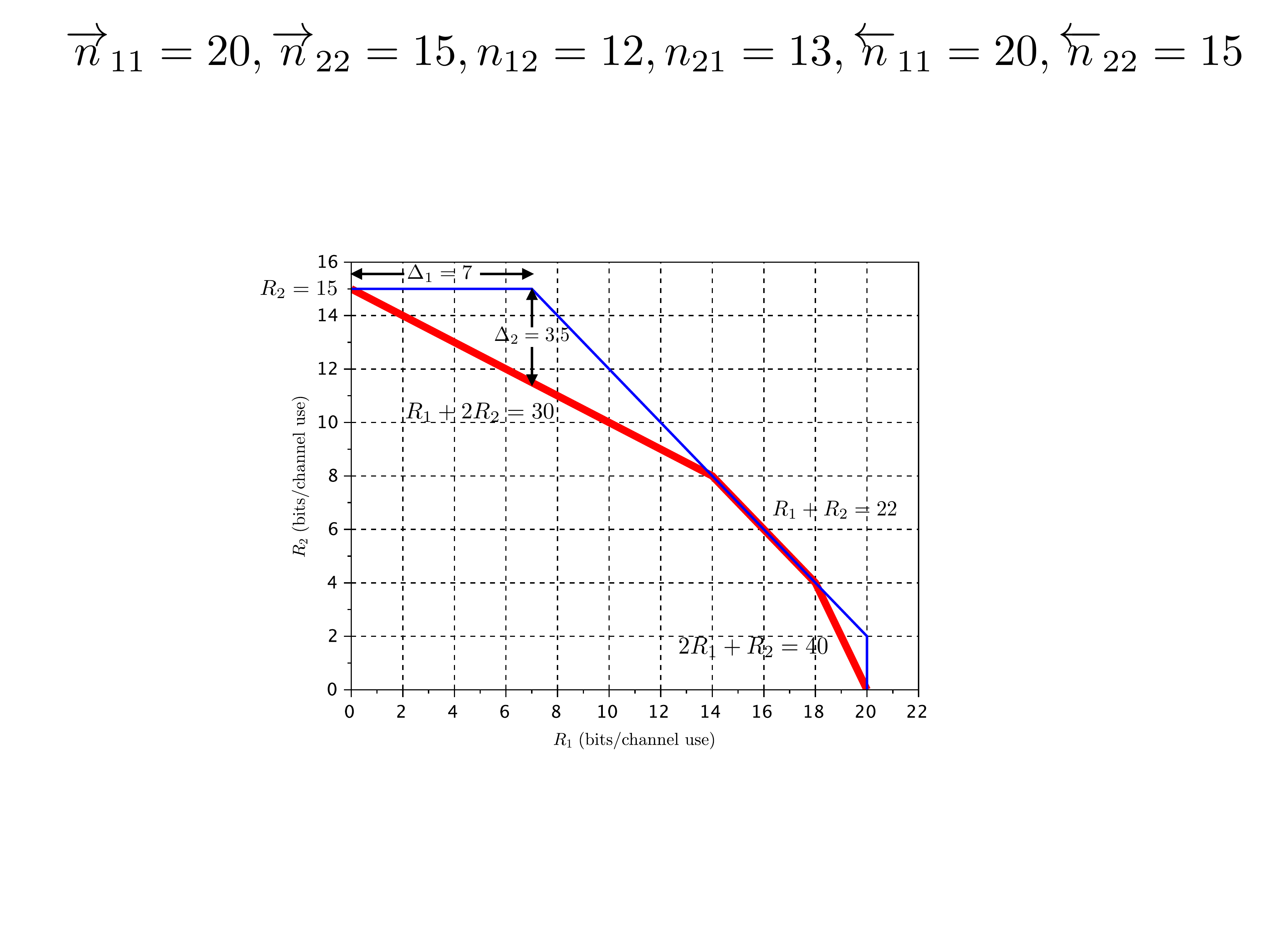,width=0.67\textwidth}}
\caption{Capacity region $\Cldicnfb(20,15,12,13,0,0)$ without feedback  (thick red line) and  $\Cldicnfb(20,15,12,13,20,15)$ with perfect channel-output feedback (thin blue line) of the Example $1$. Note that $\Delta_{1}(20,15,12,13,20,15) = 7$ bits/ch.use, $\Delta_{2}(20,15,12,13,20,15) = 3.5$ bits/ch.use and $\Sigma(20,15,12,13,20,15) = 0$ bits/ch.use. }
\label{FigExample1capc}
\end{figure}

\begin{figure}[t!]
\centerline{\epsfig{figure=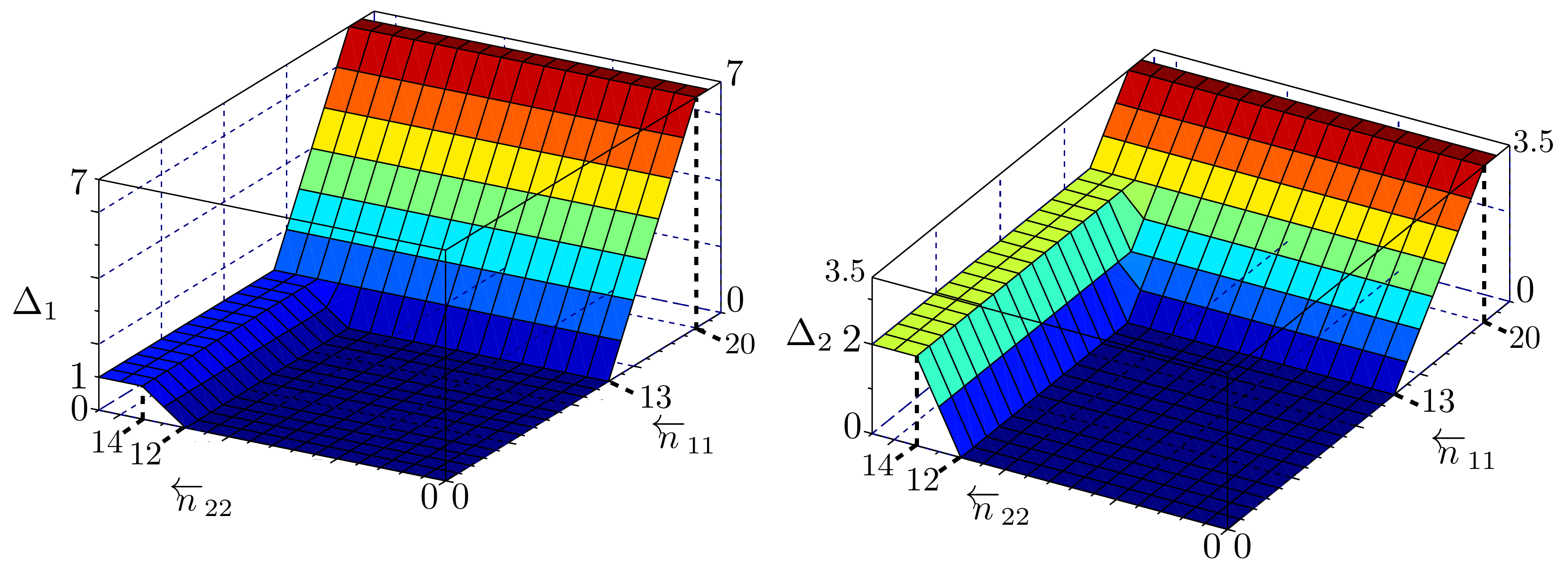,width=1\textwidth}}
\caption{Maximum improvements $\Delta_1(20,15,12,13, \cdot, \cdot)$ and $\Delta_2(20,15,12,13, \cdot, \cdot)$ of individual rates of the Example $1$.
}
\label{FigExample1a}
\end{figure}

Consider the case in which transmitter-receiver pairs $1$ and $2$ are in weak and moderate interference regimes, with $\overrightarrow{n}_{11} = 20$, $\overrightarrow{n}_{22} = 15$, $n_{12} = 12$, $n_{21} = 13$. 
In Figure~\ref{FigExample1capb} and Figure~\ref{FigExample1capc}, the capacity regions with noisy channel-output feedback and perfect channel-output feedback are plotted, respectively. 
In Figure~\ref{FigExample1a}, $\Delta_{i}(20, 15, 12, 13, \overleftarrow{n}_{11} ,\overleftarrow{n}_{22})$ with $i \in \lbrace 1, 2 \rbrace$, are plotted as functions of $\overleftarrow{n}_{11}$ and  $\overleftarrow{n}_{22}$. Therein, it is shown that:
$(a)$ Increasing parameter $\overleftarrow{n}_{11}$ beyond threshold $\overleftarrow{n}_{11}^*=13$ allows simultaneous improvement of both individual rates independently of the value of $\overleftarrow{n}_{22}$. Note that in the case of perfect channel-output feedback, i.e., $\overleftarrow{n}_{11}= \max\left(\overrightarrow{n}_{11} , n_{12} \right)$, the maximum improvement of both individual rates is simultaneously achieved even when $\overleftarrow{n}_{22}=0$.
$(b)$ Increasing parameter $\overleftarrow{n}_{22}$ beyond threshold $\overleftarrow{n}_{22}^*=12$  provides simultaneous improvement of both individual rates. However, the improvement on the individual rate $R_2$ strongly depends on the value of $\overleftarrow{n}_{11}$.
$(c)$ Finally, the sum rate does not increase by using channel-output feedback in this case.
\subsubsection*{Example 2: only one channel-output feedback link allows maximum improvement of one individual rate and the sum-rate}\label{SecExample2}

\begin{figure}[t!]
\centerline{\epsfig{figure=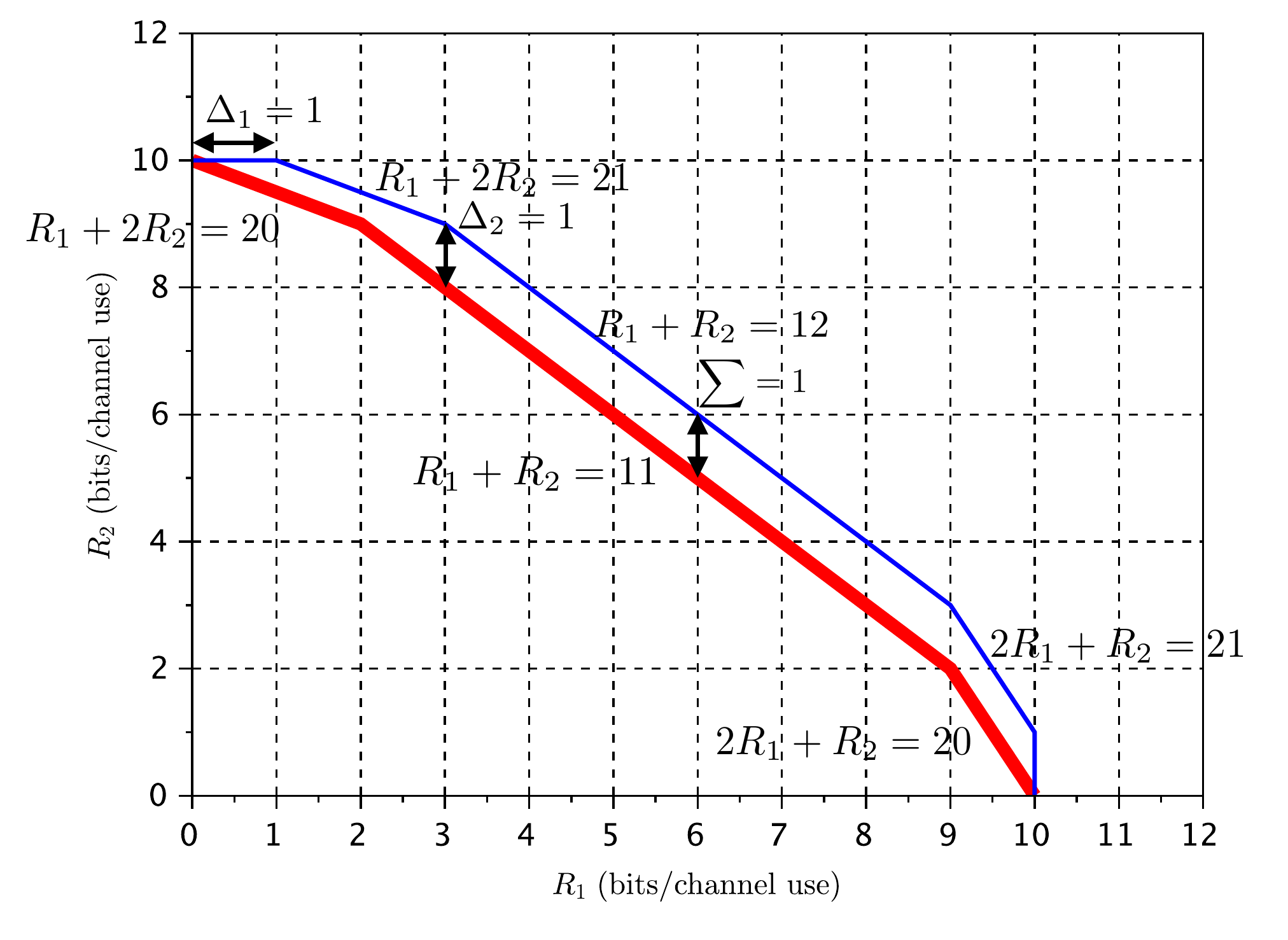,width=0.68\textwidth}}
\caption{Capacity region $\Cldicnfb(10,10,3,8,0,0)$ without feedback (thick red line) and  $\Cldicnfb(10,10,3,8,9,4)$ with noisy channel-output feedback (thin blue line) of the Example $2$. Note that $\Delta_{1}(10,10,3,8,9,4) = 1$ bit/ch.use, $\Delta_{2}(10,10,3,8,9,4) = 1$ bit/ch.use and $\Sigma(10,10,3,8,9,4) = 1$ bit/ch.use.}
\label{FigExample2capb}
\end{figure}

\begin{figure}[t!]
\centerline{\epsfig{figure=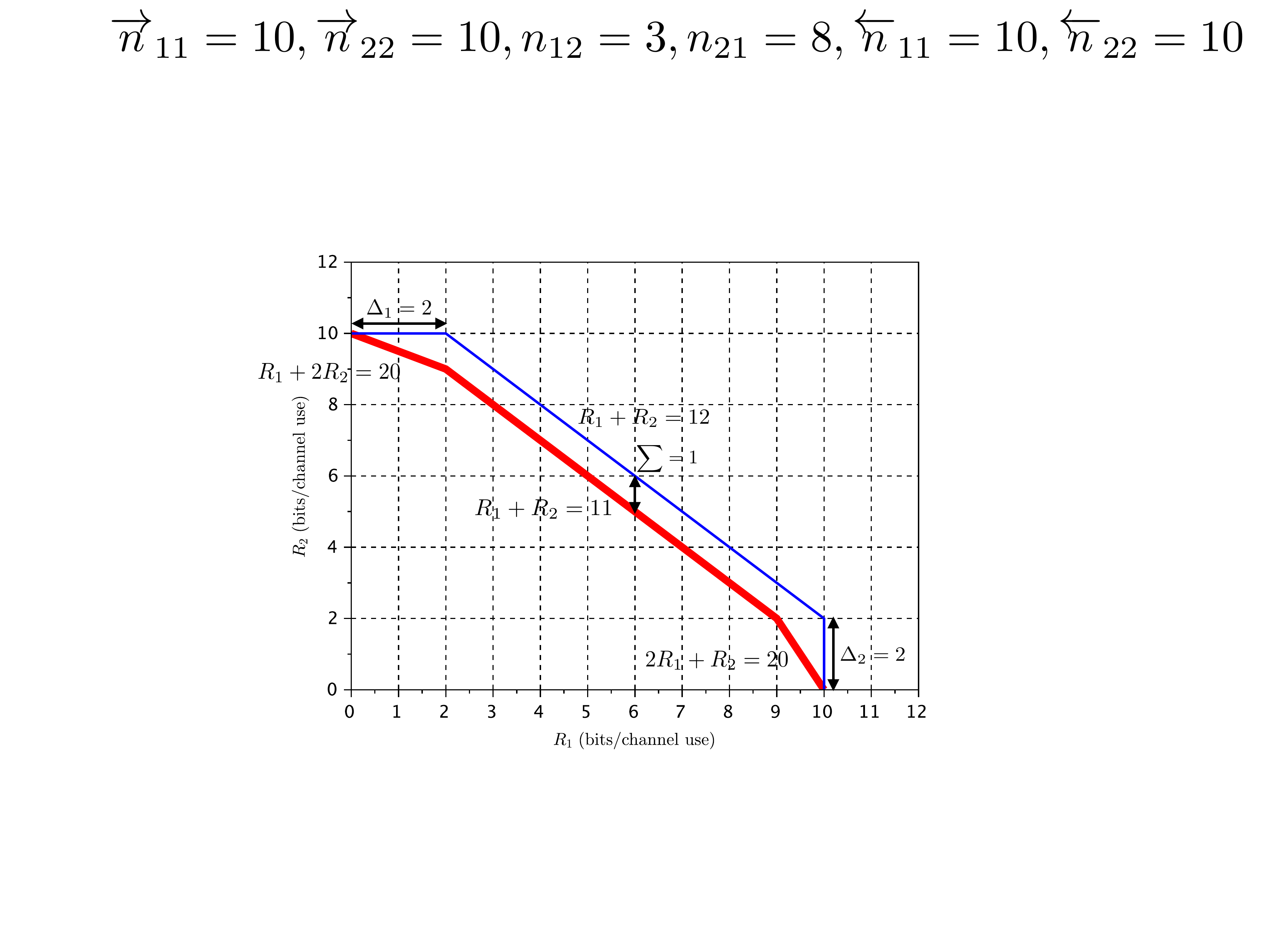,width=0.67\textwidth}}
\caption{Capacity region $\Cldicnfb(10,10,3,8,0,0)$ without feedback  (thick red line) and $\Cldicnfb(10,10,3,8,10,10)$ with perfect channel-output feedback (thin blue line) of the Example $2$.  Note that $\Delta_{1}(10,10,3,8,10,10) = 2$ bits/ch.use, $\Delta_{2}(10,10,3,8,10,10) = 2$ bits/ch.use and $\Sigma(10,10,3,8,10,10) = 1$ bit/ch.use.}
\label{FigExample2capc}
\end{figure}

\begin{figure}[h]
\centerline{\epsfig{figure=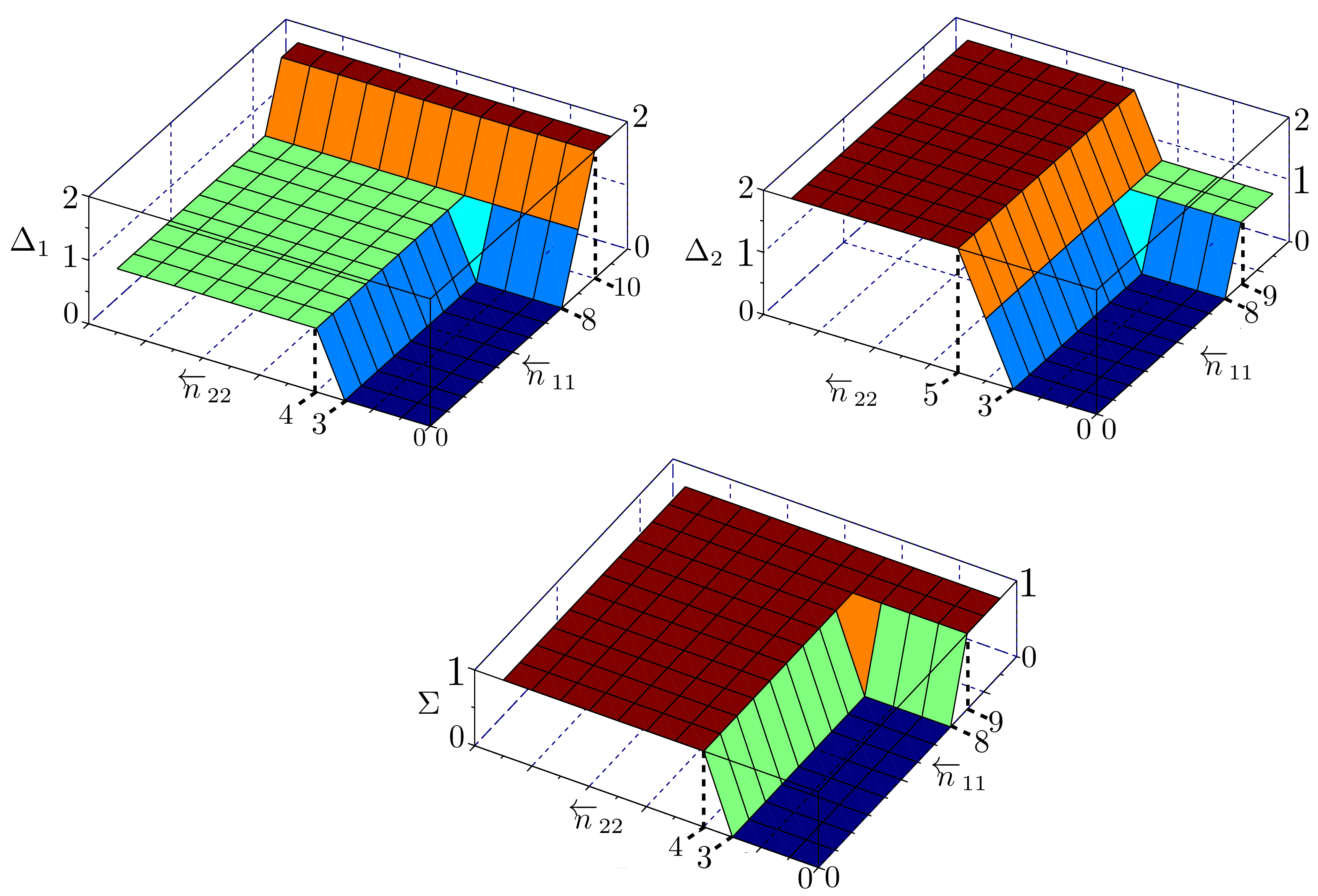,width=1\textwidth}}
\caption{Maximum improvements $\Delta_1(10,10,3,8,\cdot,\cdot)$ and $\Delta_2(10,10,3,8,\cdot,\cdot)$ of one individual rate and $\Sigma(10,10,3,8,\cdot,\cdot)$ of the sum rate of the Example $2$.
}
\label{FigExample2a}
\end{figure}

Consider the case in which transmitter-receiver pairs $1$ and $2$ are in very weak and moderate interference regimes, with $\overrightarrow{n}_{11} = 10$, $\overrightarrow{n}_{22} = 10$, $n_{12} = 3$, $n_{21} = 8$. 
In Figure~\ref{FigExample2capb} and Figure~\ref{FigExample2capc}, the capacity regions with noisy channel-output feedback and perfect channel-output feedback are plotted, respectively. 
In Figure~\ref{FigExample2a}, $\Delta_{i}(10, 10, 3, 8, \overleftarrow{n}_{11} ,\overleftarrow{n}_{22})$ with $i \in \lbrace 1, 2 \rbrace$,  are plotted as functions of $\overleftarrow{n}_{11}$ and  $\overleftarrow{n}_{22}$. 
Therein, it is shown that: 
$(a)$  Increasing $\overleftarrow{n}_{11}$ beyond threshold $\overleftarrow{n}_{11}^*=8$ or increasing $\overleftarrow{n}_{22}$ beyond threshold $\overleftarrow{n}_{22}^*=3$  allows simultaneous improvement of both individual rates. Nonetheless, maximum improvement on $R_i$ is achieved by increasing $\overleftarrow{n}_{ii}$. $(b)$ Increasing either $\overleftarrow{n}_{11}$  or $\overleftarrow{n}_{22}$ beyond thresholds $\overleftarrow{n}_{11}^{*}$ and $\overleftarrow{n}_{22}^{*}$, allows maximum improvement of the sum rate (see Figure~\ref{FigExample2a}).\\

\subsubsection*{Example 3: at least one channel-output feedback link does not have any effect over the capacity region}\label{SecExample3}
\begin{figure}[t!]
\centerline{\epsfig{figure=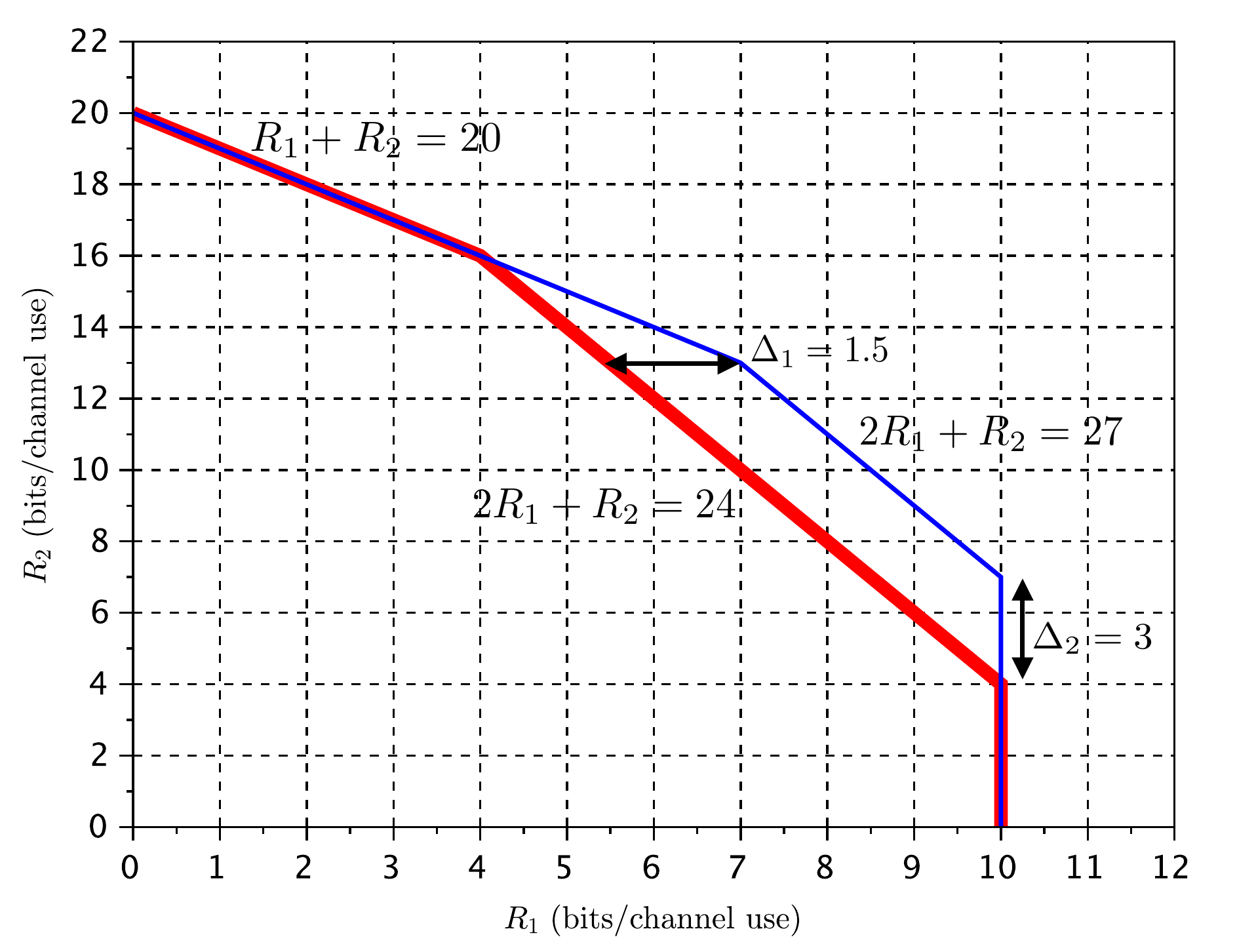,width=0.64\textwidth}}
\caption{Capacity region  $\Cldicnfb(10,20,6,12,0,0)$ without feedback (thick red line) and $\Cldicnfb(10,20,6,12,10,11)$ with noisy channel-output feedback (thin blue line) of the Example $3$.  Note that $\Delta_{1}(10,20,6,12,10,11) = 1.5$ bits/ch.use, $\Delta_{2}(10,20,6,12,10,11) = 2$ bits/ch.use and $\Sigma(10,20,6,12,10,11) = 0$ bits/ch.use.}
\label{FigExample3capb} 
\end{figure}

\begin{figure}[t!]
\centerline{\epsfig{figure=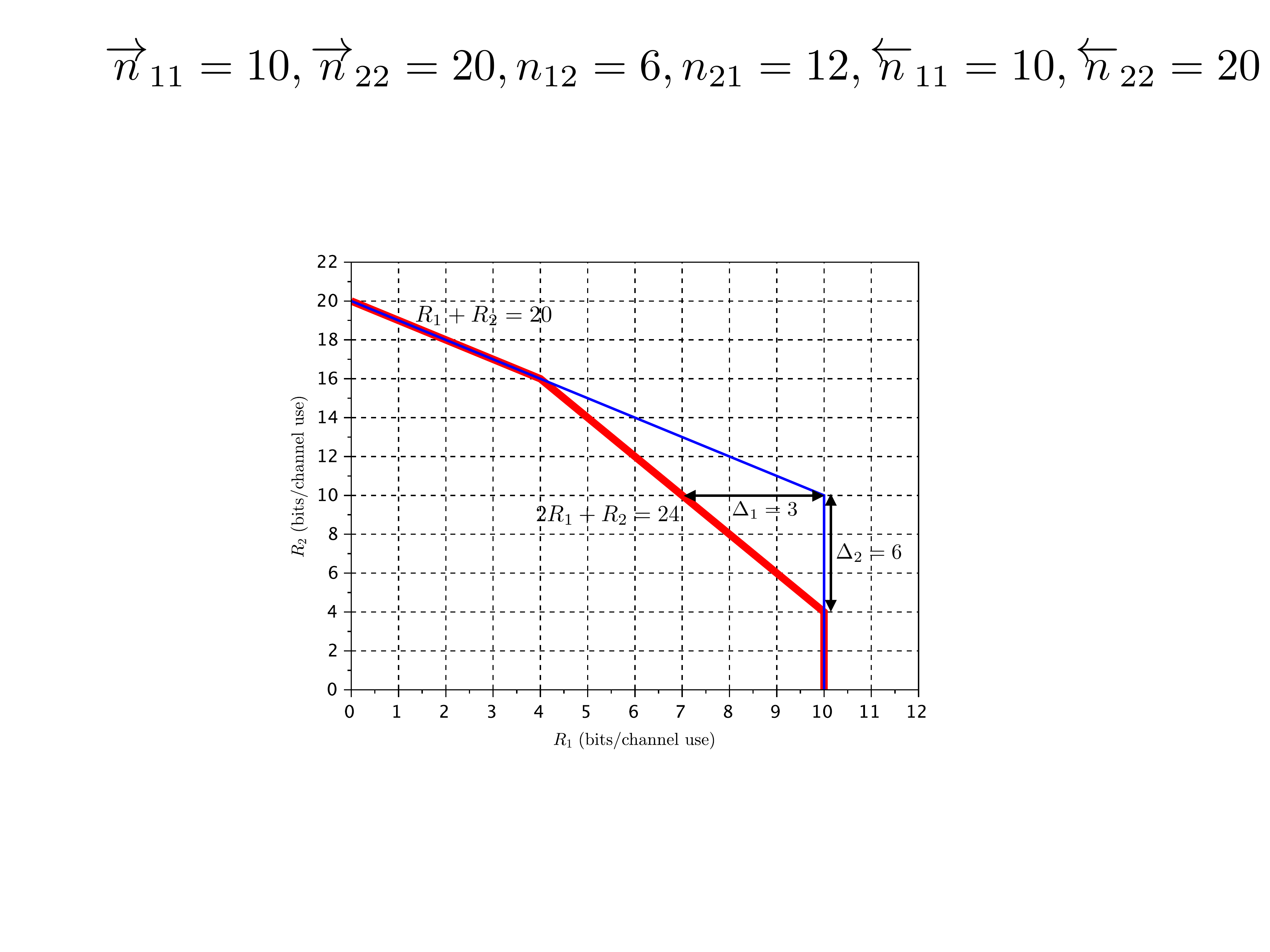,width=0.64\textwidth}}
\caption{Capacity region $\Cldicnfb(10,20,6,12,0,0)$ without feedback (thick red line) and $\Cldicnfb(10,20,6,12,10,20)$ with perfect channel-output feedback (thin blue line)  of the Example $3$.  Note that $\Delta_{1}(10,20,6,12,10,20) = 3$ bits/ch.use, $\Delta_{2}(10,20,6,12,10,20) = 6$ bits/ch.use and $\Sigma(10,20,6,12,10,20) = 0$ bits/ch.use.}
\label{FigExample3capc}
\end{figure}

\begin{figure}[h]
\centerline{\epsfig{figure=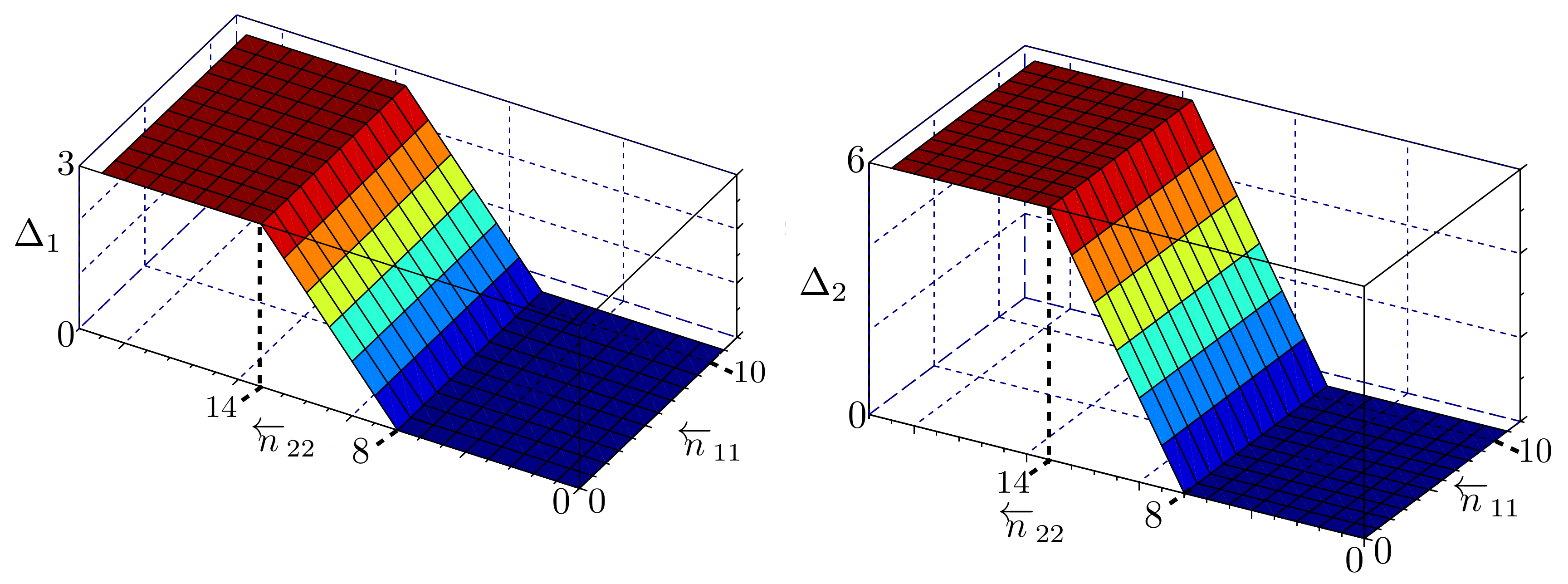,width=1\textwidth}}
\caption{Maximum improvement $\Delta_1(10,20,6,12,\cdot, \cdot)$ and $\Delta_2(10,20,6,12,\cdot, \cdot)$  of one individual rate of the Example $3$.}
\label{FigExample3}
\end{figure}
Consider the case in which transmitter-receiver pairs $1$ and $2$ are in the weak interference regime, with $\overrightarrow{n}_{11} = 10$, $\overrightarrow{n}_{22} = 20$, $n_{12} = 6$, $n_{21} = 12$. 
In Figure~\ref{FigExample3capb} and Figure~\ref{FigExample3capc}, the capacity regions with noisy channel-output feedback and perfect channel-output feedback are plotted, respectively. 
In Figure~\ref{FigExample3}, $\Delta_{i}(10, 20, 6, 12, \overleftarrow{n}_{11} ,\overleftarrow{n}_{22})$ with $i \in \lbrace 1, 2 \rbrace$, are plotted as functions of $\overleftarrow{n}_{11}$ and  $\overleftarrow{n}_{22}$. 
Therein, it is shown that:
$(a)$ Increasing parameter $\overleftarrow{n}_{11}$ does not enlarge the capacity region, independently of the value of $\overleftarrow{n}_{22}$.
$(b)$ Increasing parameter $\overleftarrow{n}_{22}$ beyond threshold $\overleftarrow{n}_{22}^*=8$ allows simultaneous improvement of both individual rates.
$(c)$ Finally, none of the parameters $\overleftarrow{n}_{11}$ or $\overleftarrow{n}_{22}$ increases the sum-rate in this case.

\subsubsection*{Example 4: the channel-output feedback of link $i$ exclusively  improves $R_j$}\label{SecExample4}
\begin{figure}[t!]
\centerline{\epsfig{figure=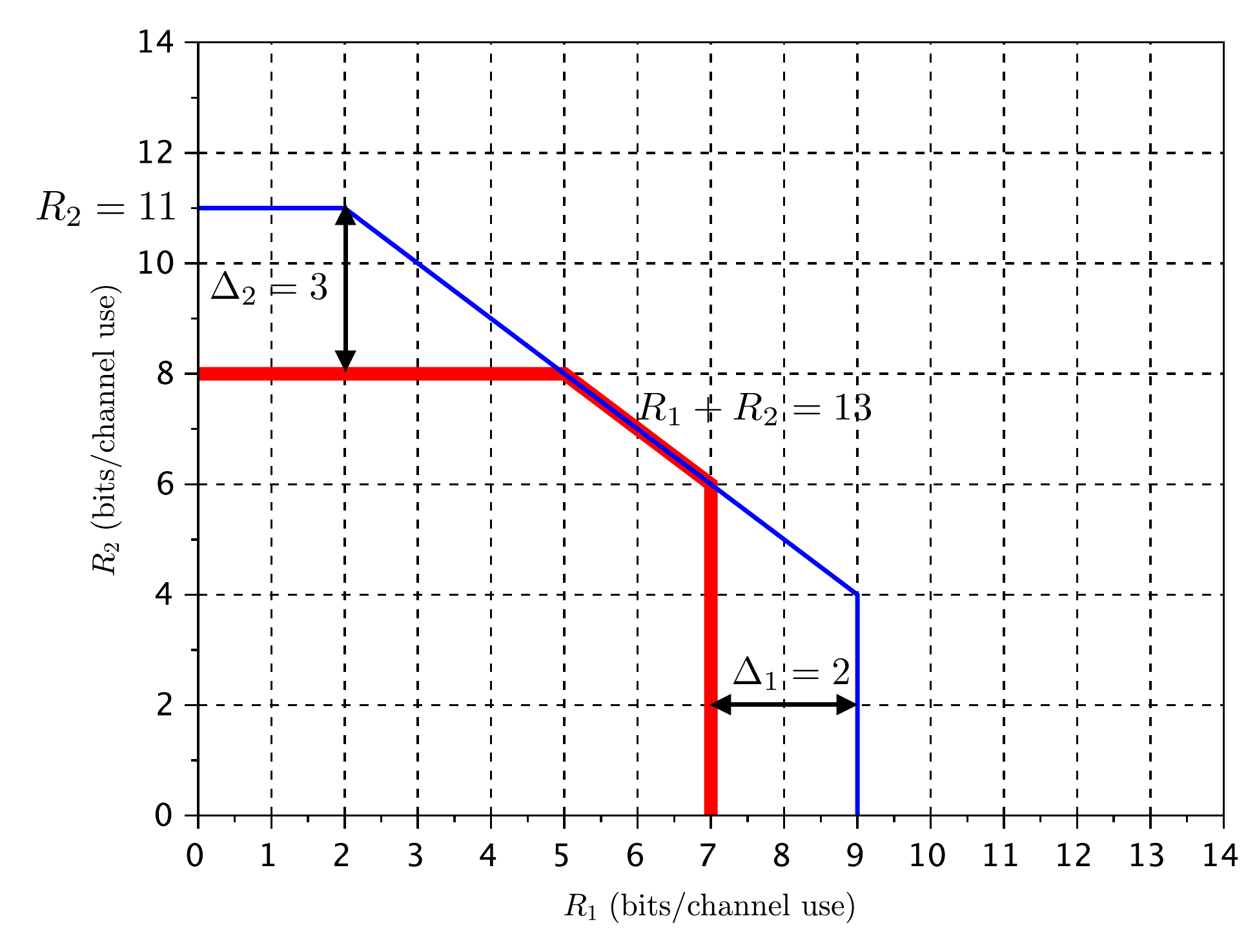,width=0.66\textwidth}}
\caption{Capacity region $\Cldicnfb(7,8,15,13,0,0)$ without feedback  (thick red line) and $\Cldicnfb(7,8,15,13,11,9)$ with noisy channel-output feedback  (thin blue line) of the Example $4$.  Note that $\Delta_{1}(7,8,15,13,11,9) = 2$ bits/ch.use, $\Delta_{2}(7,8,15,13,11,9) = 3$ bits/ch.use and $\Sigma(7,8,15,13,11,9) = 0$ bits/ch.use.}
\label{FigExample4capb}
\end{figure}

\begin{figure}[t!]
\centerline{\epsfig{figure=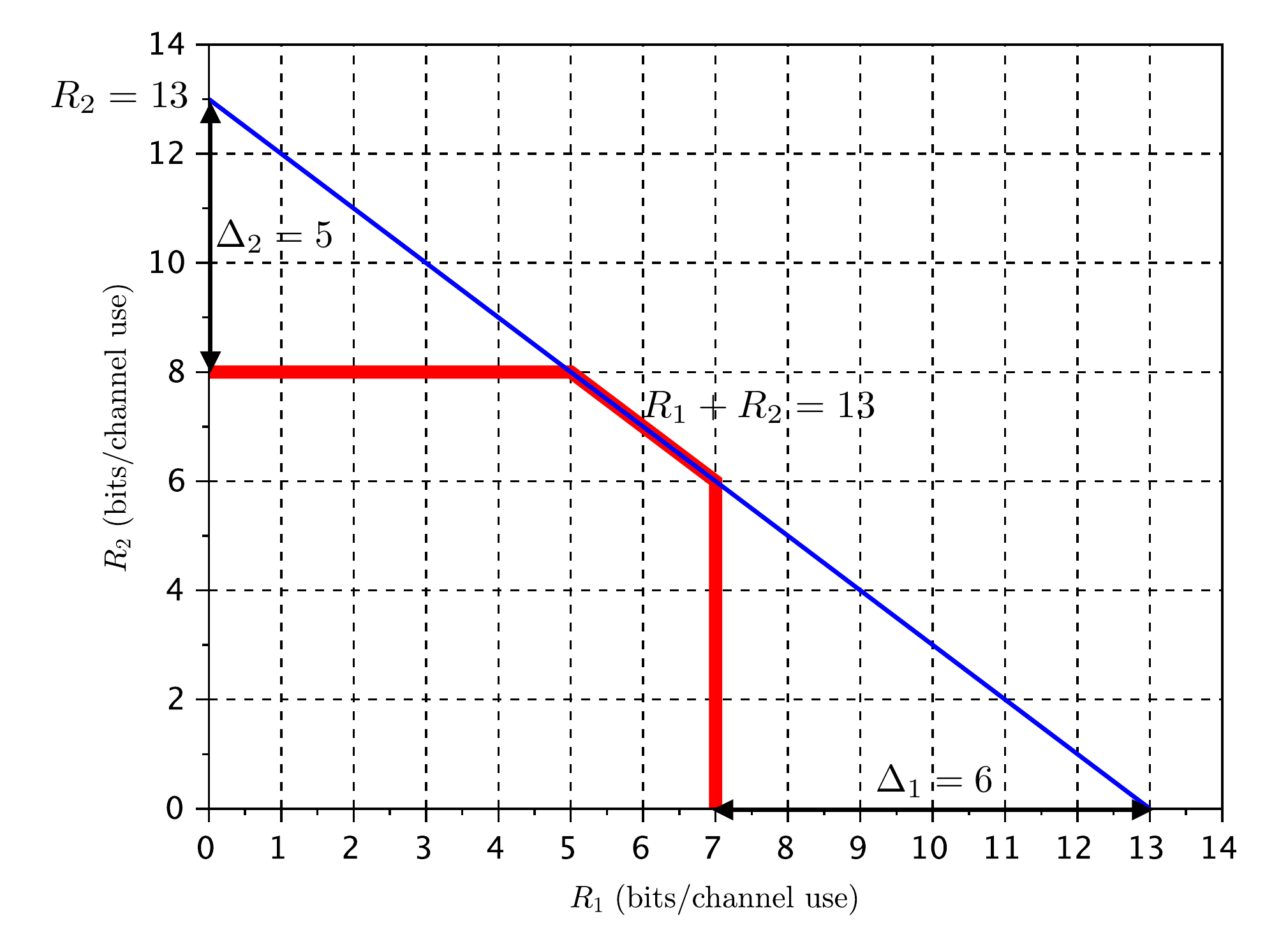,width=0.68\textwidth}}
\caption{Capacity region $\Cldicnfb(7,8,15,13,0,0)$ without feedback  (thick red line) and $\Cldicnfb(7,8,15,13,15,13)$ with perfect channel-output feedback (thin blue line) of the Example $4$. Note that $\Delta_{1}(7,8,15,13,15,13)= 6$ bits/ch.use, $\Delta_{2}(7,8,15,13,15,13) = 5$ bits/ch.use and $\Sigma(7,8,15,13,15,13) = 0$ bits/ch.use. }
\label{FigExample4capc}
\end{figure}

\begin{figure}[h]
\centerline{\epsfig{figure=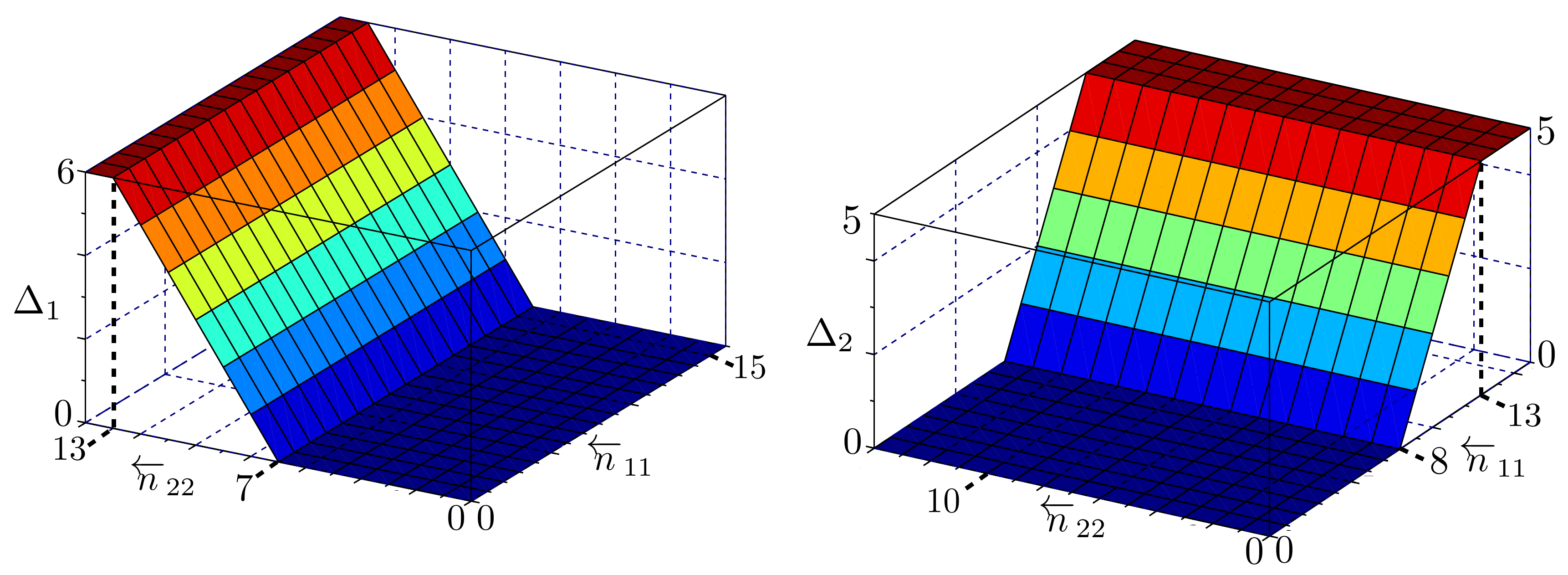,width=1\textwidth}}
\caption{Maximum improvement $\Delta_{1}(7,8,15,13,\cdot, \cdot)$ and $\Delta_{2}(7,8,15,13,\cdot, \cdot)$ of one individual rate of the Example $4$.
}
\label{FigExample4a}
\end{figure}
Consider the case in which transmitter-receiver pairs $1$ and $2$ are in the very strong and strong interference regimes, with $\overrightarrow{n}_{11} = 7$, $\overrightarrow{n}_{22} = 8$, $n_{12} = 15$, $n_{21} = 13$. 
In Figure~\ref{FigExample4capb} and Figure~\ref{FigExample4capc}, the capacity regions with noisy channel-output feedback and perfect channel-output feedback are plotted, respectively. 
In Figure~\ref{FigExample4a}, $\Delta_{i}(7, 8, 15, 13, \overleftarrow{n}_{11} ,\overleftarrow{n}_{22})$ with $i \in \lbrace 1, 2 \rbrace$, are plotted as functions of $\overleftarrow{n}_{11}$ and  $\overleftarrow{n}_{22}$. 
Therein, it is shown that:
$(a)$ Increasing parameter $\overleftarrow{n}_{11}$ beyond threshold $\overleftarrow{n}_{11}^*=8$ exclusively improves $R_2$.
$(b)$ Increasing parameter $\overleftarrow{n}_{22}$ beyond threshold $\overleftarrow{n}_{22}^*=7$ exclusively improves $R_1$.
$(c)$ None of the parameters $\overleftarrow{n}_{11}$  or $\overleftarrow{n}_{22}$  has an impact over the sum rate in this case.
Note that these observations are in line with the interpretation of channel-output feedback as an altruistic technique, as in \cite{Perlaza-TIT-2015, Perlaza-ISIT-2014a}.  This is basically because the link implementing channel-output feedback provides an alternative path to the information sent by the other link, as first suggested in \cite{Suh-TIT-2011}.

\subsubsection*{Example 5: none of the channel-output feedback links has any effect over the capacity region}\label{SecExample5}

\begin{figure}[t!]
\centerline{\epsfig{figure=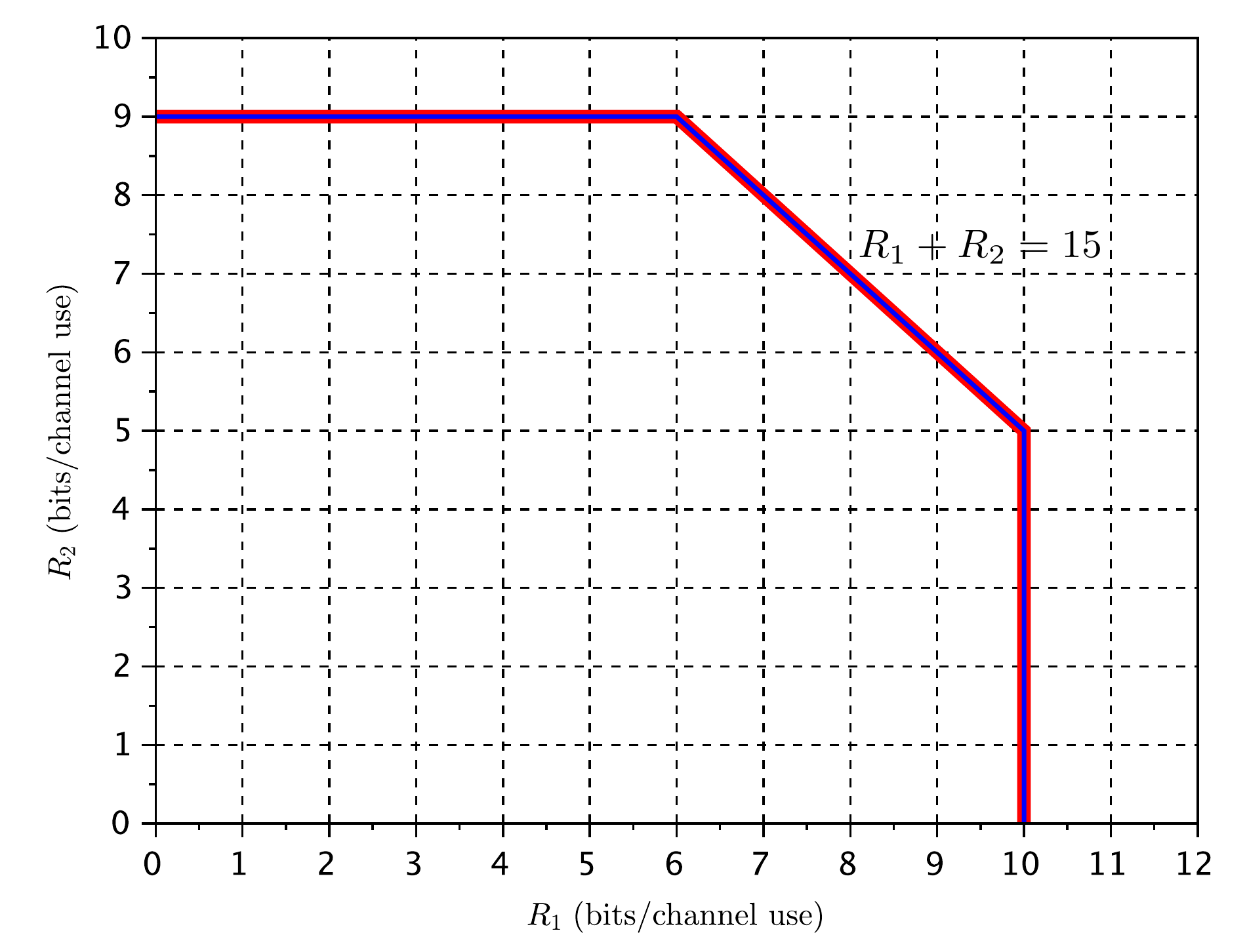,width=0.65\textwidth}}
\caption{Capacity region $\Cldicnfb(10,9,2,15,0,0)$  without feedback  (thick red line) and $\Cldicnfb(10,9,2,15,10,15)$ with perfect channel-output feedback (thin blue line) of the Example $5$. Note that  $\Cldicnfb(10,9,2,15,0,0) = \Cldicnfb(10,9,2,15,10,15)$. }
\label{FigExample5capb}
\end{figure}

Consider the case in which transmitter-receiver pairs $1$ and $2$ are in the very weak and strong interference regimes, with $\overrightarrow{n}_{11} = 10$, $\overrightarrow{n}_{22} = 9$, $n_{12} = 2$, $n_{21} = 15$. 
In Figure~\ref{FigExample5capb}, the capacity regions without channel-output feedback and with perfect channel-output feedback are plotted. Note that the capacity region of the LD-IC with and without channel-output feedback are identical.

\section{Conclusions}

In this technical report, the noisy channel-output feedback capacity of the linear deterministic interference channel has been fully characterized. 
Based on specific asymmetric examples, it is highlighted that even in the presence of noise, the benefits of channel-output feedback can be significantly relevant in terms of achievable individual rate and sum-rate improvements with respect to the case without feedback. Unfortunately, there also exist scenarios in which these benefits are totally inexistent.

\clearpage

\begin{appendices}
%

\section{Proof of Achievability} \label{AppAch-IC-NOF}

This appendix describes an achievability scheme for the IC-NOF based on a three-part message splitting, superposition coding, and backward decoding. 

\noindent
\textbf{Codebook Generation}: Fix a strictly positive joint probability distribution  
\begin{IEEEeqnarray}{rcl}
\nonumber
&P&_{U\, U_1\,U_2\, V_1\,V_2\, X_{1,P}\, X_{2,P}}(u, u_1,u_2, v_1,v_2, x_{1,P}, x_{2,P})=P_U(u) P_{U_1|U}(u_1|u) P_{U_2|U}(u_2|u) \\
\label{Eqprobdist}
& & P_{V_1|U\,U_1}(v_1|u,u_1)P_{V_2|U\,U_2}(v_2|u,u_2) P_{X_{1,P}|U\,U_1\,V_1}(x_{1,P}|u,u_1,v_1) P_{X_{2,P}|U\,U_2\,V_2}(x_{2,P}|u,u_2,v_2), \qquad
\end{IEEEeqnarray}
for all $\left(u, u_1, u_2, v_1, v_2, x_{1,P}, x_{2,P}\right) \in \left(\mathcal{X}_1\cup \mathcal{X}_2\right) \times \mathcal{X}_1 \times \mathcal{X}_2 \times \mathcal{X}_1 \times \mathcal{X}_2 \times \mathcal{X}_1 \times \mathcal{X}_2$.

\noindent
Let $R_{1,C1}$, $R_{1,C2}$, $R_{2,C1}$, $R_{2,C2}$, $R_{1,P}$, and $R_{2,P}$ be non-negative real numbers. Let also $R_{1,C}=R_{1,C1}$ $+$ $R_{1,C2}$, ${R_{2,C}=R_{2,C1}+R_{2,C2}}$, $R_{1}=R_{1,C}+R_{1,P}$, and ${R_{2}=R_{2,C}+R_{2,P}}$. 

\noindent
Generate $2^{N(R_{1,C1} + R_{2,C1})}$ i.i.d. $N$-length codewords ${\bs{u}(s,r) = \big(u_{1}(s,r), u_{2}(s,r), \ldots, u_{N}(s,r)\big)}$ according to 

\begin{equation}
P_{\bs{U}}\big(\bs{u}(s,r)\big) = \ds\prod_{i =1}^N P_{U}(u_{i}(s,r)),
\end{equation}
with $s \in \lbrace 1, 2,  \ldots, 2^{NR_{1,C1}}\rbrace$ and $r \in \lbrace 1, 2,  \ldots, 2^{NR_{2,C1}}\rbrace$. 

\noindent
For encoder $1$, generate for each codeword $\bs{u}(s,r)$, $2^{NR_{1,C1}}$ i.i.d. $N$-length codewords $\bs{u}_1(s,r,k) = \big(u_{1,1}(s,r,k), u_{1,2}(s,r,k), \ldots, u_{1,N}(s,r,k)\big)$  according to 
\begin{equation}
P_{\bs{U}_1|\bs{U}}\big(\bs{u}_1(s,r,k)|\bs{u}(s,r)\big) = \ds\prod_{i =1}^N P_{U_{1}|U}\big(u_{1,i}(s,r,k)|u_{i}(s,r)\big), 
\end{equation}
with $k \in \lbrace 1, 2,  \ldots, 2^{NR_{1,C1}}\rbrace$.  For each pair of codewords $\big(\bs{u}(s,r),\bs{u}_{1}(s,r,k)\big)$, generate $2^{NR_{1,C2}}$ i.i.d. $N$-length codewords $\bs{v}_1(s,r,k,l) = \big(v_{1,1}(s,r,k,l), v_{1,2}(s,r,k,l), \ldots, v_{1,N}(s,r,k,l)\big)$  according to 
\begin{IEEEeqnarray}{rcl}
P_{\bs{V}_1 | \bs{U}\,\bs{U}_1}\big(\bs{v}_1(s,r,k,l)|\bs{u}(s,r) ,\bs{u}_1(s,r,k)\big) = \ds\prod_{i =1}^N P_{V_{1}|U\,U_{1}}\big(v_{1,i}(s,r,k,l)|u_{i}(s,r),u_{1,i}(s,r,k)\big), \quad
\end{IEEEeqnarray}
with $l \in \lbrace 1, 2, \ldots, 2^{NR_{1,C2}}\rbrace$. For each tuple of codewords $\big(\bs{u}(s,r)$, $\bs{u}_{1}(s,r,k)$, $\bs{v}_{1}(s,r,k,l)\big)$, generate  $2^{NR_{1,P}}$ i.i.d. $N$-length codewords $\bs{x}_{1,P}(s,r,k,l,q) = \big(x_{1,P,1}(s,r,k,l,q), x_{1,P,2}(s,r,k,l,q), \ldots$, $x_{1,P,N}(s,r,k,l,q)\big)$ according to 
\begin{IEEEeqnarray}{rcl}
\nonumber
&P&_{\bs{X}_{1,P} | \bs{U}\, \bs{U}_{1} \!\, \!\bs{V}_{1}}\!\big(\bs{x}_{1,P}(s,r,k,l,q) |  \bs{u}(s,r),\!\bs{u}_{1}(s,r,k),\!\bs{v}_{1}(s,r,k,l)\!\big)\!\\
& & =\ds\prod_{i =1}^N P_{X_{1,P}|U\,U_{1}\,V_{1}}\big(x_{1,P,i}(s,r,k,l,q)|u_{i}(s,r),u_{1,i}(s,r,k),v_{1,i}(s,r,k,l)\big), 
\end{IEEEeqnarray}
with $q \in \lbrace 1, 2, \ldots, 2^{NR_{1,P}}\rbrace$. 

\begin{figure*}[t!]
 \centerline{\epsfig{figure=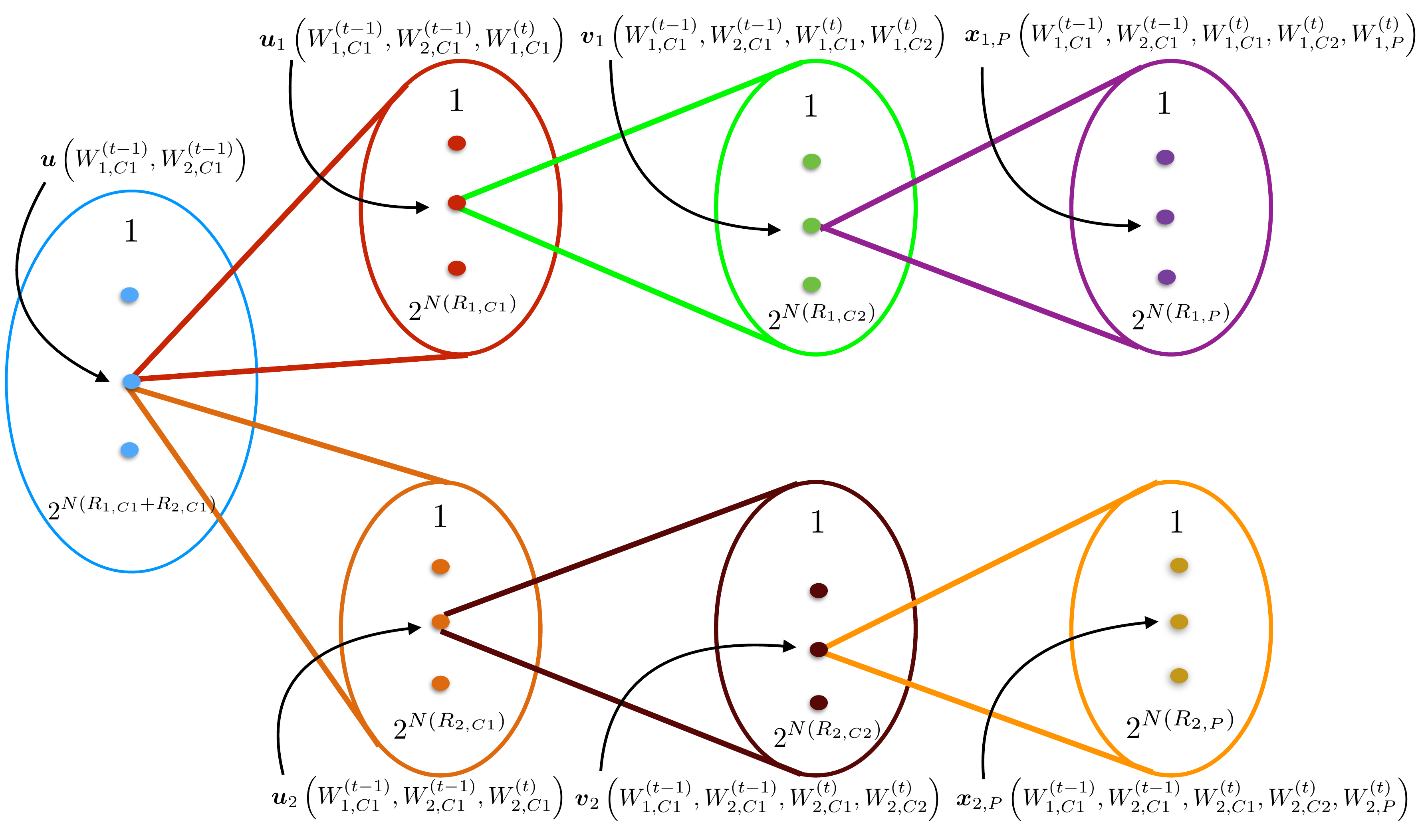,width=1\textwidth}}
 \caption{Structure of the superposition code. The codewords corresponding to the message indices $W_{1,C1}^{(t-1)}, W_{2,C1}^{(t-1)},W_{i,C1}^{(t)},W_{i,C2}^{(t)},W_{i,P}^{(t)}$ with $i \in \lbrace 1, 2 \rbrace$ as well as the block index $t$ are both highlighted. The (approximate) number of codewords for each code layer is also highlighted.} 
\label{FigSuperpos}
\end{figure*}

\noindent
For encoder $2$, generate for each codeword $\bs{u}(s,r)$,  $2^{NR_{2,C1}}$ i.i.d. $N$-length codewords  $\bs{u}_2(s,r,j) = \big(u_{2,1}(s,r,j), u_{2,2}(s,r,j), \ldots, u_{2,N}(s,r,j)\big)$  according to 
\begin{equation}
P_{\bs{U}_2|\bs{U}}\big(\bs{u}_2(s,r,j)|\bs{u}(s,r)\big) = \ds\prod_{i =1}^N P_{U_{2}|U}\big(u_{2,i}(s,r,j)|u_{i}(s,r)\big), 
\end{equation}
with $j \in \lbrace 1, 2,  \ldots, 2^{NR_{2,C1}}\rbrace$. For each pair of codewords $\big(\bs{u}(s,r),\bs{u}_{2}(s,r,j)\big)$, generate $2^{NR_{2,C2}}$ i.i.d. length-$N$ codewords $\bs{v}_2(s,r,j,m)=\big(v_{2,1}(s,r,j,m), v_{2,2}(s,r,j,m),  \ldots, v_{2,N}(s,r,j,m)\big)$ according to 
\begin{IEEEeqnarray}{rcl}
P_{\bs{V}_2 | \bs{U}\, \bs{U}_2}\big(\bs{v}_2(s,r,j,m) | \bs{u}(s,r), \bs{u}_2(s,r,j)\big) = \ds\prod_{i =1}^N P_{V_{2} |  U\, U_{2}}(v_{2,i}(s,r,j,m) |  u_{i}(s,r), u_{2,i}(s,r,j)),  \qquad
\end{IEEEeqnarray}
with $m \in \lbrace 1, 2,  \ldots, 2^{NR_{2,C2}}\rbrace$. For each tuple of codewords $\big(\bs{u}(s,r)$, $\bs{u}_{2}(s,r,j),\bs{v}_{2}(s,r,j,m)\big)$, generate  $2^{NR_{2,P}}$ i.i.d. $N$-length codewords $\bs{x}_{2,P}(s,r,j,m,b) \! = \! \big(\! x_{2,P,1}(s,r,j,m,b) \!, \! x_{2,P,2}(s,r,j,m,b) \!, \! \ldots$,  $ x_{2,P,N}(s,r,j,m,b) \!\big)$ according to 
\begin{IEEEeqnarray}{rcl}
\nonumber
&P&_{\bs{X}_{2,P}  |  \bs{U}\,\bs{U}_{2}\! \, \! \bs{V}_{2}} \!\big( \! \bs{x}_{2,P}(s,r,j,m,b)|  \bs{u}(s,r),  \bs{u}_{2}(s,r,j), \! \bs{v}_{2}(s,r,j,m) \! \big) \!  \\
& & = \ds\prod_{i =1}^N P_{X_{2,P} | U\, U_{2}\, V_{2}}\big(x_{2,P,i}(s,r,j,m,b) | u_{i}(s,r), u_{2,i}(s,r,j), v_{2,i}(s,r,j,m,b)\big), 
\end{IEEEeqnarray}
with $b \in \lbrace 1, 2,  \ldots, 2^{NR_{2,P}}\rbrace$.  
The resulting code structure is shown in Figure~\ref{FigSuperpos}.

\noindent
\textbf{Encoding}: Denote by $W_{i}^{(t)} \in \lbrace 1, 2, \ldots, 2^{NR_{i}} \rbrace$  the message index of transmitter $i \in \lbrace 1,2 \rbrace$ during block $t \in \lbrace 1, 2,  \ldots, T \rbrace$, with $T$ the total number of blocks. Let $W_{i}^{(t)}$ be composed by the message index $W_{i,C}^{(t)} \in \lbrace 1, 2,  \ldots, 2^{NR_{i,C}} \rbrace$ and message index $W_{i,P}^{(t)} \in \lbrace 1$, $2$, $  \ldots, 2^{NR_{i,P}} \rbrace$. That is, $W_{i}^{(t)}=\left(W_{i,C}^{(t)},W_{i,P}^{(t)}\right)$.  The message index $W_{i,P}^{(t)}$ must be reliably decoded at receiver $i$. Let also $W_{i,C}^{(t)}$ be composed by the message indices $W_{i,C1}^{(t)} \in \lbrace 1, 2,  \ldots, 2^{NR_{i,C1}} \rbrace$ and $W_{i,C2}^{(t)} \in \lbrace 1, 2,  \ldots, 2^{NR_{i,C2}} \rbrace$. That is, $W_{i,C}^{(t)}=\Big(W_{i,C1}^{(t)}$,$W_{i,C2}^{(t)}\Big)$.  The message index $W_{i,C1}^{(t)}$ must be reliably decoded by the other transmitter (via feedback) and by the non-intended receiver, but not necessarily by the intended receiver. The message index $W_{i,C2}^{(t)}$ must be reliably decoded by the non-intended receiver, but not necessarily by the intended receiver. 

\noindent
Consider Markov encoding over $T$ blocks. At encoding step $t$, with $t \in \lbrace 1, 2,  \ldots, T \rbrace$, transmitter $1$ sends the codeword:
 \begin{IEEEeqnarray}{rcl}
 \nonumber
 \bs{x}_1^{(t)} &=& \Theta_1 \Bigg(\! \bs{u}\Big(\!W_{1,C1}^{(t-1)}, W_{2,C1}^{(t-1)} \!\Big), \! \bs{u}_1\Big(\! W_{1,C1}^{(t-1)}, W_{2,C1}^{(t-1)},W_{1,C1}^{(t)} \!\Big), \bs{v}_1\Big(W_{1,C1}^{(t-1)}, W_{2,C1}^{(t-1)},W_{1,C1}^{(t)}, W_{1,C2}^{(t)}\Big), \\
\label{Eqtransmittercodeword}
 & & \bs{x}_{1,P}\Big(W_{1,C1}^{(t-1)}, W_{2,C1}^{(t-1)}, W_{1,C1}^{(t)}, W_{1,C2}^{(t)},W_{1,P}^{(t)}\Big)\Bigg), 
 \end{IEEEeqnarray}
 where,  $\Theta_1: \left(\mathcal{X}_1\cup \mathcal{X}_2\right)^{N} \times \mathcal{X}_1^{N} \times \mathcal{X}_1^{N} \times \mathcal{X}_1^{N} \rightarrow \mathcal{X}_1^{N}$ is a function that transforms the codewords $\bs{u}\Big(W_{1,C1}^{(t-1)}$, $W_{2,C1}^{(t-1)}\Big)$, $\bs{u}_1\Big(W_{1,C1}^{(t-1)}, W_{2,C1}^{(t-1)},W_{1,C1}^{(t)}\Big)$, $\bs{v}_1\Big(W_{1,C1}^{(t-1)}, W_{2,C1}^{(t-1)},W_{1,C1}^{(t)}, W_{1,C2}^{(t)}\Big) \ $, and $\bs{x}_{1,P}\Big(W_{1,C1}^{(t-1)}$, $W_{2,C1}^{(t-1)}$, $W_{1,C1}^{(t)}$, $W_{1,C2}^{(t)}$, $W_{1,P}^{(t)}\Big)$ into the N-dimensional vector $\bs{x}_1^{(t)}$ of channel inputs. The indices $W_{1,C1}^{(0)} = W_{1,C1}^{(T)} = s^*$ and  $W_{2,C1}^{(0)} = W_{2,C1}^{(T)} = r^*$, and the pair $(s^*,r^*) \in  \lbrace 1, 2,  \ldots, 2^{N \, R_{1,C1}} \rbrace \times \lbrace 1, 2,  \ldots, 2^{NR_{2,C1}} \rbrace$  are pre-defined and known by both receivers and transmitters. It is worth noting that the message index  $W_{2,C1}^{(t-1)}$ is obtained by transmitter $1$ from the feedback signal $\overleftarrow{\bs{y}}_{1}^{(t-1)}$ at the end of the previous encoding step $t-1$.

\noindent
Transmitter $2$ follows a similar encoding scheme.

\noindent
\textbf{Decoding}: Both receivers decode their message indices at the end of block $T$ in a backward decoding fashion. At each decoding step $t$, with $t \in \lbrace 1, 2,  \ldots, T \rbrace$, receiver $1$ obtains the message indices $\big(\widehat{W}_{1,C1}^{(T-t)}$, $\widehat{W}_{2,C1}^{(T-t)}$, $\widehat{W}_{1,C2}^{(T-(t-1))}$, $\widehat{W}_{1,P}^{(T-(t-1))}$, $\widehat{W}_{2,C2}^{(T-(t-1))}\big) \in \lbrace 1$, $2,  \ldots $, $ 2^{NR_{1,C1}}\rbrace \times \lbrace 1$, $ 2,  \ldots, 2^{NR_{2,C1}} \rbrace  \times  \lbrace 1$, $ 2,  \ldots, 2^{NR_{1,C2}} \rbrace \times  \lbrace 1$, $ 2,  \ldots, 2^{NR_{1,P}}\rbrace \times  \lbrace 1$, $2,  \ldots, 2^{NR_{2,C2}} \rbrace $ from the channel output $\overrightarrow{\bs{y}}_1^{(T-(t-1))}$. The tuple $\Big(\widehat{W}_{1,C1}^{(T-t)}$, $\widehat{W}_{2,C1}^{(T-t)}$, $\widehat{W}_{1,C2}^{(T-(t-1))}$, $\widehat{W}_{1,P}^{(T-(t-1))}$, $\widehat{W}_{2,C2}^{(T-(t-1))}\Big)$ is the unique tuple that satisfies
\begin{IEEEeqnarray}{ll}
\nonumber
\Big( & \bs{u}\left(\widehat{W}_{1,C1}^{(T-t)}, \widehat{W}_{2,C1}^{(T-t)}\right), \bs{u}_1\left(\widehat{W}_{1,C1}^{(T-t)}, \widehat{W}_{2,C1}^{(T-t)}, W_{1,C1}^{(T-(t-1))}\right),  \\
\nonumber
& \bs{v}_1 \left(\widehat{W}_{1,C1}^{(T-t)}, \widehat{W}_{2,C1}^{(T-t)}, W_{1,C1}^{(T-(t-1))}, \widehat{W}_{1,C2}^{(T-(t-1))} \right), \\
\nonumber
& \bs{x}_{1,P}\Big(\widehat{W}_{1,C1}^{(T-t)}, \widehat{W}_{2,C1}^{(T-t)}, W_{1,C1}^{(T-(t-1))}, \widehat{W}_{1,C2}^{(T-(t-1))}, \widehat{W}_{1,P}^{(T-(t-1))}\Big),\\
\nonumber
&  \bs{u}_2\left(\widehat{W}_{1,C1}^{(T-t)}, \widehat{W}_{2,C1}^{(T-t)}, W_{2,C1}^{(T-(t-1))}\right), \bs{v}_2\left(\widehat{W}_{1,C1}^{(T-t)}, \widehat{W}_{2,C1}^{(T-t)}, W_{2,C1}^{(T-(t-1))},\widehat{W}_{2,C2}^{(T-(t-1))}\right), \\
\label{EqDecodingW1cW2cW1p}
& \overrightarrow{\bs{y}}_1^{(T-(t-1))} \Big) \in \mathcal{T}_{\big[U \ U_1 \ V_1  \ X_{1,P} \ U_2 \ V_2  \ \overrightarrow{Y}_1\big]}^{(N, e)}, 
\end{IEEEeqnarray}
where $W_{1,C1}^{(T-(t-1))}$ and $W_{2,C1}^{(T-(t-1))}$ are assumed to be perfectly decoded in the previous decoding step $t-1$. The set $\mathcal{T}_{\big[U \ U_1 \ V_1  \ X_{1,P} \ U_2 \ V_2  \ \overrightarrow{Y}_1\big]}^{(N, e)}$ represents the set of jointly typical sequences of the random variables $U, U_1, V_1, X_{1,P}, U_2, V_2$, and $\overrightarrow{Y}_1$, with $e>0$.
Receiver $2$ follows a similar decoding scheme.

\noindent
\textbf{Probability of Error Analysis}: An error might occur during encoding step $t$ if the message index $W_{2,C1}^{(t-1)}$ is not correctly decoded at transmitter $1$. From the asymptotic equipartion property (AEP) \cite{Cover-Book-1991}, it follows that the message index $W_{2,C1}^{(t-1)}$ can be reliably decoded at transmitter $1$ during encoding step $t$, under the condition:
\begin{eqnarray}
\nonumber
R_{2,C1} & \leqslant & I\left( \overleftarrow{Y}_1 ; U_2  | U, U_1, V_1, X_1 \right) \\
\label{EqConditionNoError1}
&=& I\left( \overleftarrow{Y}_1 ; U_2  | U, X_1  \right).
\end{eqnarray} 
An error might occur during the (backward) decoding step $t$ if the message indices $W_{1,C1}^{(T-t)}$, $W_{2,C1}^{(T-t)}$, $W_{1,C2}^{(T-(t-1))}, W_{1,P}^{(T-(t-1))}$, and $W_{2,C2}^{(T-(t-1))}$ are not decoded correctly given that the message indices $W_{1,C1}^{(T-(t-1))}$ and $W_{2,C1}^{(T-(t-1))}$ were correctly decoded in the previous decoding step $t-1$. 
These errors might arise for two reasons: $(i)$ there does not exist a tuple $\Big(\widehat{W}_{1,C1}^{(T-t)}$, $\widehat{W}_{2,C1}^{(T-t)}, \widehat{W}_{1,C2}^{(T-(t-1))},\widehat{W}_{1,P}^{(T-(t-1))},\widehat{W}_{2,C2}^{(T-(t-1))}\Big)$  that satisfies \eqref{EqDecodingW1cW2cW1p}, or $(ii)$ there exist several tuples $\Big(\widehat{W}_{1,C1}^{(T-t)}, \widehat{W}_{2,C1}^{(T-t)}, \widehat{W}_{1,C2}^{(T-(t-1))},\widehat{W}_{1,P}^{(T-(t-1))}, \widehat{W}_{2,C2}^{(T-(t-1))}\Big)$ that simultaneously satisfy \eqref{EqDecodingW1cW2cW1p}. 
From the asymptotic equipartion property (AEP) \cite{Cover-Book-1991}, the probability of an error due to $(i)$ tends to zero when $N$ grows to infinity. Consider the error due to $(ii)$ and define the event $E_{(s, r, l, q, m)}$ that describes the case in which the codewords $\big(\bs{u}(s,r)$, $\bs{u}_1(s,r,W_{1,C1}^{(T-(t-1))})$, $\bs{v}_1(s,r,W_{1,C1}^{(T-(t-1))},l)$, $\bs{x}_{1,P}(s,r,W_{1,C1}^{(T-(t-1))},l,q)$, $\bs{u}_2(s,r,W_{2,C1}^{(T-(t-1))})$, and $\bs{v}_2(s,r,W_{2,C1}^{(T-(t-1))},m)\big)$ are jointly typical with $ \overrightarrow{\bs{y}}_1^{(T-(t-1))}$ during decoding step $t$. 
Assume now that the codeword to be decoded at decoding step $t$ corresponds to the indices $(s,r,l,q,m) = (1,1,1,1,1)$ without loss of generality due to the symmetry of the code. Then, the probability of error due to $(ii)$ during decoding step $t$, can be  bounded as follows:

\begin{IEEEeqnarray}{lcl}
\nonumber
P_e & = &\pr{\ds\bigcup_{(s,r,l,q,m) \neq (1,1,1,1,1)} E_{(s,r,l,q,m)} }\\
\label{EqConditionNoError2a}
& \leqslant & \ds\sum_{\scriptscriptstyle (s, r, l, q, m) \in \mathcal{T}} \pr{ E_{(s,r,l,q,m)}}, 
\end{IEEEeqnarray}
with $\mathcal{T}=\Big\lbrace \lbrace 1$,$ 2,  \ldots 2^{NR_{1,C1}} \rbrace \times \lbrace 1$,$ 2,  \ldots 2^{NR_{2,C1}} \rbrace \times \lbrace 1$,$ 2,  \ldots 2^{NR_{1,C2}} \rbrace \times \lbrace 1$,$ 2,  \ldots 2^{NR_{1,P}} \rbrace \times \lbrace 1$ , $ 2,  \ldots 2^{NR_{2,C2}} \rbrace \Big\rbrace \setminus \lbrace (1,1,1,1,1) \rbrace$.

\noindent
From AEP \cite{Cover-Book-1991}, it follows that

\begin{IEEEeqnarray}{lcl}
\nonumber
P_e& \leqslant & 2^{N (R_{2,C2} - I(\overrightarrow{Y}_1;V_2 | U, U_1, U_2, V_1, X_1) + 2\epsilon) }+2^{N (R_{1,P} - I(\overrightarrow{Y}_1;X_1 | U, U_1, U_2, V_1, V_2) + 2\epsilon) }\\
 \nonumber
 & & +2^{N (R_{2,C2} +R_{1,P} - I(\overrightarrow{Y}_1;V_2, X_1 | U, U_1, U_2, V_1) + 2\epsilon) }+2^{N (R_{1,C2} - I(\overrightarrow{Y}_1; V_1, X_1 | U, U_1, U_2, V_2) + 2\epsilon) } \\
 \nonumber
 & & +2^{N (R_{1,C2} +R_{2,C2} - I(\overrightarrow{Y}_1;V_1,V_2, X_1 | U, U_1, U_2) + 2\epsilon) }+2^{N (R_{1,C2} +R_{1,P} - I(\overrightarrow{Y}_1; V_1, X_1 | U, U_1, U_2, V_2) + 2\epsilon) }\\
 \nonumber
 & & +2^{N (R_{1,C2} + R_{1,P}+R_{2,C2} - I(\overrightarrow{Y}_1;V_1, V_2, X_1 | U, U_1, U_2) + 2\epsilon) }+2^{N (R_{2,C1}  - I(\overrightarrow{Y}_1; U, U_1, U_2, V_1, V_2, X_1) + 2\epsilon) }\\ 
\nonumber
 & & +2^{N (R_{2,C1} + R_{2,C2} - I(\overrightarrow{Y}_1; U, U_1, U_2, V_1, V_2, X_1) + 2\epsilon) }+2^{N (R_{2,C1} +R_{1,P} - I(\overrightarrow{Y}_1; U, U_1, U_2, V_1, V_2, X_1) + 2\epsilon) } \\
  \nonumber
& & +2^{N (R_{2,C1} + R_{1,P}+R_{2,C2} - I(\overrightarrow{Y}_1; U, U_1, U_2, V_1, V_2, X_1) + 2\epsilon) } +2^{N (R_{2,C1} +R_{1,C2}  - I(\overrightarrow{Y}_1; U, U_1, U_2, V_1, V_2, X_1) + 2\epsilon) }\\
\nonumber
& & +2^{N (R_{2,C1} +R_{1,C2}+R_{2,C2} - I(\overrightarrow{Y}_1; U, U_1, U_2, V_1, V_2, X_1) + 2\epsilon) }\\
\nonumber
& & +2^{N (R_{2,C1} +R_{1,C2}+R_{1,P}  - I(\overrightarrow{Y}_1; U, U_1, U_2, V_1, V_2, X_1) + 2\epsilon) }\\
\nonumber
& & +2^{N (R_{2,C} +R_{1,C2}+R_{1,P} - I(\overrightarrow{Y}_1; U, U_1, U_2, V_1, V_2, X_1) + 2\epsilon) }+2^{N (R_{1,C1} - I(\overrightarrow{Y}_1; U, U_1, U_2, V_1, V_2, X_1) + 2\epsilon) }\\
\nonumber
& & +2^{N (R_{1,C1} +R_{2,C2} - I(\overrightarrow{Y}_1; U, U_1, U_2, V_1, V_2, X_1) + 2\epsilon) } +2^{N (R_{1,C1} +R_{1,P} - I(\overrightarrow{Y}_1; U, U_1, U_2, V_1, V_2, X_1) + 2\epsilon) }
\end{IEEEeqnarray}
\begin{IEEEeqnarray}{lcl}
\nonumber
& & +2^{N (R_{1,C1} +R_{1,P}+R_{2,C2} - I(\overrightarrow{Y}_1; U, U_1, U_2, V_1, V_2, X_1) + 2\epsilon) } +2^{N (R_{1,C1} +R_{1,C2} - I(\overrightarrow{Y}_1; U, U_1, U_2, V_1, V_2, X_1) + 2\epsilon) } \\
\nonumber
& & +2^{N (R_{1,C1} +R_{1,C2}+R_{2,C2} - I(\overrightarrow{Y}_1; U, U_1, U_2, V_1, V_2, X_1) + 2\epsilon) } \\
\nonumber
& & +2^{N (R_{1,C1} +R_{1,C2}+R_{1,P} - I(\overrightarrow{Y}_1; U, U_1, U_2, V_1, V_2, X_1) + 2\epsilon) } \\
\nonumber
& & +2^{N (R_{1,C1} +R_{1,C2}+R_{1,P}+R_{2,C2} - I(\overrightarrow{Y}_1; U, U_1, U_2, V_1, V_2, X_1) + 2\epsilon) } \\
\nonumber
& & +2^{N (R_{1,C1} +R_{2,C1} - I(\overrightarrow{Y}_1; U, U_1, U_2, V_1, V_2, X_1) + 2\epsilon) }+2^{N (R_{1,C1} +R_{2,C1}+R_{2,C2}  - I(\overrightarrow{Y}_1; U, U_1, U_2, V_1, V_2, X_1) + 2\epsilon) }\\
\nonumber
& & +2^{N (R_{1,C1} +R_{2,C1}+R_{1,P} - I(\overrightarrow{Y}_1; U, U_1, U_2, V_1, V_2, X_1) + 2\epsilon) } \\
\nonumber
& & +2^{N (R_{1,C1} +R_{2,C1}+R_{1,P}+R_{2,C2} - I(\overrightarrow{Y}_1; U, U_1, U_2, V_1, V_2, X_1) + 2\epsilon) }\\
\nonumber
& & +2^{N (R_{1,C1} +R_{2,C1}+R_{1,C2} - I(\overrightarrow{Y}_1; U, U_1, U_2, V_1, V_2, X_1) + 2\epsilon) }\\
\nonumber
& & +2^{N (R_{1,C1} +R_{2,C1}+R_{1,C2} +R_{2,C2}  - I(\overrightarrow{Y}_1; U, U_1, U_2, V_1, V_2, X_1) + 2\epsilon) }\\
\nonumber
& & +2^{N (R_{1,C1} +R_{2,C1}+R_{1,C2} +R_{1,P} - I(\overrightarrow{Y}_1; U, U_1, U_2, V_1, V_2, X_1) + 2\epsilon) }+2^{N (R_{1} +R_{2,C} - I(\overrightarrow{Y}_1; U, U_1, U_2, V_1, V_2, X_1) + 2\epsilon) }.\\
\label{EqConditionNoError2b}
\end{IEEEeqnarray}
The same analysis of the probability of error holds for transmitter-receiver pair $2$.
Hence, in general, from \eqref{EqConditionNoError1} and \eqref{EqConditionNoError2b}, reliable decoding holds under the following conditions for transmitter $i \in \lbrace1,2 \rbrace$, with $j \in  \lbrace1,2 \rbrace\setminus\lbrace i \rbrace$:

\begin{subequations}
\label{EqRateRegion-z}
\begin{IEEEeqnarray}{rcl}
\nonumber
R_{j,C1}  & \leqslant &  I\left( \overleftarrow{Y}_i ; U_j  | U, U_i, V_i,X_i  \right) \\ 
\nonumber
&=& I\left( \overleftarrow{Y}_i ; U_j  | U, X_i  \right)\\
\label{EqRateRegion0}
& \triangleq &  \theta_{1,i}, \\
\nonumber
R_{i} + R_{j,C}  & \leqslant &  I(\overrightarrow{Y}_i; U,U_i, U_j,V_i, V_j, X_i) \\
\nonumber
&=& I(\overrightarrow{Y}_i; U, U_j,V_j, X_i) \\
\label{EqRateRegion1}
&\triangleq&  \theta_{2,i}, \\
\nonumber
R_{j,C2}  & \leqslant & I(\overrightarrow{Y}_i; V_j | U, U_i, U_j, V_i, X_i)\\
\nonumber
&=& I(\overrightarrow{Y}_i; V_j | U, U_j, X_i) \\
\label{EqRateRegion2}
&\triangleq&  \theta_{3,i}, \\
\nonumber
R_{i,P}    &\leqslant &  I(\overrightarrow{Y}_i; X_i | U, U_i, U_j,V_i, V_j) \\
\label{EqRateRegion3}
&\triangleq& \theta_{4,i}, \\
\nonumber
R_{i,P}+R_{j,C2}  & \leqslant & I(\overrightarrow{Y}_i; V_j, X_i | U, U_i, U_j, V_i) \\
\label{EqRateRegion4}
&\triangleq& \theta_{5,i}, \\
\nonumber
R_{i,C2}+R_{i,P}  & \leqslant & I(\overrightarrow{Y}_i; V_i,X_i | U, U_i, U_j, V_j) \\
\nonumber
&=& I(\overrightarrow{Y}_i; X_i | U, U_i, U_j, V_j) \\
\label{EqRateRegion5}
&\triangleq& \theta_{6,i},  \mbox{ and }\\
\nonumber
R_{i,C2}+R_{i,P}+R_{j,C2} & \leqslant & I(\overrightarrow{Y}_i; V_i, V_j, X_i | U, U_i, U_j) \\
\nonumber
&=& I(\overrightarrow{Y}_i; V_j, X_i | U, U_i, U_j) \\
\label{EqRateRegion6}
&\triangleq& \theta_{7,i}.
\end{IEEEeqnarray}
\end{subequations}
Taking into account that $R_i=R_{i,C1}+R_{i,C2}+R_{i,P}$, a Fourier-Motzkin elimination process in \eqref{EqRateRegion-z} yields:
\begin{subequations}
\label{EqRateRegion2b}
\begin{IEEEeqnarray}{rcl}
\label{EqRateRegion21}
R_{1}  & \leqslant & \min\left(\theta_{2,1},\theta_{6,1}+\theta_{1,2},\theta_{4,1}+\theta_{1,2}+\theta_{3,2}\right),  \\ 
\label{EqRateRegion22}
R_{2}   & \leqslant & \min\left(\theta_{2,2},\theta_{1,1}+a_{6,2},\theta_{1,1}+\theta_{3,1}+\theta_{4,2}\right), \\
\nonumber
R_{1}+R_{2}  & \leqslant & \min(\theta_{2,1}+\theta_{4,2}, \theta_{2,1}+a_{6,2}, \theta_{4,1}+\theta_{2,2}, \theta_{6,1}+\theta_{2,2}, \theta_{1,1}+\theta_{3,1}+\theta_{4,1}+\theta_{1,2}+\theta_{5,2},  \\
\nonumber
& & \theta_{1,1}+\theta_{7,1}+\theta_{1,2}+\theta_{5,2}, \theta_{1,1}+\theta_{4,1}+\theta_{1,2}+\theta_{7,2}, \theta_{1,1}+\theta_{5,1}+\theta_{1,2}+\theta_{3,2}+\theta_{4,2}, \\
\label{EqRateRegion23}
& & \theta_{1,1}+\theta_{5,1}+\theta_{1,2}+\theta_{5,2}, \theta_{1,1}+\theta_{7,1}+\theta_{1,2}+\theta_{4,2}), \\
\nonumber
2R_{1}+R_{2}  & \leqslant & \min(\theta_{2,1}+\theta_{4,1}+\theta_{1,2}+\theta_{7,2}, \theta_{1,1}+\theta_{4,1}+\theta_{7,1}+2\theta_{1,2}+\theta_{5,2}, \theta_{2,1}+\theta_{4,1}+\theta_{1,2}+\theta_{5,2}), \\
\label{EqRateRegion24}\\
\nonumber
R_{1}+2R_{2}  & \leqslant & \min(\theta_{1,1}+\theta_{5,1}+\theta_{2,2}+\theta_{4,2}, \theta_{1,1}+\theta_{7,1}+\theta_{2,2}+\theta_{4,2}, 2\theta_{1,1}+\theta_{5,1}+\theta_{1,2}+\theta_{4,2}+\theta_{7,2}), \\
\label{EqRateRegion25}
\end{IEEEeqnarray}
\end{subequations}
where $\theta_{l,i}$ are defined in \eqref{EqRateRegion-z} with $(l,i) \in \lbrace1,  \ldots, 7 \rbrace \times \lbrace 1, 2 \rbrace$.

\noindent
In the LD-IC-NOF model, the channel input of transmitter $i$ at each channel use is a $q$-dimensional vector $\bs{X}_{i} \in \lbrace 0, 1 \rbrace^{q}$ with $i \in \lbrace 1, 2 \rbrace$ and $q$ as defined in \eqref{Eqq}. Following this observation, the random variables $U$, $U_i$, $V_i$, and $X_{i,P}$ described in \eqref{Eqprobdist} in the codebook generation are also vectors, and thus, in this subsection, they are denoted by $\bs{U}$, $\bs{U}_i$, $\bs{V}_i$ and $\bs{X}_{i,P}$, respectively.

\noindent
The random variables $\bs{U}_i$, $\bs{V}_i$, and $\bs{X}_{i,P}$ are assumed to be mutually independent and uniformly distributed over the sets $\lbrace 0,1\rbrace^{\left(n_{ji}-\left(\max\left(\overrightarrow{n}_{jj},n_{ji}\right)-\overleftarrow{n}_{jj}\right)^+\right)^+}$, 
$\lbrace 0,1\rbrace^{\left(\min\left(n_{ji},\left(\max\left(\overrightarrow{n}_{jj},n_{ji}\right)-\overleftarrow{n}_{jj}\right)^+\right)\right)}$ and 
$\lbrace 0,1\rbrace^{\left(\overrightarrow{n}_{ii}-n_{ji}\right)^+}$, respectively.
Note that the random variables $\bs{U}_i$, $\bs{V}_i$, and $\bs{X}_{i,P}$ have the following dimensions:
\begin{subequations}
\label{Eqdim}
\begin{IEEEeqnarray}{lcl}
\label{EqdimUi}
\dim \bs{U}_i &=& \left(n_{ji}-\left(\max\left(\overrightarrow{n}_{jj},n_{ji}\right)-\overleftarrow{n}_{jj}\right)^+\right)^+,\\
\label{EqdimVi}
\dim \bs{V}_i &=& \min \left(n_{ji},\left(\max\left(\overrightarrow{n}_{jj},n_{ji}\right)-\overleftarrow{n}_{jj}\right)^+\right), \mbox{ and } \qquad\\
\label{EqdimXiP}
\dim \bs{X}_{i,P} &=&\left(\overrightarrow{n}_{ii}-n_{ji}\right)^+.
\end{IEEEeqnarray}
\end{subequations}  
These dimensions satisfy the following condition: 
\begin{equation}
\dim \bs{U}_i + \dim \bs{V}_i + \dim \bs{X}_{i,P} = \max\left(\overrightarrow{n}_{ii},n_{ji}\right) \leqslant q.
\end{equation}

\noindent
Note that the random variable $\bs{U}$ in \eqref{Eqprobdist} is not used, and therefore, is a constant. The input symbol of transmitter $i$ during channel use $n$ is ${\bs{X}_{i} = \left(\bs{U}_i^{\sfT}, \bs{V}_i^{\sfT}, \bs{X}_{i,P}^{\sfT}, \left(0, \ldots, 0\right)\right)^{\sfT}}$, where $\left(0, \ldots, 0\right)$ is put to meet the dimension constraint $\dim \bs{X}_{i} = q$. 
Hence, during block $t \in \lbrace 1, 2,  \ldots, T  \rbrace$, the codeword $\bs{X}_i^{(t)}$ in the LD-IC-NOF is a $q \times N$ matrix, i.e., $\bs{X}_i^{(t)} = \left( \bs{X}_{i,1}, \bs{X}_{i,2} \ldots, \bs{X}_{i,N} \right) \in \lbrace 0,1\rbrace^{q \times N}$.

\begin{figure*}[ht!]
 \centerline{\epsfig{figure=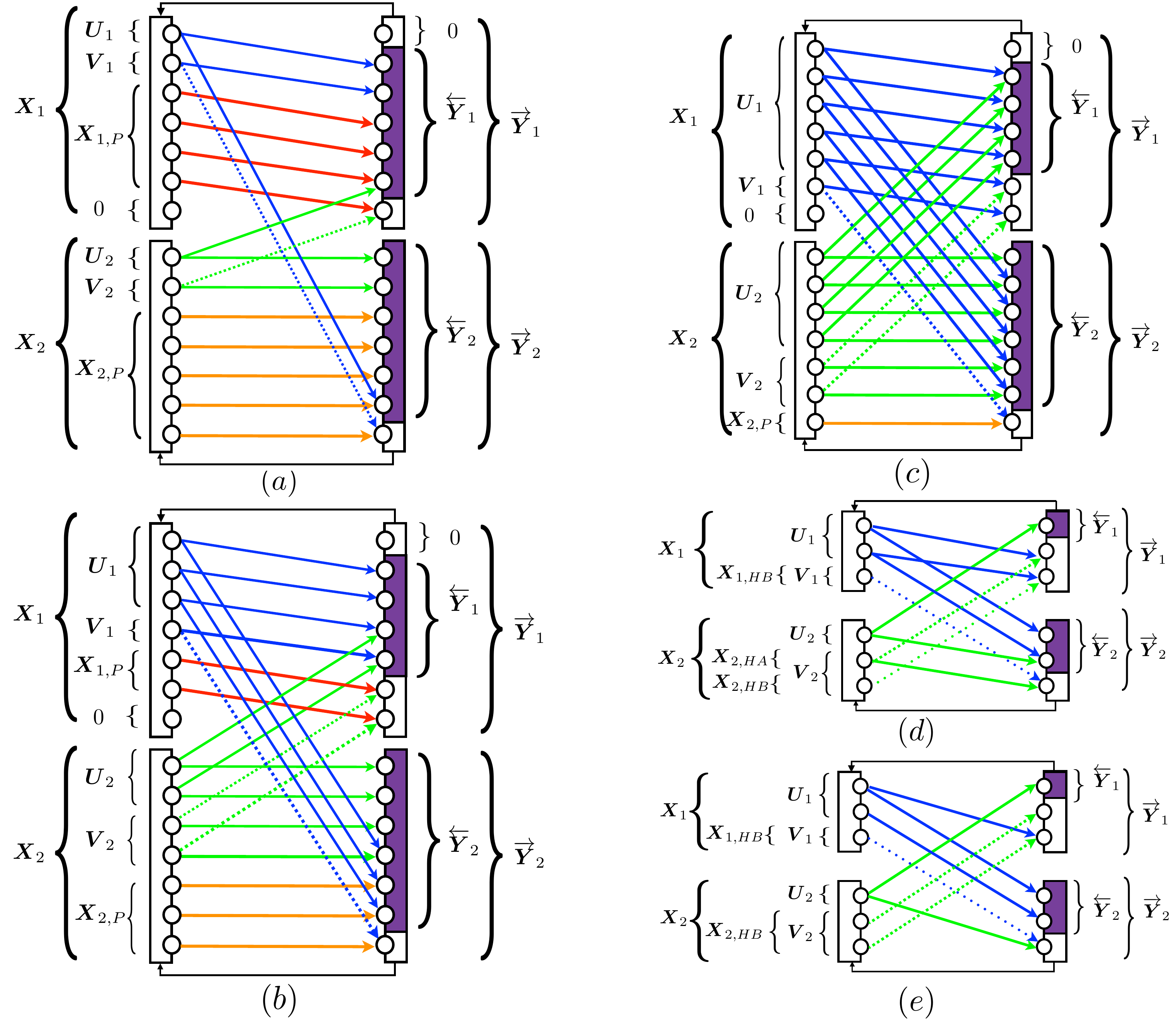,width=0.9\textwidth}}
 \caption{The auxiliary random variables and their relation with signals when channel-output feedback is considered in $(a)$ very weak interference regime, $(b)$ weak interference regime, $(c)$ moderate interference regime, $(d)$ strong interference regime and $(e)$ very strong interference regime.}
\label{FigachievabilityMessagesLDICNFB}
\end{figure*}

\noindent
The intuition behind this choice is based on the following observations: 
$(a)$ The vector $\bs{U}_i$ represents the bits in $\bs{X}_i$ that can be observed by transmitter $j$ via feedback but no necessarily by receiver $i$;
$(b)$ The vector $\bs{V}_i$ represents the bits in $\bs{X}_i$ that can be observed by receiver $j$ but no necessarily by receiver $i$; and finally,
$(c)$ The vector $\bs{X}_{i,P}$ is a notational artefact to denote the bits of $\bs{X}_i$ that are neither in $\bs{U}_i$ nor $\bs{V}_i$. In particular, the bits in $\bs{X}_{i,P}$ are only observed by receiver $i$, as shown in Figure~\ref{FigachievabilityMessagesLDICNFB}.
This intuition justifies the dimensions described in \eqref{Eqdim}.

\noindent
Considering this particular code structure, the following holds for the terms $\theta_{l,i}$, with $(l,i) \in \lbrace 1, \ldots, 7 \rbrace \times \lbrace 1, 2 \rbrace$, in \eqref{EqRateRegion-z}: 
\begin{subequations}
\label{EqRateRegionLDM}
\begin{IEEEeqnarray}{rcl}
\nonumber
\theta_{1,i}&=& I\Big( \overleftarrow{\bs{Y}}_i ; \bs{U}_j  | \bs{U}, \bs{X}_i  \Big) \\
\nonumber
&\stackrel{(a)}{=}& H\left( \overleftarrow{\bs{Y}}_i  | \bs{U}, \bs{X}_i \right) \\
\nonumber
&=& H\left(\bs{U}_j\right) \\
\label{EqRateRegion26}
&=& \left(n_{ij}-\left(\max\left(\overrightarrow{n}_{ii},n_{ij}\right)-\overleftarrow{n}_{ii}\right)^+\right)^+; \\
\nonumber
\theta_{2,i}&=&I\Big(\overrightarrow{\bs{Y}}_i; \bs{U}, \bs{U}_j,\bs{V}_j, \bs{X}_i\Big) \\
\nonumber
&\stackrel{(b)}{=}&  H\left(\overrightarrow{\bs{Y}}_i\right) \\
\label{EqRateRegion27}
&=&  \max\left(\overrightarrow{n}_{ii},n_{ij}\right); 
\end{IEEEeqnarray}
\begin{IEEEeqnarray}{rcl}
\nonumber
\theta_{3,i}&=&I\Big(\overrightarrow{\bs{Y}}_i; \bs{V}_j | \bs{U}, \bs{U}_j, \bs{X}_i\Big)  \\
\nonumber
&\stackrel{(b)}{=}& H\left(\overrightarrow{\bs{Y}}_i | \bs{U}, \bs{U}_j, \bs{X}_i\right) \\
\nonumber
&=& H\left(\bs{V}_j\right) \\
\label{EqRateRegion28}
&=& \min \left(n_{ij},\left(\max\left(\overrightarrow{n}_{ii},n_{ij}\right)-\overleftarrow{n}_{ii}\right)^+\right); \\
\nonumber
\theta_{4,i}&=&I\Big(\overrightarrow{\bs{Y}}_i; \bs{X}_i | \bs{U}, \bs{U}_i, \bs{U}_j, \bs{V}_i, \bs{V}_j\Big) \\
\nonumber
\nonumber
&\stackrel{(b)}{=}&  H\left(\overrightarrow{\bs{Y}}_i | \bs{U}, \bs{U}_i, \bs{U}_j, \bs{V}_i, \bs{V}_j\right) \\
\nonumber
&=&  H\left(\bs{X}_{i,P}\right) \\
\label{EqRateRegion29}
&=& \left(\overrightarrow{n}_{ii}-n_{ji}\right)^+; \mbox{ and }  \\
\nonumber
\theta_{5,i}&=&I\Big(\overrightarrow{\bs{Y}}_i; \bs{V}_j, \bs{X}_i | \bs{U}, \bs{U}_i, \bs{U}_j, \bs{V}_i\Big)  \\
\nonumber
 &\stackrel{(b)}{=}&  H\left(\overrightarrow{\bs{Y}}_i | \bs{U}, \bs{U}_i, \bs{U}_j, \bs{V}_i\right) \\
\nonumber
&=& \max\left(\dim \bs{X}_{i,P}, \dim \bs{V}_j \right) \\ 
\label{EqRateRegion30}
&=&  \max \Big(\left(\overrightarrow{n}_{ii}-n_{ji}\right)^+, \min\left(n_{ij},\left(\max\left(\overrightarrow{n}_{ii},n_{ij}\right)-
\overleftarrow{n}_{ii}\right)^+\right)\Big),
\end{IEEEeqnarray}
where
(a) follows from the fact that $H\left( \overleftarrow{\bs{Y}}_i  | \bs{U}, \bs{U}_j, \bs{X}_i  \right)=0$; and 
(b) follows from the fact that $H(\overrightarrow{\bs{Y}}_i|\bs{U}, \bs{U}_j,\bs{V}_j, \bs{X}_i)=0$.

\noindent
For the calculation of the last two mutual information terms in inequalities \eqref{EqRateRegion5} and \eqref{EqRateRegion6}, special notation is used. Let for instance the vector $\bs{V}_i$ be the concatenation 
of the vectors $\bs{X}_{i,HA}$ and $\bs{X}_{i,HB}$, i.e., $\bs{V}_i=\left(\bs{X}_{i,HA}, \bs{X}_{i,HB}\right)$. The vector $\bs{X}_{i,HA}$ is the part of $\bs{V}_i$ that is available in both receivers. The vector  $\bs{X}_{i,HB}$ is the part of $\bs{V}_i$ that is exclusively available in receiver $j$ (see Figure~\ref{FigachievabilityMessagesLDICNFB}). Note that  ${H\left(\bs{V}_i\right)=H\left(\bs{X}_{i,HA}\right)+H\left(\bs{X}_{i,HB}\right)}$. Note also that the vectors $\bs{X}_{i,HA}$ and $\bs{X}_{i,HB}$ possess the following dimensions: 
\begin{IEEEeqnarray}{rcl}
\nonumber
\dim \bs{X}_{i,HA} &=& \min\left(n_{ji},\left(\max\left(\overrightarrow{n}_{jj},n_{ji}\right)-\overleftarrow{n}_{jj}\right)^+\right)\!-\!\min\big(\left(n_{ji}\!-\!\overrightarrow{n}_{ii}\right)^+,\left(\max\left(\overrightarrow{n}_{jj},n_{ji}\right)\!-\!\overleftarrow{n}_{jj}\right)^+\big)\\
\nonumber
\dim \bs{X}_{i,HB} &=& \min\big(\left(n_{ji}-\overrightarrow{n}_{ii}\right)^+,\left(\max\left(\overrightarrow{n}_{jj},n_{ji}\right)-\overleftarrow{n}_{jj}\right)^+\big).
\end{IEEEeqnarray}
Using this notation, the following holds:
\begin{IEEEeqnarray}{rcl}
\nonumber
\theta_{6,i}&=&I\Big(\overrightarrow{\bs{Y}}_i; \bs{X}_i | \bs{U}, \bs{U}_i, \bs{U}_j, \bs{V}_j\Big) \\
\nonumber
&\stackrel{(c)}{=}& H\left(\overrightarrow{\bs{Y}}_i | \bs{U}, \bs{U}_i, \bs{U}_j, \bs{V}_j\right)\\
\nonumber
&= & H\left(\bs{X}_{i,HA},\bs{X}_{i,P}\right)\\
\nonumber
&=& \dim \bs{X}_{i,HA}+ \dim \bs{X}_{i,P}\\
\nonumber
&=&  \min\left(n_{ji},\left(\max\left(\overrightarrow{n}_{jj},n_{ji}\right)-\overleftarrow{n}_{jj}\right)^+\right)-\min\big(\left(n_{ji}-\overrightarrow{n}_{ii}\right)^+,\left(\max\left(\overrightarrow{n}_{jj},n_{ji}\right)-\overleftarrow{n}_{jj}\right)^+\big)\\
\label{EqRateRegion31}
& &+\left(\overrightarrow{n}_{ii}-n_{ji}\right)^+;   \textrm{ and }  \\
\nonumber
\theta_{7,i}&=&I\Big(\overrightarrow{\bs{Y}}_i; \bs{V}_j, \bs{X}_i | \bs{U}, \bs{U}_i, \bs{U}_j\Big) \\
\nonumber
&=& I\left(\overrightarrow{\bs{Y}}_i; \bs{X}_i | \bs{U}, \bs{U}_i, \bs{U}_j\right)+I\left(\overrightarrow{\bs{Y}}_i; \bs{V}_j  | \bs{U}, \bs{U}_i, \bs{U}_j, \bs{X}_i\right) \\
\nonumber
&=& I\left(\overrightarrow{\bs{Y}}_i; \bs{X}_i | \bs{U}, \bs{U}_i, \bs{U}_j\right)+I\left(\overrightarrow{\bs{Y}}_i; \bs{V}_j  | \bs{U}, \bs{U}_j, \bs{X}_i\right) \\
\nonumber
&\stackrel{(c)}{=}& H\left(\overrightarrow{\bs{Y}}_i | \bs{U}, \bs{U}_i, \bs{U}_j\right)
\end{IEEEeqnarray}
\begin{IEEEeqnarray}{rcl}
\nonumber
&=& \max\left(H\left(\bs{V}_j\right), H\left(\bs{X}_{i,HA}\right)+H\left(\bs{X}_{i,P}\right)\right) \\ 
\nonumber
&=& \max\left(\dim \bs{V}_j, \dim \bs{X}_{i,HA}+\dim \bs{X}_{i,P}\right) \\ 
\nonumber
&=&  \max\big(\min\big(n_{ij},\left(\max\left(\overrightarrow{n}_{ii},n_{ij}\right)-\overleftarrow{n}_{ii}\right)^+\big), \min\big(n_{ji},\left(\max\left(\overrightarrow{n}_{jj},n_{ji}\right)-\overleftarrow{n}_{jj}\right)^+\big)\\
\label{EqRateRegion32}
& & -\min\big(\left(n_{ji}-\overrightarrow{n}_{ii}\right)^+,\left(\max\left(\overrightarrow{n}_{jj},n_{ji}\right)-\overleftarrow{n}_{jj}\right)^+\big)+\left(\overrightarrow{n}_{ii}-n_{ji} \right)^+\big);
\end{IEEEeqnarray}
\end{subequations}
where
(c) follows from the fact that $H(\overrightarrow{\bs{Y}}_i|\bs{U}, \bs{U}_j,\bs{V}_j, \bs{X}_i)=0$.

\noindent
Plugging \eqref{EqRateRegionLDM} into  \eqref{EqRateRegion2b}  (after some trivial manipulations) yields the system of inequalities in Theorem~\ref{TheoremANFBLDMCap}.  

\noindent
The sum-rate bound in \eqref{EqRateRegion23} can be simplified as follows:
\begin{IEEEeqnarray}{rcl}
\label{EqSumrateRegion2bsimplified}
R_{1}+R_{2}   &\leqslant&  \min(\theta_{2,1}+\theta_{4,2}, \theta_{4,1}+\theta_{2,2}, \theta_{1,1}+\theta_{5,1}+\theta_{1,2}+\theta_{5,2}). 
\end{IEEEeqnarray}
Note that this follows from the realization that $\max(\theta_{2,1}+\theta_{4,2}, \theta_{4,1}+\theta_{2,2}, \theta_{1,1}+\theta_{5,1}+\theta_{1,2}+\theta_{5,2})\leqslant\min(\theta_{2,1}+a_{6,2}, \theta_{6,1}+\theta_{2,2}, \theta_{1,1}+\theta_{3,1}+\theta_{4,1}+\theta_{1,2}+\theta_{5,2}, \theta_{1,1}+\theta_{7,1}+\theta_{1,2}+\theta_{5,2},\theta_{1,1}+\theta_{4,1}+\theta_{1,2}+\theta_{7,2},  \theta_{1,1}+\theta_{5,1}+\theta_{1,2}+\theta_{3,2}+\theta_{4,2}, \theta_{1,1}+\theta_{7,1}+\theta_{1,2}+\theta_{4,2})$.

\clearpage

 \section{Proof of Converse} \label{App-C-LD-IC-NOF}

This appendix provides a converse proof for Theorem~\ref{TheoremANFBLDMCap}. 
Inequalities \eqref{EqRiV2} and \eqref{EqRi+Rj-1-V2} correspond to the minimum cut-set bound \cite{Shannon-IRETIT-1956} and the sum-rate bound for the case of the two-user LD-IC-POF. The proofs of these bounds are presented in  \cite{Suh-TIT-2011}. 
The rest of this appendix provides a proof of the inequalities \eqref{EqRi-2-V2}, \eqref{EqRi+Rj-2-V2} and \eqref{Eq2Ri+Rj-V2}.
 
\begin{figure*}[t]
 \centerline{\epsfig{figure=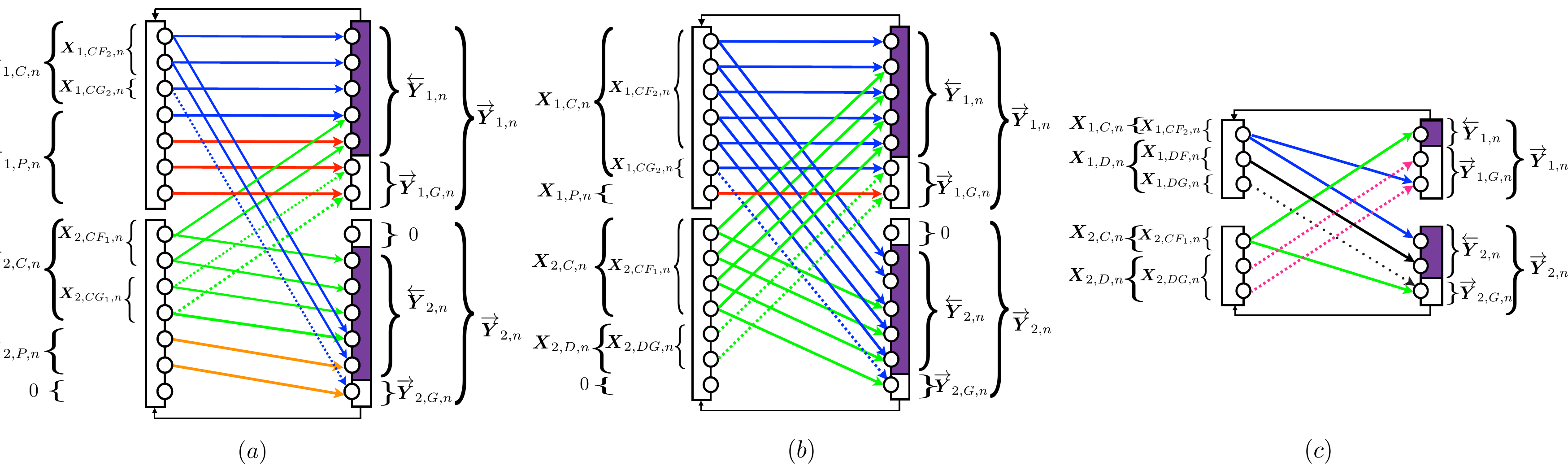,width=1.0\textwidth}}
 \caption{Example of the notation of the channel inputs and the channel outputs when channel-output feedback is considered.}
\label{FigMessagesLDICNFB}
\vspace{-0.2in}
\end{figure*}

\noindent
\textbf{Notation.}  For all $i \in \lbrace 1,2 \rbrace$, the channel input $\bs{X}_{i,n}$ of the LD-IC-NOF in \eqref{EqLDICsignals} for any channel use $n \in \lbrace 1, 2,  \ldots, N\rbrace$  is  a $q$-dimensional vector, with $q$ in \eqref{Eqq}, that can be written as the concatenation of four vectors: $\bs{X}_{i,C,n}$, $\bs{X}_{i,P,n}$, $\bs{X}_{i,D,n}$, and $\bs{X}_{i,Q,n}$, i.e., $\bs{X}_{i,n} = \Big(\bs{X}_{i,C,n}^{\sfT},\bs{X}_{i,P,n}^{\sfT}, \bs{X}_{i,D,n}^{\sfT},\bs{X}_{i,Q,n}^{\sfT}\Big)^{\sfT}$, as shown in Figure~\ref{FigMessagesLDICNFB}. Note that this notation is independent of the feedback parameters $\overleftarrow{n}_{11}$ and $\overleftarrow{n}_{22}$, and it holds for all $n\in\{1, 2, \ldots,N\}$. More specifically, 
\begin{subequations}

\noindent
$\bs{X}_{i,C,n}$ represents the bits of $\bs{X}_{i,n}$ that are observed by both receivers. Then, 
\begin{IEEEeqnarray}{lcl}
\label{EqdimXic}
\dim \bs{X}_{i,C,n} & = & \min\left( \overrightarrow{n}_{ii}, n_{ji} \right) ; 
\end{IEEEeqnarray}

\noindent
$\bs{X}_{i,P,n}$ represents the bits of $\bs{X}_{i,n}$ that are observed only at receiver $i$. Then, 
\begin{IEEEeqnarray}{lcl}
\label{EqdimXip}
\dim\bs{X}_{i,P,n} & = & (\overrightarrow{n}_{ii} - n_{ji})^+; 
\end{IEEEeqnarray}

\noindent
$\bs{X}_{i,D,n}$ represents the bits of $\bs{X}_{i,n}$ that are observed only at receiver $j$. Then, 
\begin{IEEEeqnarray}{lcl}
\label{EqdimXid}
\dim \bs{X}_{i,D,n} & = & (n_{ji} - \overrightarrow{n}_{ii})^+;\mbox{ and }
\end{IEEEeqnarray}

\noindent
$\bs{X}_{i,Q,n}=\left(0,\ldots,0\right)^{\sfT}$  is included for dimensional matching of the model in \eqref{EqLDICsignalsc}. Then,
\begin{IEEEeqnarray}{lcl}
\label{EqdimXid}
\dim \bs{X}_{i,Q,n} & = & q - \max\left( \overrightarrow{n}_{ii}, n_{ji} \right).
\end{IEEEeqnarray}
The bits $\bs{X}_{i,Q,n}$ are fixed and thus do not carry any information. Hence, the following holds:
\begin{IEEEeqnarray}{lcl}
\nonumber
H\left( \bs{X}_{i,n} \right) & = &H\big( \bs{X}_{i,C,n}, \bs{X}_{i,P,n}, \bs{X}_{i,D,n}, \bs{X}_{i,Q,n} \big)\\
\nonumber
& = &H\big( \bs{X}_{i,C,n}, \bs{X}_{i,P,n}, \bs{X}_{i,D,n} \big)\\
\label{EqHXi}
& \leqslant & \dim\bs{X}_{i,C,n} + \dim\bs{X}_{i,P,n} + \dim\bs{X}_{i,D,n}. \qquad
\end{IEEEeqnarray}
Note that vectors $\bs{X}_{i,P,n}$ and $\bs{X}_{i,D,n}$ do not exist simultaneously. The former exists when $\overrightarrow{n}_{ii} > n_{ji}$, while the latter exists when $\overrightarrow{n}_{ii} < n_{ji}$. Moreover, the dimension of $\bs{X}_{i,n}$ satisfies
\begin{IEEEeqnarray}{lcl}
\nonumber
\dim \bs{X}_{i,n} & = & \dim \bs{X}_{i,C,n} + \dim \bs{X}_{i,P,n} + \dim \bs{X}_{i,D,n} + \dim \bs{X}_{i,Q,n}\\
\label{EqdimXi}
& = & q.
\end{IEEEeqnarray}
\end{subequations}
For the case in which feedback is taken into account an alternative notation is adopted.
Let $\bs{X}_{i,D,n}$ be written in terms of $\bs{X}_{i,DF,n}$ and $\bs{X}_{i,DG,n}$, i.e., $\bs{X}_{i,D,n}=\Big(\bs{X}_{i,DF,n}^{\sfT},\bs{X}_{i,DG,n}^{\sfT}\Big)^{\sfT}$.
The vector $\bs{X}_{i,DF,n}$ represents the bits of $\bs{X}_{i,D,n}$ that are above the noise level in the feedback link from receiver $j$ to transmitter $j$;
and $\bs{X}_{i,DG,n}$ represents the bits of $\bs{X}_{i,D,n}$ that are below the noise level in the feedback link from receiver $j$ to transmitter $j$, as shown in Figure~\ref{FigMessagesLDICNFB}.  
The dimension of vectors $\bs{X}_{i,DF,n}$ and $\bs{X}_{i,DG,n}$ are given by
\begin{subequations}
\begin{IEEEeqnarray}{lcl}
\label{dimXiDFG}
\nonumber
\dim  \bs{X}_{i,DF,n} &=&  \min\Big(\left(n_{ji}-\overrightarrow{n}_{ii}\right)^+,\Big(\overleftarrow{n}_{jj}-\overrightarrow{n}_{ii}-\min\left(\left(\overrightarrow{n}_{jj}-n_{ji}\right)^+,n_{ij}\right)\\
\label{dimXiDF-1}
& & -\left(\left(\overrightarrow{n}_{jj}-n_{ij}\right)^+-n_{ji}\right)^+\Big)^+\Big) \mbox{ and }\\
\dim \bs{X}_{i,DG,n} &=&\dim \bs{X}_{i,D,n}  - \dim \bs{X}_{i,DF,n}.
\end{IEEEeqnarray}
\end{subequations}
Let $\bs{X}_{i,C,n}$ be written in terms of $\bs{X}_{i,CF_j,n}$ and $\bs{X}_{i,CG_j,n}$, i.e., ${\bs{X}_{i,C,n}=\Big(\bs{X}_{i,CF_j,n}^{\sfT}, \bs{X}_{i,CG_j,n}^{\sfT}\Big)^{\sfT}}$.
The vector $\bs{X}_{i,CF_j,n}$ represents the bits of $\bs{X}_{i,C,n}$ that are above the noise level in the feedback link from receiver $j$ to transmitter $j$; 
and $\bs{X}_{i,CG_j,n}$ represents the bits of $\bs{X}_{i,C,n}$ that are below the noise level in the feedback link from receiver $j$ to transmitter $j$, as shown in Figure~\ref{FigMessagesLDICNFB}.  
Let also, the dimension of vector $\left(\bs{X}_{i,CF_j,n}^{\sfT}, \bs{X}_{i,DF,n}^{\sfT} \right)$ be defined as follows:

\begin{IEEEeqnarray}{lll}
\label{EqdimXicfjXidf}
\dim \left(\left(\bs{X}_{i,CF_j,n}^{\sfT}, \bs{X}_{i,DF,n}^{\sfT} \right)\right) &=& \left(\min\left(\overleftarrow{n}_{jj}, \max\left(\overrightarrow{n}_{jj},n_{ji}\right)\right)-\left(\overrightarrow{n}_{jj}-n_{ji}\right)^+\right)^+. \quad
\end{IEEEeqnarray}
The dimension of vectors $\bs{X}_{i,CF_j,n}$ and $\bs{X}_{i,CG_j,n}$ can be obtained as follows:
\begin{subequations}
\begin{IEEEeqnarray}{lcl}
\nonumber
\dim \bs{X}_{i,CF_j,n} &=&\dim \left(\left(\bs{X}_{i,CF_j,n}^{\sfT}, \bs{X}_{i,DF,n}^{\sfT} \right)\right)  - \dim \bs{X}_{i,DF,n}, \\
\textrm{ and }\label{dimXiCFjn}\\
\label{dimXiCGjn}
\dim \bs{X}_{i,CG_j,n} &=&\dim \bs{X}_{i,C,n}  - \dim \bs{X}_{i,CF_j,n}.
\end{IEEEeqnarray}
\end{subequations}

\noindent
More generally, when needed,  the vector $\bs{X}_{iF_k,n}$ is used to represent the bits of $\bs{X}_{i,n}$  that are above the noise level in the feedback link from receiver $k$ to transmitter $k$, with $k \in\{1,2\}$. The vector $\bs{X}_{iG_k,n}$ is used to represent the bits of $\bs{X}_{i,n}$ that are below the noise level in the feedback link from receiver $k$ to transmitter $k$.

\noindent
The vector $\bs{X}_{i,U,n}$ is used to represent the bits of vector  $\bs{X}_{i,n}$ that interfere with bits of $\bs{X}_{j,C,n}$ at receiver $j$ and those bits of $\bs{X}_{i,n}$ that are observed by receiver $j$ and do not interfere any bits from transmitter $j$. An example is shown in Figure~\ref{FigXiu}. 

\noindent
Based on its definition, the dimension of vector $\bs{X}_{i,U,n}$ is
\begin{IEEEeqnarray}{lcl}
\label{EqHXtopi2}
\dim \bs{X}_{i,U,n} &=& \min\left(\overrightarrow{n}_{jj},n_{ij} \right)-\min\left(\left(\overrightarrow{n}_{jj}-n_{ji}\right)^+,n_{ij} \right)+\left(n_{ji}-\overrightarrow{n}_{jj}\right)^+.
\end{IEEEeqnarray}

\noindent
Finally, for all $i \in \lbrace 1, 2 \rbrace$, with $j \in \lbrace 1, 2 \rbrace \setminus \lbrace i \rbrace$, the channel output $\overrightarrow{\bs{Y}}_{i,n}$ of the LD-IC-NOF in \eqref{EqLDICsignals} for any channel use $n \in \lbrace 1, 2, \ldots, N\rbrace$  is  a $q$-dimensional vector, with $q$ in \eqref{Eqq}, that can be written as the concatenation of three vectors: $\overrightarrow{\bs{Y}}_{i,Q,n}$, $\overleftarrow{\bs{Y}}_{i,n}$, and $\overrightarrow{\bs{Y}}_{i,G,n}$, i.e., ${\overrightarrow{\bs{Y}}_{i,n} = \Big(\overrightarrow{\bs{Y}}_{i,Q,n}^{\sfT}, \overleftarrow{\bs{Y}}_{i,n}^{\sfT}, \overrightarrow{\bs{Y}}_{i,G,n}^{\sfT}\Big)^{\sfT}}$, as shown in Figure~\ref{FigMessagesLDICNFB}.  More specifically, the vector $\overleftarrow{\bs{Y}}_{i,n}$ contains the bits that are above the noise level in the feedback link from receiver $i$ to transmitter $i$. Then, 
\begin{subequations}
\begin{IEEEeqnarray}{lcl}
\label{EqdimYib}
\dim \overleftarrow{\bs{Y}}_{i,n}  & = & \min\Big( \overleftarrow{n}_{ii}, \max\left(\overrightarrow{n}_{ii},n_{ij}\right)\Big).
\end{IEEEeqnarray}

\noindent
The vector $\overrightarrow{\bs{Y}}_{i,G,n}$ contains the bits that are below the noise level in the feedback link from receiver $i$ to transmitter $i$. Then, 
\begin{IEEEeqnarray}{lcl}
\label{EqdimYig}
\dim \overrightarrow{\bs{Y}}_{i,G,n} & = & \Big(\max\left(\overrightarrow{n}_{ii},n_{ij}\right)-\overleftarrow{n}_{ii} \Big)^+.
\end{IEEEeqnarray}

\noindent
The vector $\overrightarrow{\bs{Y}}_{i,Q,n}=\left(0, \ldots, 0 \right)$ is included for dimensional matching with the model in \eqref{EqLDICsignalsc}. Then, 

\noindent
\begin{IEEEeqnarray}{lcl}
\nonumber
H\left( \overrightarrow{\bs{Y}}_{i,n} \right) & = &H\big(\overrightarrow{\bs{Y}}_{i,Q,n}, \overleftarrow{\bs{Y}}_{i,n}, \overrightarrow{\bs{Y}}_{i,G,n} \big)\\
\nonumber
& = &H\big(\overleftarrow{\bs{Y}}_{i,n}, \overrightarrow{\bs{Y}}_{i,G,n} \big)\\
\label{EqHYi}
& \leqslant & \dim \overleftarrow{\bs{Y}}_{i,n} + \dim \overrightarrow{\bs{Y}}_{i,G,n}.
\end{IEEEeqnarray}
\end{subequations}

\noindent
The dimension of $\overrightarrow{\bs{Y}}_{i,n}$ satisfies $\dim \overrightarrow{\bs{Y}}_{i,n}  =  q$.

\noindent
Using this notation, the proof continues as follows: 

\noindent
\textbf{Proof of \eqref{EqRi-2-V2}:}
First, consider $n_{ji} \leqslant \overrightarrow{n}_{ii}$, i.e., vector $\bs{X}_{i,P,n}$ exists and vector $\bs{X}_{i,D,n}$ does not exist.  From the assumption that the message index $W_i$ is i.i.d. following a uniform distribution over the set $\mathcal{W}_i$, the following holds for any $k \in \lbrace 1, 2, \ldots, N \rbrace$: 
\begin{IEEEeqnarray}{lcl}
\nonumber
NR_i &=& H\left(W_i\right)\\
\nonumber
&\stackrel{(a)}{=}& H\left(W_i|W_j\right)\\
\nonumber
&\stackrel{(b)}{\leqslant}& I\left(W_i;\overrightarrow{\bs{Y}}_i,\overleftarrow{\bs{Y}}_j|W_j\right)+N\delta(N)\\
\nonumber
&=& H\left(\overrightarrow{\bs{Y}}_i,\overleftarrow{\bs{Y}}_j|W_j\right)+N\delta(N)\\
\nonumber
&\stackrel{(c)}{=}& \sum_{n=1}^{N}H\Big(\overrightarrow{\bs{Y}}_{i,n},\overleftarrow{\bs{Y}}_{j,n}|W_j,\overrightarrow{\bs{Y}}_{i,(1:n-1)},\overleftarrow{\bs{Y}}_{j,(1:n-1)}, \bs{X}_{j,n}\Big)+N\delta(N)\\
\nonumber
&\leqslant& \sum_{n=1}^{N}H\Big(\bs{X}_{i,n},\overleftarrow{\bs{Y}}_{j,n}|\bs{X}_{j,n}\Big)+N\delta(N)\\
\nonumber
&\leqslant& \sum_{n=1}^{N}H\left(\bs{X}_{i,n}\right)+N\delta(N)\\
\nonumber
&=& NH\left(\bs{X}_{i,k}\right)+N\delta(N), \\
\label{EqRi-42}
&\leqslant & N\left(\dim \bs{X}_{i,C,k} + \dim \bs{X}_{i,P,k}\right)+N\delta(N),
\end{IEEEeqnarray}
where, 
(a) follows from the fact that $W_1$ and $W_2$ are independent; 
(b) follows from Fano's inequality; and 
(c) follows from the fact that $\bs{X}_{j,n}=f_j^{(n)}\left(W_j,\overleftarrow{\bs{Y}}_{j,(1:n-1)}\right)$.

\noindent
Second, consider the case in which ${n_{ji} > \overrightarrow{n}_{ii}}$. In this case the vector $\bs{X}_{i,P,n}$ does not exist and the vector $\bs{X}_{i,D,n}$ exists. From the assumption that the message index $W_i$ is i.i.d. following a uniform distribution over the set $\mathcal{W}_i$, hence the following holds:  

\begin{IEEEeqnarray}{lcl}
\nonumber
NR_i &=& H\left(W_i\right)\\
\nonumber
&\stackrel{(a)}{=}& H\left(W_i|W_j\right)\\
\nonumber
&\stackrel{(b)}{\leqslant}& I\left(W_i;\overrightarrow{\bs{Y}}_i,\overleftarrow{\bs{Y}}_j|W_j\right)+N\delta(N)\\
\nonumber
&=& H\left(\overrightarrow{\bs{Y}}_i,\overleftarrow{\bs{Y}}_j|W_j\right)+N\delta(N)
\end{IEEEeqnarray}
\begin{IEEEeqnarray}{lcl}
\nonumber
&\stackrel{(c)}{=}& \sum_{n=1}^{N}H\Big(\overrightarrow{\bs{Y}}_{i,n},\overleftarrow{\bs{Y}}_{j,n}|W_j,\overrightarrow{\bs{Y}}_{i,(1:n-1)},\overleftarrow{\bs{Y}}_{j,(1:n-1)}, \bs{X}_{j,n}\Big)+N\delta(N)\\
\nonumber
&\leqslant& \sum_{n=1}^{N}H\Big(\bs{X}_{i,C,n}, \bs{X}_{i,CF_j,n}, \bs{X}_{i,DF,n}\Big)+N\delta(N)\\
\nonumber
&=& \sum_{n=1}^{N}H\Big(\bs{X}_{i,C,n}, \bs{X}_{i,DF,n}\Big)+N\delta(N)\\
\nonumber
&=& NH\Big(\bs{X}_{i,C,k}, \bs{X}_{i,DF,k}\Big)+N\delta(N), \\
\label{EqRi-23}
&\leqslant & N\left(\dim \bs{X}_{i,C,k} + \dim \bs{X}_{i,DF,k}\right)+N\delta(N).
\end{IEEEeqnarray}

\noindent
Then,  \eqref{EqRi-42} and \eqref{EqRi-23} can be expressed as one inequality in the asymptotic regime, as follows:
\begin{IEEEeqnarray}{lcl}
\label{EqRi-44}
R_i &\leqslant&\dim \bs{X}_{i,C,k}+\dim \bs{X}_{i,P,k}+\dim \bs{X}_{i,DF,k},
\end{IEEEeqnarray}
which holds for any $k \in \{1, 2, \ldots,N\}$.

\noindent
Plugging \eqref{EqdimXic}, \eqref{EqdimXip}, and \eqref{dimXiDF-1} in \eqref{EqRi-44}, and after some trivial manipulations, the following holds: 
\begin{figure*}[t]
 \centerline{\epsfig{figure=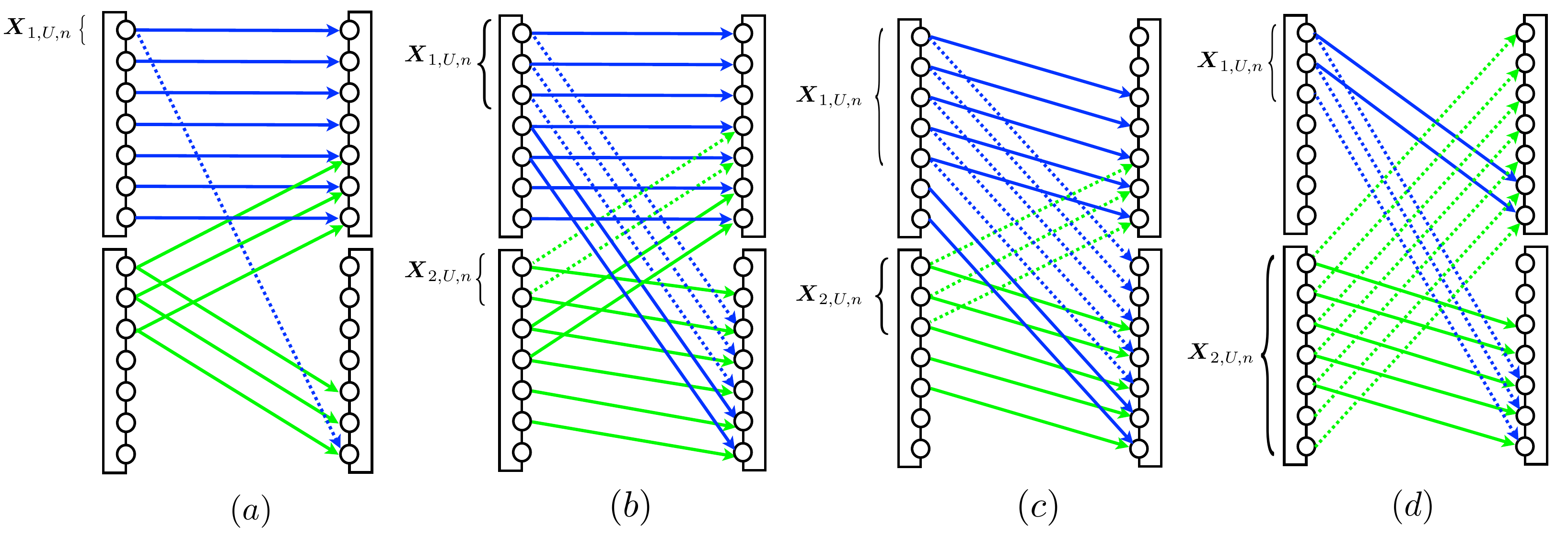,width=1.0\textwidth}}
 \caption{Vector $\bs{X}_{i,U,n}$ in different combination of interference regimes.}
\label{FigXiu}
\end{figure*}
\begin{IEEEeqnarray}{lcl}
\nonumber
R_i &\leqslant& \min \! \Big( \! \max \! \left(\overrightarrow{n}_{ii},n_{ji}\right),\max\Big(\overrightarrow{n}_{ii},\overleftarrow{n}_{jj} \! - \! \left(\overrightarrow{n}_{jj} \! - \! n_{ji}\right)^+\Big)\Big).\\
\label{EqRi-46}
\end{IEEEeqnarray}

\noindent
This completes the proof of \eqref{EqRi-2-V2}. 

\noindent
\textbf{Proof of \eqref{EqRi+Rj-2-V2}:}  From the assumption that the message indices $W_1$ and $W_2$  are i.i.d. following a uniform distribution over the sets $\mathcal{W}_1$ and $\mathcal{W}_2$ respectively, the following holds for any $k \in \lbrace 1, 2, \ldots, N \rbrace$: 
\begin{IEEEeqnarray}{rcl}
\nonumber
N\left(R_1+R_2\right) &=& H\left(W_1\right)+H\left(W_2\right)\\
\nonumber
&\stackrel{(a)}{\leqslant}& I\left(W_1;\overrightarrow{\bs{Y}}_1,\overleftarrow{\bs{Y}}_1\right)+I\left(W_2;\overrightarrow{\bs{Y}}_2,\overleftarrow{\bs{Y}}_2\right)+N\delta(N)\\
\nonumber
&\leqslant&H\left(\overrightarrow{\bs{Y}}_1\right)-H\left(\overleftarrow{\bs{Y}}_1|W_1\right)-H\left(\bs{X}_{2,C}|W_1,\overleftarrow{\bs{Y}}_1,\bs{X}_1\right)+H\left(\overrightarrow{\bs{Y}}_2\right)-H\left(\overleftarrow{\bs{Y}}_2|W_2\right)\\
\nonumber
& & -H\left(\bs{X}_{1,C}|W_2,\overleftarrow{\bs{Y}}_2,\bs{X}_2\right)+N\delta(N) 
\end{IEEEeqnarray}
\begin{IEEEeqnarray}{rcl}
\nonumber
&=&H\left(\overrightarrow{\bs{Y}}_1\right)-H\left(\overleftarrow{\bs{Y}}_1|W_1\right)-H\left(\bs{X}_{2,C},\bs{X}_{1,U}|W_1,\overleftarrow{\bs{Y}}_1,\bs{X}_1\right)+H\left(\overrightarrow{\bs{Y}}_2\right)-H\left(\overleftarrow{\bs{Y}}_2|W_2\right) \\
\nonumber
& & -H\left(\bs{X}_{1,C},\bs{X}_{2,U}|W_2,\overleftarrow{\bs{Y}}_2,\bs{X}_2\right)+N\delta(N)\\
\nonumber
&=&H\left(\overrightarrow{\bs{Y}}_1\right)+\Big[I\left(\bs{X}_{2,C},\bs{X}_{1,U};W_1,\overleftarrow{\bs{Y}}_1\right)-H\left(\bs{X}_{2,C},\bs{X}_{1,U}\right)\Big]+H\left(\overrightarrow{\bs{Y}}_2\right)\\
\nonumber
& & +\left[I\left(\bs{X}_{1,C},\bs{X}_{2,U};W_2,\overleftarrow{\bs{Y}}_2\right)-H\left(\bs{X}_{1,C},\bs{X}_{2,U}\right)\right]-H\left(\overleftarrow{\bs{Y}}_1|W_1\right)-H\left(\overleftarrow{\bs{Y}}_2|W_2\right)\\
\nonumber
& & +N\delta(N)\\
\nonumber
&\stackrel{(b)}{=}&H\left(\overrightarrow{\bs{Y}}_1|\bs{X}_{1,C},\bs{X}_{2,U}\right)-H\left(\bs{X}_{1,C},\bs{X}_{2,U}|\overrightarrow{\bs{Y}}_1\right)+H\left(\overrightarrow{\bs{Y}}_2|\bs{X}_{2,C},\bs{X}_{1,U}\right)\\
\nonumber
& & -H\left(\bs{X}_{2,C},\bs{X}_{1,U}|\overrightarrow{\bs{Y}}_2\right)+I\left(\bs{X}_{2,C},\bs{X}_{1,U};W_1,\overleftarrow{\bs{Y}}_1\right)+I\left(\bs{X}_{1,C},\bs{X}_{2,U};W_2,\overleftarrow{\bs{Y}}_2\right)\\
\nonumber
& &-H\left(\overleftarrow{\bs{Y}}_1|W_1\right)-H\left(\overleftarrow{\bs{Y}}_2|W_2\right)+N\delta(N)\\
\nonumber
&\leqslant& H\left(\overrightarrow{\bs{Y}}_1|\bs{X}_{1,C},\bs{X}_{2,U}\right)+H\left(\overrightarrow{\bs{Y}}_2|\bs{X}_{2,C},\bs{X}_{1,U}\right)+I\left(\bs{X}_{2,C},\bs{X}_{1,U};W_1,\overleftarrow{\bs{Y}}_1\right)\\
\nonumber
& & +I\left(\bs{X}_{1,C},\bs{X}_{2,U};W_2,\overleftarrow{\bs{Y}}_2\right)-H\left(\overleftarrow{\bs{Y}}_1|W_1\right)-H\left(\overleftarrow{\bs{Y}}_2|W_2\right)+N\delta(N)\\
\nonumber
&\leqslant& H\left(\overrightarrow{\bs{Y}}_1|\bs{X}_{1,C},\bs{X}_{2,U}\right)+H\left(\overrightarrow{\bs{Y}}_2|\bs{X}_{2,C},\bs{X}_{1,U}\right)+I\left(\bs{X}_{2,C},\bs{X}_{1,U},W_2,\overleftarrow{\bs{Y}}_2;W_1,\overleftarrow{\bs{Y}}_1\right)\\
\nonumber
& & +I\left(\bs{X}_{1,C},\bs{X}_{2,U},W_1,\overleftarrow{\bs{Y}}_1;W_2,\overleftarrow{\bs{Y}}_2\right)-H\left(\overleftarrow{\bs{Y}}_1|W_1\right)-H\left(\overleftarrow{\bs{Y}}_2|W_2\right)+N\delta(N)\\
\nonumber
&=& H\left(\overrightarrow{\bs{Y}}_1|\bs{X}_{1,C},\bs{X}_{2,U}\right)+H\left(\overrightarrow{\bs{Y}}_2|\bs{X}_{2,C},\bs{X}_{1,U}\right)+I\left(W_2;W_1,\overleftarrow{\bs{Y}}_1\right)\\
\nonumber
& & +I\left(\bs{X}_{2,C},\bs{X}_{1,U},\overleftarrow{\bs{Y}}_2;W_1, \overleftarrow{\bs{Y}}_1|W_2\right)+I\left(W_1;W_2,\overleftarrow{\bs{Y}}_2\right)\\
\nonumber
& &  +I\left(\bs{X}_{1,C},\bs{X}_{2,U},\overleftarrow{\bs{Y}}_1;W_2,\overleftarrow{\bs{Y}}_2|W_1\right)-H\left(\overleftarrow{\bs{Y}}_1|W_1\right)-H\left(\overleftarrow{\bs{Y}}_2|W_2\right)+N\delta(N)\\
\nonumber
&\stackrel{(c)}{=}& H\left(\overrightarrow{\bs{Y}}_1|\bs{X}_{1,C},\bs{X}_{2,U}\right)+H\left(\overrightarrow{\bs{Y}}_2|\bs{X}_{2,C},\bs{X}_{1,U}\right)+H\left(W_1\right)+H\left(\overleftarrow{\bs{Y}}_1|W_1\right) \\
\nonumber
& & -H\left(W_1|W_2\right)-H\left(\overleftarrow{\bs{Y}}_1|W_2,W_1\right)+H\left(\bs{X}_{2,C},\bs{X}_{1,U},\overleftarrow{\bs{Y}}_2|W_2\right)+H\left(W_2\right)\\
\nonumber
& & +H\left(\overleftarrow{\bs{Y}}_2|W_2\right)-H\left(W_2|W_1\right)-H\left(\overleftarrow{\bs{Y}}_2|W_1,W_2\right)+H\left(\bs{X}_{1,C},\bs{X}_{2,U},\overleftarrow{\bs{Y}}_1|W_1\right)\\
\nonumber
& & -H\left(\overleftarrow{\bs{Y}}_1|W_1\right)-H\left(\overleftarrow{\bs{Y}}_2|W_2\right)+N\delta(N)\\
\nonumber
&\leqslant& H\left(\overrightarrow{\bs{Y}}_1|\bs{X}_{1,C},\bs{X}_{2,U}\right)+H\left(\overrightarrow{\bs{Y}}_2|\bs{X}_{2,C},\bs{X}_{1,U}\right)+H\left(\bs{X}_{2,C},\bs{X}_{1,U},\overleftarrow{\bs{Y}}_2|W_2\right)\\
\nonumber
& & +H\left(\bs{X}_{1,C},\bs{X}_{2,U},\overleftarrow{\bs{Y}}_1|W_1\right)+N\delta(N)\\
\nonumber
&=& \sum_{n=1}^{N}\Big[H\left(\overrightarrow{\bs{Y}}_{1,n}|\bs{X}_{1,C},\bs{X}_{2,U},\overrightarrow{\bs{Y}}_{1,(1:n-1)}\right)+H\left(\overrightarrow{\bs{Y}}_{2,n}|\bs{X}_{2,C},\bs{X}_{1,U},\overrightarrow{\bs{Y}}_{2,(1:n-1)}\right)\\
\nonumber
& & +H\Big(\bs{X}_{2,C,n},\bs{X}_{1,U,n},\overleftarrow{\bs{Y}}_{2,n}|W_2,\bs{X}_{2,C,(1:n-1)}, \bs{X}_{1,U,(1:n-1)}, \overleftarrow{\bs{Y}}_{2,(1:n-1)}\Big) \\
\nonumber
& & +H\Big(\bs{X}_{1,C,n},\bs{X}_{2,U,n},\overleftarrow{\bs{Y}}_{1,n}|W_1,\bs{X}_{1,C,(1:n-1)}, \bs{X}_{2,U,(1:n-1)},\overleftarrow{\bs{Y}}_{1,(1:n-1)}\Big)\Big]+N\delta(N)\\
\nonumber
&\stackrel{(d)}{=}& \sum_{n=1}^{N}\Big[H\left(\overrightarrow{\bs{Y}}_{1,n}|\bs{X}_{1,C},\bs{X}_{2,U},\overrightarrow{\bs{Y}}_{1,(1:n-1)}\right)+H\left(\overrightarrow{\bs{Y}}_{2,n}|\bs{X}_{2,C},\bs{X}_{1,U},\overrightarrow{\bs{Y}}_{2,(1:n-1)}\right)\\
\nonumber
& & +H\Big(\bs{X}_{2,C,n},\bs{X}_{1,U,n},\overleftarrow{\bs{Y}}_{2,n}|W_2,\bs{X}_{2,C,(1:n-1)}, \bs{X}_{1,U,(1:n-1)},\overleftarrow{\bs{Y}}_{2,(1:n-1)},\bs{X}_{2,(1:n)}\Big) \\
\nonumber
& & +H\Big(\bs{X}_{1,C,n},\bs{X}_{2,U,n},\overleftarrow{\bs{Y}}_{1,n}|W_1,\bs{X}_{1,C,(1:n-1)}, \bs{X}_{2,U,(1:n-1)},\overleftarrow{\bs{Y}}_{1,(1:n-1)},\bs{X}_{1,(1:n)}\Big)\Big]+N\delta(N) \\
\nonumber
&\stackrel{(e)}{\leqslant}& \sum_{n=1}^{N}\Big[H\left(\overrightarrow{\bs{Y}}_{1,n}|\bs{X}_{1,C,n},\bs{X}_{2,U,n}\right)+H\left(\overrightarrow{\bs{Y}}_{2,n}|\bs{X}_{2,C,n},\bs{X}_{1,U,n}\right)+H\left(\bs{X}_{1,U,n},\overleftarrow{\bs{Y}}_{2,n}|\bs{X}_{2,n}\right)\\
\nonumber
& & +H\left(\bs{X}_{2,U,n},\overleftarrow{\bs{Y}}_{1,n}|\bs{X}_{1,n}\right)\Big]+N\delta(N)
\end{IEEEeqnarray}
\begin{IEEEeqnarray}{lcl}
\nonumber
&\leqslant& \sum_{n=1}^{N}\Big[H\left(\bs{X}_{1,P,n}\right)+H\left(\bs{X}_{2,P,n}\right)+H\left(\bs{X}_{1,U,n},\overleftarrow{\bs{Y}}_{2,n}|\bs{X}_{2,n}\right)+H\left(\bs{X}_{2,U,n},\overleftarrow{\bs{Y}}_{1,n}|\bs{X}_{1,n}\right)\Big]+N\delta(N)\\
\nonumber
&\stackrel{(e)}{\leqslant}& N\Big[H\left(\bs{X}_{1,P,k}\right)+H\left(\bs{X}_{2,P,k}\right)+H\left(\bs{X}_{1,U,k}\right)+H\left(\overleftarrow{\bs{Y}}_{2,k}|\bs{X}_{2,k},\bs{X}_{1,U,k}\right)+H\left(\bs{X}_{2,U,k}\right)\\
\nonumber
& & +H\left(\overleftarrow{\bs{Y}}_{1,k}|\bs{X}_{1,k},\bs{X}_{2,U,k}\right)\Big]+N\delta(N), \\
\nonumber
&=& N\Big[H\left(\bs{X}_{1,P,k}\right)+H\left(\bs{X}_{2,P,k}\right)+H\left(\bs{X}_{1,U,k}\right)+H\left(\bs{X}_{1,CF_2,k},\bs{X}_{1,DF,k}|\bs{X}_{2,k},\bs{X}_{1,U,k}\right)+H\left(\bs{X}_{2,U,k}\right)\\
\nonumber
& & +H\left(\bs{X}_{2,CF_1,k},\bs{X}_{2,DF,k}|\bs{X}_{1,k},\bs{X}_{2,U,k}\right)\Big]+N\delta(N)\\
\nonumber
&\leqslant& N\Big[H\left(\bs{X}_{1,P,k}\right)+H\left(\bs{X}_{2,P,k}\right)+H\left(\bs{X}_{1,U,k}\right)+H\left(\bs{X}_{1,CF_2,k},\bs{X}_{1,DF,k}|\bs{X}_{1,U,k}\right)+H\left(\bs{X}_{2,U,k}\right)\\
\nonumber
& & +H\left(\bs{X}_{2,CF_1,k},\bs{X}_{2,DF,k}|\bs{X}_{2,U,k}\right)\Big]+N\delta(N), \\
\nonumber
&\leqslant& N \Big[ \dim \bs{X}_{1,P,k}+\dim \bs{X}_{2,P,k}+\dim \bs{X}_{1,U,k}+\Big(\dim \left(\bs{X}_{1,CF_2,k},\bs{X}_{1,DF,k}\right) - \dim \bs{X}_{1,U,k}\Big)^+\\
\label{EqR1+R2-32c}
& & +\dim \bs{X}_{2,U,k}+\Big(\dim \left(\bs{X}_{2,CF_1,k},\bs{X}_{2,DF,k}\right)- \dim \bs{X}_{2,U,k}\Big)^+\Big]+N\delta(N).
\end{IEEEeqnarray}
where, 
(a) follows from Fano's inequality; 
(b) follows from the fact that $H(Y)-H(X)=H(Y|X)-H(X|Y)$; 
(c) follows from the fact that $H\Big(\bs{X}_{i,C}$, $\bs{X}_{j,U}$, $\overleftarrow{\bs{Y}}_i|W_i$, $W_j$, $\overleftarrow{\bs{Y}}_j\Big)=0$; 
(d) follows from the fact that $\bs{X}_{i,n}=f_i^{(n)}\left(W_i,\overleftarrow{\bs{Y}}_{i,(1:n-1)}\right)$; and 
(e) follows from the fact that conditioning reduces the entropy.

\noindent
Plugging \eqref{EqdimXip}, \eqref{EqdimXicfjXidf}, and \eqref{EqHXtopi2} in \eqref{EqR1+R2-32c} and after some trivial manipulations, the following holds in the asymptotic regime:  
\begin{IEEEeqnarray}{lcl}
\nonumber
R_1+R_2  &\leqslant&  \max\Big(\left(\overrightarrow{n}_{11}-{n}_{12} \right)^+, n_{21}, \overrightarrow{n}_{11}-\left(\max\left(\overrightarrow{n}_{11},n_{12}\right)-\overleftarrow{n}_{11}\right)^+\Big)\\
\label{EqR1+R2-33c}
& &+\max\Big(\left(\overrightarrow{n}_{22}-{n}_{21} \right)^+, n_{12}, \overrightarrow{n}_{22}-\left(\max\left(\overrightarrow{n}_{22},n_{21}\right)-\overleftarrow{n}_{22}\right)^+\Big).
\end{IEEEeqnarray}

\noindent
This completes the proof of \eqref{EqRi+Rj-2-V2}. 

\noindent
\textbf{Proof of \eqref{Eq2Ri+Rj-V2}:} From the assumption that the message indices $W_i$ and $W_j$  are i.i.d. following a uniform distribution over the sets $\mathcal{W}_i$ and $\mathcal{W}_j$ respectively, for all $i \in \lbrace1,2\rbrace$, with $j \in \lbrace1,2 \rbrace \setminus\lbrace i \rbrace$, the following holds for any $k \in \lbrace 1, 2, \ldots, N \rbrace$: 

\begin{IEEEeqnarray}{lcl}
\nonumber
N \big(2 R_i + R_j\big) &=& 2H\left(W_i\right)+H\left(W_j\right)\\
\nonumber
&\stackrel{(a)}{\leqslant}& I\left(W_i;\overrightarrow{\bs{Y}}_i,\overleftarrow{\bs{Y}}_i\right)+I\left(W_i;\overrightarrow{\bs{Y}}_i,\overleftarrow{\bs{Y}}_j|W_j\right)+I\left(W_j;\overrightarrow{\bs{Y}}_j,\overleftarrow{\bs{Y}}_j\right)+N\delta(N) \\
\nonumber
&\stackrel{(b)}{=}& H\left(\overrightarrow{\bs{Y}}_i\right)-H\left(\overleftarrow{\bs{Y}}_i|W_i\right)-H\left(\overrightarrow{\bs{Y}}_i|W_i,\overleftarrow{\bs{Y}}_i\right)+H\left(\overrightarrow{\bs{Y}}_i|W_j,\overleftarrow{\bs{Y}}_j\right)+H\left(\overrightarrow{\bs{Y}}_j\right)\\
\nonumber
& & -H\left(\overrightarrow{\bs{Y}}_j|W_j,\overleftarrow{\bs{Y}}_j\right)+N\delta(N)\\
\nonumber
&=& H\left(\overrightarrow{\bs{Y}}_i\right)-H\left(\overleftarrow{\bs{Y}}_i|W_i\right)-H\left(\bs{X}_{j,C},\bs{X}_{j,D}|W_i,\overleftarrow{\bs{Y}}_i\right)+H\left(\overrightarrow{\bs{Y}}_i|W_j,\overleftarrow{\bs{Y}}_j\right)+H\left(\overrightarrow{\bs{Y}}_j\right)\\
\nonumber
& & -H\left(\bs{X}_{i,C}, \bs{X}_{i,D}|W_j,\overleftarrow{\bs{Y}}_j\right)+N\delta(N) \\
\nonumber
&\leqslant& H\left(\overrightarrow{\bs{Y}}_i\right)-H\left(\overleftarrow{\bs{Y}}_i|W_i\right)-H\left(\bs{X}_{j,C},\bs{X}_{i,U}|W_i,\overleftarrow{\bs{Y}}_i\right)+H\left(\overrightarrow{\bs{Y}}_i|W_j,\overleftarrow{\bs{Y}}_j\right)+H\left(\overrightarrow{\bs{Y}}_j\right)\\
\nonumber
& & -H\left(\bs{X}_{i,C}|W_j,\overleftarrow{\bs{Y}}_j\right)+N\delta(N) \\
\nonumber
&\leqslant& H\left(\overrightarrow{\bs{Y}}_i\right)-H\left(\overleftarrow{\bs{Y}}_i|W_i\right)+\big[I\left(\bs{X}_{j,C},\bs{X}_{i,U};W_i,\overleftarrow{\bs{Y}}_i\right)-H\left(\bs{X}_{j,C},\bs{X}_{i,U}\right)  \big]\\
\nonumber
& & +H\left(\overrightarrow{\bs{Y}}_i,\bs{X}_{i,C}|W_j,\overleftarrow{\bs{Y}}_j\right)+H\left(\overrightarrow{\bs{Y}}_j\right)-H\left(\bs{X}_{i,C}|W_j,\overleftarrow{\bs{Y}}_j\right)+N\delta(N) 
\end{IEEEeqnarray}
\begin{IEEEeqnarray}{lcl}
\nonumber
&=& H\left(\overrightarrow{\bs{Y}}_i\right)-H\left(\overleftarrow{\bs{Y}}_i|W_i\right)+\Big[I\left(\bs{X}_{j,C},\bs{X}_{i,U};W_i,\overleftarrow{\bs{Y}}_i\right)-H\left(\bs{X}_{j,C},\bs{X}_{i,U}\right)  \Big]\\
\nonumber
& & +H\left(\overrightarrow{\bs{Y}}_i|W_j,\overleftarrow{\bs{Y}}_j,\bs{X}_{i,C}\right)+H\left(\overrightarrow{\bs{Y}}_j\right)+N\delta(N)\\
\nonumber
&\leqslant& H\left(\overrightarrow{\bs{Y}}_i\right)-H\left(\overleftarrow{\bs{Y}}_i|W_i\right)+\Big[I\left(\bs{X}_{j,C},\bs{X}_{i,U};W_i,\overleftarrow{\bs{Y}}_i\right)-H\left(\bs{X}_{j,C},\bs{X}_{i,U}\right)  \Big]\\
\nonumber
& & +H\left(\overrightarrow{\bs{Y}}_i|W_j,\overleftarrow{\bs{Y}}_j,\bs{X}_{i,C}\right)+H\left(\overrightarrow{\bs{Y}}_j,\bs{X}_{j,C},\bs{X}_{i,U}\right)+N\delta(N)\\
\nonumber
&\stackrel{(c)}{=}& H\left(\overrightarrow{\bs{Y}}_i\right)-H\left(\overleftarrow{\bs{Y}}_i|W_i\right)+I\left(\bs{X}_{j,C},\bs{X}_{i,U};W_i,\overleftarrow{\bs{Y}}_i\right)+H\left(\overrightarrow{\bs{Y}}_i|W_j,\overleftarrow{\bs{Y}}_j,\bs{X}_{i,C}\right)\\
\nonumber
& & +H\left(\overrightarrow{\bs{Y}}_j|\bs{X}_{j,C},\bs{X}_{i,U}\right)+N\delta(N) \\
\nonumber
&\leqslant& H\left(\overrightarrow{\bs{Y}}_i\right)-H\left(\overleftarrow{\bs{Y}}_i|W_i\right)+I\left(\bs{X}_{j,C},\bs{X}_{i,U},W_j,\overleftarrow{\bs{Y}}_j;W_i,\overleftarrow{\bs{Y}}_i\right)\\
\nonumber
& & +H\left(\overrightarrow{\bs{Y}}_i|W_j,\overleftarrow{\bs{Y}}_j,\bs{X}_{i,C}\right)+H\left(\overrightarrow{\bs{Y}}_j|\bs{X}_{j,C},\bs{X}_{i,U}\right)+N\delta(N) \\
\nonumber
&\stackrel{(d)}{=}& H\left(\overrightarrow{\bs{Y}}_i\right)-H\left(\overleftarrow{\bs{Y}}_i|W_j,W_i,\right)+H\left(\bs{X}_{j,C},\bs{X}_{i,U},\overleftarrow{\bs{Y}}_j|W_j\right)\\
\nonumber
& & +H\left(\overrightarrow{\bs{Y}}_i|W_j,\overleftarrow{\bs{Y}}_j,\bs{X}_{i,C}\right)+H\left(\overrightarrow{\bs{Y}}_j|\bs{X}_{j,C},\bs{X}_{i,U}\right)+N\delta(N)\\
\nonumber
&\leqslant& H\left(\overrightarrow{\bs{Y}}_i\right)+H\left(\bs{X}_{j,C},\bs{X}_{i,U},\overleftarrow{\bs{Y}}_j|W_j\right)+H\left(\overrightarrow{\bs{Y}}_i|W_j,\overleftarrow{\bs{Y}}_j,\bs{X}_{i,C}\right)\\
\nonumber
& & +H\left(\overrightarrow{\bs{Y}}_j|\bs{X}_{j,C},\bs{X}_{i,U}\right)+N\delta(N)\\
\nonumber
&\leqslant& \sum_{n=1}^N \Big[H\left(\overrightarrow{\bs{Y}}_{i,n}\right)\\
\nonumber
& & +H\Big(\bs{X}_{j,C,n},\bs{X}_{i,U,n},\overleftarrow{\bs{Y}}_{j,n}|W_j, \bs{X}_{j,C,(1:n-1)},\bs{X}_{i,U,(1:n-1)},\overleftarrow{\bs{Y}}_{j,(1:n-1)}\Big)\\
\nonumber
& & +H\left(\overrightarrow{\bs{Y}}_{i,n}|W_j,\overleftarrow{\bs{Y}}_{j},\bs{X}_{i,C},\overrightarrow{\bs{Y}}_{i,(1:n-1)}\right)+H\left(\overrightarrow{\bs{Y}}_{j,n}|\bs{X}_{j,C},\bs{X}_{i,U},\overrightarrow{\bs{Y}}_{j,(1:n-1)}\right)\Big]\\
\nonumber
& & +N\delta(N)\\
\nonumber
&=& \sum_{n=1}^N \Big[H\left(\overrightarrow{\bs{Y}}_{i,n}\right)\\
\nonumber
& & +H\Big(\bs{X}_{j,C,n},\bs{X}_{i,U,n},\overleftarrow{\bs{Y}}_{j,n}|W_j, \bs{X}_{j,C,(1:n-1)},\bs{X}_{i,U,(1:n-1)},\overleftarrow{\bs{Y}}_{j,(1:n-1)},\bs{X}_{j,(1:n)}\Big)\\
\nonumber
& & +H\left(\overrightarrow{\bs{Y}}_{i,n}|W_j,\overleftarrow{\bs{Y}}_{j},\bs{X}_{i,C},\overrightarrow{\bs{Y}}_{i,(1:n-1)},\bs{X}_{j,(1:n)}\right) \\
\nonumber
& & +H\left(\overrightarrow{\bs{Y}}_{j,n}|\bs{X}_{j,C},\bs{X}_{i,U},\overrightarrow{\bs{Y}}_{j,(1:n-1)}\right)\Big]+N\delta(N) \\
\nonumber
&\leqslant& \sum_{n=1}^N \Big[H\left(\overrightarrow{\bs{Y}}_{i,n}\right)+H\left(\bs{X}_{i,U,n}|\bs{X}_{j,n}\right)+H\left(\overleftarrow{\bs{Y}}_{j,n}|\bs{X}_{j,n},\bs{X}_{i,U,n}\right)+H\left(\overrightarrow{\bs{Y}}_{i,n}|\bs{X}_{i,C,n},\bs{X}_{j,n}\right)\\
\nonumber
& &  +H\left(\overrightarrow{\bs{Y}}_{j,n}|\bs{X}_{j,C,n},\bs{X}_{i,U,n}\right)\Big]+N\delta(N) \\
\nonumber
&\leqslant& N\Big[H\left(\overrightarrow{\bs{Y}}_{i,k}\right)+H\left(\bs{X}_{i,U,k}\right)+H\left(\overleftarrow{\bs{Y}}_{j,k}|\bs{X}_{j,k},\bs{X}_{i,U,k}\right)+H\left(\bs{X}_{i,P,k}\right)+H\left(\bs{X}_{j,P,k}\right)\Big]+N\delta(N) \\
\nonumber
&=& N\Big[H\left(\overrightarrow{\bs{Y}}_{i,k}\right)+H\left(\bs{X}_{i,U,k}\right)+H\left(\bs{X}_{i,CF_j,k},\bs{X}_{i,DF,k} |\bs{X}_{i,U,k}\right)+H\left(\bs{X}_{i,P,k}\right)+H\left(\bs{X}_{j,P,k}\right)\Big]\\
\nonumber
& & +N\delta(N), \\
\nonumber
&\leqslant& N\Big[\dim \overleftarrow{\bs{Y}}_{i,k}+\dim \overrightarrow{\bs{Y}}_{i,G,k}+\dim \bs{X}_{i,U,k}+\left(\dim\left(\bs{X}_{i,CF_j,k},\bs{X}_{i,DF,k}\right) - \dim \bs{X}_{i,U,k}\right)^+\\
\label{Eq2Ri+Rjc38}
& &+\dim \bs{X}_{i,P,k}+\dim\bs{X}_{j,P,k}\big]+N\delta(N),
\end{IEEEeqnarray}
where, 
(a) follows from Fano's inequality; 
(b) follows from the fact that $H\left(\overrightarrow{\bs{Y}}_i,\overleftarrow{\bs{Y}}_j|W_i,W_j\right)=0$; 
(c) follows from the fact that $H(Y|X)=H(X,Y)-H(X)$; and
(d) follows from the fact that $H\Big(\bs{X}_{j,C}$, $\bs{X}_{i,U}$, $\overleftarrow{\bs{Y}}_j|W_j$, $W_i$, $\overleftarrow{\bs{Y}}_i\Big)=0$.

\noindent
Plugging \eqref{EqdimXip}, \eqref{EqdimXicfjXidf}, \eqref{EqHXtopi2}, \eqref{EqdimYib}, and \eqref{EqdimYig} in \eqref{Eq2Ri+Rjc38} and after some trivial manipulations, the following holds in the asymptotic regime: 
\begin{IEEEeqnarray}{lcl}
\nonumber
2 R_i+R_j &\leqslant&  \max\left(\overrightarrow{n}_{ii},{n}_{ji} \right)+\left(\overrightarrow{n}_{ii}-{n}_{ij} \right)^+\\
\label{Eq2Ri+Rjc39}
& & +\max\Big(\left(\overrightarrow{n}_{jj}-{n}_{ji} \right)^+, n_{ij}, \overrightarrow{n}_{jj}-\left(\max\left(\overrightarrow{n}_{jj},{n}_{ji} \right)-\overleftarrow{n}_{jj}\right)^+\Big).
\end{IEEEeqnarray}

\noindent
This completes the proof of \eqref{Eq2Ri+Rj-V2}.

\end{appendices}

\clearpage

\bibliographystyle{IEEEtran}
\bibliography{IT-GT}

\begin{thebibliography}{1}
\providecommand{\url}[1]{#1}
\csname url@samestyle\endcsname
\providecommand{\newblock}{\relax}
\providecommand{\bibinfo}[2]{#2}
\providecommand{\BIBentrySTDinterwordspacing}{\spaceskip=0pt\relax}
\providecommand{\BIBentryALTinterwordstretchfactor}{4}
\providecommand{\BIBentryALTinterwordspacing}{\spaceskip=\fontdimen2\font plus
\BIBentryALTinterwordstretchfactor\fontdimen3\font minus
  \fontdimen4\font\relax}
\providecommand{\BIBforeignlanguage}[2]{{%
\expandafter\ifx\csname l@#1\endcsname\relax
\typeout{** WARNING: IEEEtran.bst: No hyphenation pattern has been}%
\typeout{** loaded for the language `#1'. Using the pattern for}%
\typeout{** the default language instead.}%
\else
\language=\csname l@#1\endcsname
\fi
#2}}
\providecommand{\BIBdecl}{\relax}
\BIBdecl

\bibitem{Bresler-ETT-2008}
G.~Bresler and D.~N.~C. Tse, ``The two user {G}aussian interference channel: A
  deterministic view,'' \emph{European Transactions on Telecommunications},
  vol.~19, no.~4, pp. 333--354, Apr. 2008.

\bibitem{Suh-TIT-2011}
C.~Suh and D.~N.~C. Tse, ``Feedback capacity of the {G}aussian interference
  channel to within 2 bits,'' \emph{IEEE Transactions on Information Theory},
  vol.~57, no.~5, pp. 2667--2685, May. 2011.

\bibitem{SyQuoc-TIT-2015}
S.-Q. Le, R.~Tandon, M.~Motani, and H.~V. Poor, ``Approximate capacity region
  for the symmetric {G}aussian interference channel with noisy feedback,''
  \emph{IEEE Transactions on Information Theory}, vol.~61, no.~7, pp.
  3737--3762, Jul. 2015.

\bibitem{Sahai-TIT-2013}
A.~Sahai, V.~Aggarwal, M.~Yuksel, and A.~Sabharwal, ``Capacity of all nine
  models of channel output feedback for the two-user interference channel,''
  \emph{IEEE Transactions on Information Theory}, vol.~59, no.~11, pp.
  6957--6979, 2013.

\bibitem{Tuninetti-ISIT-2007}
D.~Tuninetti, ``On interference channel with generalized feedback ({IFC-GF}),''
  in \emph{Proc. of International Symposium on Information Theory (ISIT)},
  Nice, France, Jun. 2007, pp. 2661--2665.

\bibitem{Perlaza-TIT-2015}
S.~M. Perlaza, R.~Tandon, H.~V. Poor, and Z.~Han, ``Perfect output feedback in
  the two-user decentralized interference channel,'' \emph{IEEE Transactions on
  Information Theory}, vol.~61, no.~10, pp. 5441--5462, Oct. 2015.

\bibitem{Perlaza-ISIT-2014a}
S.~M. Perlaza, R.~Tandon, and H.~V. Poor, ``Symmetric decentralized
  interference channels with noisy feedback,'' in \emph{Proc. IEEE Intl.
  Symposium on Information Theory (ISIT)}, Honolulu, HI, USA, Jun. 2014.

\bibitem{Cover-Book-1991}
T.~M. Cover and J.~A. Thomas, \emph{Elements of Information Theory}.\hskip 1em
  plus 0.5em minus 0.4em\relax Hoboken, NJ, USA: Wiley-Interscience, 1991.

\bibitem{Shannon-IRETIT-1956}
C.~E. Shannon, ``The zero-error capacity of a noisy channel,'' \emph{IRE
  Transactions on Information Theory}, vol.~2, no.~3, pp. 8--19, Sep. 1956.

\end{thebibliography}

\end{document}